\documentclass[twocolumn, trackchanges]{aastex701}
\usepackage{amsmath}

\newcommand{\teff}{$T_{\rm eff}$}

\newcommand{\logg}{$\log g$}

\newcommand{\feh}{$\rm{[Fe/H]}$}

\newcommand{\vmic}{$\rm{v_{t}}$} %dlo: fixed it, bro ;)

\newcommand{\sm}{$M_{\odot}$}

\newcommand{\vsini}{$v \sin i$}

\newcommand{\be}{\begin{equation}}
\newcommand{\ee}{\end{equation}}
\newcommand{\ben}{\begin{eqnarray}}
\newcommand{\een}{\end{eqnarray}}
\newcommand{\bfg}{\begin{figure}}
\newcommand{\efg}{\end{figure}}

\newcommand{\li}{$A$(Li)}
\newcommand{\ber}{$A$(Be)}
\newcommand{\fe}{$A$(Fe)}

\newcommand{\g}{$G$}

\newcommand{\Tc}{$T_{\rm{C}}$}

\usepackage{physics}
\usepackage{tabularx}
\usepackage{array}
\usepackage{placeins}

\shortauthors{Winnick et al.}
\begin{document}

\title{Extreme Lithium Depletion in Solar Twins: Challenging Non-Standard Mixing Models \footnote{This paper includes data gathered with the 6.5 m Magellan Telescopes located at Las Campanas Observatory (LCO), Chile. This research is based on data collected at the Subaru Telescope, which is operated by the National Astronomical Observatory of Japan (NAOJ). We are honored and grateful for the opportunity to observe the Universe from Maunakea, which has cultural, historical, and natural significance in Hawai'i.}}

\correspondingauthor{I. Winnick}

\author[0009-0004-8511-3721]{Isabelle Winnick}
\affiliation{Pomona College, 333 N College Way, Claremont, CA 91711, USA}
\affiliation{The Observatories of the Carnegie Institution for Science, 813 Santa Barbara Street, Pasadena, CA 91101, USA}
\email{iwaa2021@mymail.pomona.edu}

\author[orcid=0000-0001-9261-8366, gname=Jhon, sname=Yana Galarza]{Jhon Yana Galarza}
\altaffiliation{Carnegie Fellow}
\affiliation{The Observatories of the Carnegie Institution for Science, 813 Santa Barbara Street, Pasadena, CA 91101, USA}
\affiliation{Departamento de Astronomía, Facultad de Ciencias Físicas y Matemáticas Universidad de Concepción, Av. Esteban Iturra s/n Barrio Universitario, Casilla 160-C, Chile}
\email{jyanagalarza@carnegiescience.edu}

\author[0000-0001-6533-6179]{Henrique Reggiani}
\affiliation{Gemini South, Gemini Observatory, NSF's NOIRLab, Casilla 603, La Serena, Chile}
\email{henrique.reggiani@noirlab.edu}

\author[0000-0003-2059-470X]{Thiago Ferreira}
\affiliation{Department of Astronomy, Yale University, 219 Prospect St., New Haven, CT 06511, USA}
\email{thiago.dossantos@yale.edu}

\author[0000-0001-8365-5982]{Isabelle Baraffe}
\affiliation{Department of Physics and Astronomy, University of Exeter, Exeter, EX4 4QL, UK}
\affiliation{École Normale Supérieure de Lyon, CRAL (UMR CNRS 5574), Université de Lyon 1, 69007, Lyon, France}
\email{I.Baraffe@exeter.ac.uk}

\author[0000-0002-1387-2954]{Diego Lorenzo-Oliveira}
\affiliation{Laborat\'orio Nacional de Astrof\'isica, Rua Estados Unidos 154, 37504-364, Itajubá - MG, Brazil}
\email{diegolorenzo.astro@gmail.com}

\author[0009-0006-8679-6282]{Micaela Oyague}
\affiliation{Seminario Permanente de Astronomía y Ciencias Espaciales, Facultad de Ciencias Físicas, Universidad Nacional Mayor de San Marcos, Avenida Venezuela s/n, Lima 15081, Perú}
\email{micaela.oyague@unmsm.edu.pe}

\author{Rita Valle}
\affiliation{Seminario Permanente de Astronomía y Ciencias Espaciales, Facultad de Ciencias Físicas, Universidad Nacional Mayor de San Marcos, Avenida Venezuela s/n, Lima 15081, Perú}
\email{rita.valle@unmsm.edu.pe}

\author{Renzo Trujillo Diaz}
\affiliation{Seminario Permanente de Astronomía y Ciencias Espaciales, Facultad de Ciencias Físicas, Universidad Nacional Mayor de San Marcos, Avenida Venezuela s/n, Lima 15081, Perú}
\email{renzorjtd@gmail.com}

\author[0000-0003-0347-276X]{Nathan Leigh}
\affiliation{Departamento de Astronomía, Facultad de Ciencias Físicas y Matemáticas Universidad de Concepción, Av. Esteban Iturra s/n Barrio Universitario, Casilla 160-C, Chile}
\affiliation{Department of Astrophysics, American Museum of Natural History, Central Park West and 79th Street, New York, NY, 10024, USA}
\email{nleigh@amnh.org}

\author[0000-0003-4749-8486]{Matias Flores Trivigno}
\affiliation{Instituto de Ciencias Astron\'omicas, de la Tierra y del Espacio (ICATE), Espa\~na Sur 1512, CC 49, 5400 San Juan, Argentina}
\affiliation{Facultad de Ciencias Exactas, F\'isicas y Naturales, Universidad Nacional de San Juan, San Juan, Argentina}
\affiliation{Consejo Nacional de Investigaciones Cient\'ificas y T\'ecnicas (CONICET), Argentina}
\email{mattiasgft@gmail.com}

\author[0000-0002-7795-0018]{Ricardo López-Valdivia}
\affiliation{Universidad Nacional Autónoma de México, Instituto de Astronomía, AP 106, Ensenada 22800, BC, México}
\email{rlopezv@astro.unam.mx}

\author[0000-0002-0715-709X]{Gabriela Carvalho Silva}
\affiliation{Departamento de Astronomia do IAG/USP, Universidade de São Paulo, Rua do Matão 1226, 05508-090 São Paulo, SP, Brazil}
\email{gabriela.carvalho.silva@usp.br}

\author[0000-0002-5084-168X]{Eder Martioli}
\affiliation{Laborat\'orio Nacional de Astrof\'isica, Rua Estados Unidos 154, 37504-364, Itajubá - MG, Brazil}
\affiliation{Instituto Nacional de Pesquisas Espaciais/MCTI, Av. dos Astronautas, 1758, S\~{a}o Jos\'{e} dos Campos, SP, Brazil}
\email{emartioli@lna.br}

\author[0000-0002-0537-4146]{Hélio D. Perottoni}
\affiliation{Observatório Nacional (ON), MCTI, Rua Gal. José Cristino 77, Rio de Janeiro 20921-400, RJ, Brazil}
\email{hperottoni@on.br}

\begin{abstract}

Lithium (Li) is a powerful tracer of stellar mixing, gradually depleted in solar twins by non-standard transport below the convective zone. Here, we identify six new solar twins with exceptionally low Li levels that are not explained by current non-standard mixing models and, together with our previously reported anomalous solar twin HIP 8522, suggest a distinct population marked by a violent evolutionary past. Employing high-resolution spectra ($R=60,000 - 165,000$), we infer precise stellar parameters and chemical compositions, including Li abundances. We consider possible scenarios generating enhanced mixing, including planetary engulfment, blue straggler stars (BSSs), and early episodic accretion. Our planet engulfment simulations indicate that only one star may have engulfed an exoplanet, rapidly depleting Li via thermohaline convection. In the BSS scenario, radial velocity data rule out binary mass transfer, revealing no stellar companions but instead two new exoplanets. If these stars are field BSSs, a binary merger is likely though uncertain given that current BSS models focus mostly on stars in open clusters. Using pre-main-sequence episodic accretion models, we find that solar-mass stars can experience enhanced Li depletion without significant beryllium (Be) depletion. This is consistent with the Be abundances measured in two of our stars and represents the most plausible scenario, pending Be measurements for the remaining stars. These unique stars, together with HIP 8522, represent exceptional cases for testing stellar evolution models and probing internal mixing processes in Sun-like stars.

\end{abstract}



%%%%%%%%%%%%%%%%%%%%%%
\section{Introduction} 
\renewcommand{\thefootnote}{\arabic{footnote}} % Use numbers

Lithium (Li) and Beryllium (Be) can provide major insights into the workings of our Universe, from studies of Big Bang nucleosynthesis to stellar evolution models. $^7$Li, its most abundant isotope, was largely produced in the extreme environment of the Big Bang through nucleosynthesis and the subsequent decay of $^7$Be \citep[e.g.,][]{Wagoner:1967ApJ...148....3W, Burles:1999astro.ph..3300B, Steigman:2007ARNPS..57..463S, Khatri:2011AstL...37..367K, Fields:2020JCAP...03..010F}. The most well documented site of fresh Li enrichment is Galactic cosmic ray spallation, a process by which high-energy protons and neutrons collide with CNO nuclei in the interstellar medium, breaking these abundant heavier elements into Li and Be \citep{Reeves:1970Natur.226..727R, Meneguzzi:1971A&A....15..337M}. However, Li may have other possible sources such as novae \citep{Arnould:1975A&A....42...55A, Starrfield:1978ApJ...222..600S, Jose:1998ApJ...494..680J}, core-collapse supernovae \citep{Woosley:1990ApJ...356..272W}, Asymptotic Giant Branch (AGB) stars \citep{Scalo:1975ApJ...196..805S, Sackmann:1992ApJ...392L..71S, Smith:1989ApJ...345L..75S, Smith:1990ApJ...361L..69S}, and Red Giant Branch stars \citep{Cameron:1955ApJ...121..144C, Cameron:1971ApJ...164..111C}. We refer to \citet{Romano:2021A&A...653A..72R} and  \citet{Borisov:2024A&A...691A.142B} and references therein for more details on these possible sources.

Li is an exceptionally fragile element, destroyed in stellar interiors via the proton capture process $^7$Li$(p, \alpha$)$^{4}$He \citep{Burbidge:1957RvMP...29..547B} at temperatures reaching $2.5 \times 10^{6}$ K. Measurements of Li, along with Be and boron---destroyed at 3.5 and $5 \times 10^{6}$ K, respectively---can serve as a powerful probe of stellar astrophysics by tracing the depth of material transport and providing constraints on mixing mechanisms \citep[e.g.,][]{Herbig:1965ApJ...141..588H, Wallerstein:1965ApJ...141..610W, Conti:1966ApJ...146..383C, Merchant:1966ApJ...143..336M, Danziger:1967ApJ...150..733D, 
Wallerstein:1969ARA&A...7...99W,
Zappala:1972ApJ...172...57Z, Boesgaard:1986ApJ...302L..49B, Boesgaard:1998ApJ...492..727B, Melendez:2010Ap&SS.328..193M, Carlos:2019MNRAS.tmp..667C, Tucci:2015A&A...576L..10T, Boesgaard:2022ApJ...927..118B, Boesgaard:2023ApJ...943...40B, Martos:2023MNRAS.522.3217M,  Reggiani:2025ApJ...984..108R}. 

Standard stellar models predict that solar-mass stars deplete some of their surface Li content during the pre-main sequence (PMS) when the convective zone is extended and Li is pulled further into the stellar interior, with the extent of Li depletion decreasing with mass and increasing with age and metallicity \citep{Cummings:2017AJ....153..128C, Jeffries:2017MNRAS.464.1456J}. During the main-sequence (MS) phase, the base of the convective envelope of solar-mass stars reaches only $2 \times 10^{6}$  K. Therefore, standard stellar models, which consider only convective transport, do not predict significant Li burning during the MS lifetime \citep{Dantona:1984A&A...138..431D, Jeffries:2021MNRAS.500.1158J}. 

However, in our own solar system, the Sun is depleted in Li by over two orders of magnitude compared to the pristine meteoritic value (representative of the Sun’s Li composition at the Zero Age Main Sequence, ZAMS) and is only one-hundredth as abundant as on Earth \citep[e.g.,][]{Greenstein:1951, Anders:1989, Asplund2009}. This suggests that there must be an additional mechanism, referred to as `extra mixing', responsible for transporting and burning Li below the base of the convective zone, where temperatures can reach the Li destruction threshold. Numerous measurements of other solar-mass stars have confirmed their gradual Li depletion after the ZAMS. First suggested by \citet{Herbig:1965ApJ...141..588H}, early studies supported that Li depletion is a promising indicator of age in MS stars \citep{Conti:1966ApJ...146..383C, Merchant:1966ApJ...143..336M, Danziger:1967ApJ...150..733D, Wallerstein:1969ARA&A...7...99W, Skumanich:1972ApJ...171..565S, Zappala:1972ApJ...172...57Z}. Furthermore, recent observational surveys of solar twins\footnote{Following the criteria adopted by \citet{Ramirez:2014A&A...572A..48R}, we define solar twins as stars with parameters relative to the Sun of \teff\ = 5777 $\pm$ 100 K, \logg\ = 4.44 $\pm$ 0.10 dex, and \feh\ = 0.00 $\pm$ 0.10 dex.}, demonstrate a particularly strong correlation between Li and age.
\citep{Melendez:2010Ap&SS.328..193M, Monroe2013, Carlos2016, YanaGalarza2016, Carlos:2019MNRAS.tmp..667C, Martos:2023MNRAS.522.3217M, Rathsam2023}.

%\footnote{Solar twins are stars with stellar parameters within \teff\ = 5777 $\pm$ 100 K, \logg\ = 4.44 $\pm$ 0.1 dex, and \feh\ = 0.0 $\pm$ 0.1 dex \citep{Ramirez:2014A&A...572A..48R}.}

Although standard stellar models fail to explain this post-ZAMS Li depletion, non-standard models incorporate a variety of additional mixing mechanisms to account for the transport of Li below the base of the convective envelope in solar-mass stars. Some proposed mixing mechanisms include rotationally induced mixing \citep{Pinsonneault1989, Chaboyer1995, Charbonnel2005, DoNascimento2009}, rotation-driven turbulent diffusion \citep{Denissenkov2010}, microscopic diffusion and gravitational settling \citep{Michaud2004}, convective overshooting of fast and narrow downflows \citep{Freytag1996, Schlattl1999, Xiong2007, Xiong:2007ChA&A..31..244X, Baraffe:2017ApJ...845L...6B}, internal gravity waves \citep{Charbonnel2005}, and convective settling \citep{Andrassy2015}. Other alternative sources of depletion are the removal of Li from the stellar atmosphere via mass loss \citep{Hobbs1989}, and thermohaline mixing, by which metal-rich accreted material destabilizes the thermohaline convection and drives mixing into the radiative interior \citep{Theado2012, Sevilla2022}. 

To better understand the complex behavior of Li in the Universe, it is often compared with stellar metallicity \citep[e.g., see Fig. 6 of][]{Ryan:2001ApJ...549...55R}. Metal-poor MS stars preserve primordial Li, with a nearly constant abundance of \li\ $\sim$ 2.32 dex up to [Fe/H] $\lesssim -0.5$ dex. This is known as the Spite Plateau \citep{Spite:1982Natur.297..483S}, and it suggests that most Li was produced during the Big Bang. At higher metallicities ([Fe/H] $>-0.5$ dex), the \li\ shows large scatter, from $\sim$1.0 dex to $\sim$3.5 dex, likely due to a combination of galactic Li production and stellar depletion from internal mixing. Remarkably, some stars with metallicity [Fe/H] $\lesssim -0.5$ dex show significant Li depletion (\li\ $\sim$ 1 dex) below the Spite Plateau. Over the last three decades, these stars have been recognized as blue stragglers (BSSs) of various types \citep{Pritchet:1991ApJ...373..105P, Hobbs:1991PASP..103..431H, Glaspey:1994AJ....108..271G, Carney:2005AJ....129..466C}. Additionally, \citet{Boesgaard:2007ApJ...667.1196B} found Be depletion in these low-Li stars. To our knowledge, they represent the first low-Li stars identified in the Milky Way's halo and thick disk, preceding the findings we present in this paper.

BSSs were first identified by \citet{Sandage:1953AJ.....58...61S} in the globular cluster (GC) M3. Since their discovery, BSSs have been recognized as some of the most fascinating class of stars observed in the Galaxy. A defining characteristic of BSSs is their position on the colour-magnitude diagram (CMD), where they appear brighter and bluer than the main-sequence turn-off. Because BSSs are typically more massive than main-sequence stars, their formation is often attributed to processes that increase their initial mass \citep{Shara:1997ApJ...489L..59S, Gilliland:1998ApJ...507..818G, DeMarco:2005ApJ...632..894D, Beccari:2006ApJ...652L.121B}. The most widely supported mechanisms for their formation include Roche lobe overflow binary mass transfer \citep{McCrea:10.1093/mnras/128.2.147, Zinn:1976ApJ...209..734Z}, either mass transfer or merger within the inner binary of a triple system \citep{Perets:2009ApJ...697.1048P, Portegies:2019ApJ...876L..33P, Ferraro:2020RLSFN..31...19F, Leigh:2020MNRAS.496.1819L, Wang:2024arXiv241010314W}, and direct stellar collisions \citep{Hills:1976ApL....17...87H, Chatterjee:2013ApJ...777..106C, Perets:2015ASSL..413..251P}. In dense environments like GCs, mass transfer and collisions might act simultaneously, with collisions dominating in the central region and mass transfer in the outer regions \citep{Leigh_2007}. In open clusters (OCs) and galactic fields, mass transfer and merger mechanisms dominate over collisions \citep{Ferraro:2020RLSFN..31...19F}. 

These stars are not exclusive to clusters. \citet{Preston:1994AJ....108.2267P} identified the first field blue straggler (FBSS), a metal-poor star bluer than the main-sequence turnoff of GCs, and therefore analogous to cluster BSS. Later, \citet{Preston:2000AJ....120.1014P} found that about 60\% of a sample of 62 FBSSs are binary systems, suggesting that mass transfer is the dominant formation channel, a conclusion later supported by \citet{Sneden_2003}. Abundance topics are of particular interest for FBSSs. Li and Be are extremely depleted, likely due to mixing processes resulting from collisions, mergers, or mass transfer \citep[e.g.,][]{Schirbel:2015AA...584A.116S, Rathsam:2025AA...693A..26R}. Among all BSS formation channels, binary mass transfer leaves the most identifiable chemical signatures. For example, some BSSs show significant depletion in carbon and oxygen \citep{Ferraro:2006ApJ...647L..53F, Lovisi:2010ApJ...719L.121L}, while others display surface enrichment in $s$-process elements such as barium, produced during the thermally pulsing phase of AGB evolution \citep{Milliman:2015AJ....150...84M, Nine:2024ApJ...970..187N}.

However, the three formation channels of BSSs are not the only pathways to Li depletion. A more recent proposal involves early episodic accretion, a process linked to the pre-ZAMS. In the simplest model of star formation that involves steady accretion, stars form in dense regions of molecular clouds, and for low-mass stars this begins when a small protostar gathers gas from the surrounding collapsing cloud at accretion rates of $\sim2\times 10^{-6}$ \sm\ yr$^{-1}$. Because the cloud contains significant angular momentum, the gas does not fall straight onto the protostar. Instead, it forms a rotating disk around the young star \citep{Pringle1981, Shu:1987ARA&A..25...23S}. For the gas in the disk to move inward and reach the protostar, angular momentum needs to be transported outward via a mechanism that creates viscosity within the disk such as gravitational instabilities \citep{Lodato:2004MNRAS.351..630L} and magneto-rotational instability \citep{Balbus:1998RvMP...70....1B}. As the gas moves inward, it loses gravitational energy, which is converted into radiation and heat.

Even though the general concept of accretion is well understood, the detailed process by which gas falls onto young stars remains unclear. \citet{Kenyon:1990AJ.....99..869K} found that most protostars have luminosities significantly lower than expected from steady accretion. This so-called `luminosity problem' was later confirmed by data from the Spitzer Space Telescope \citep{Werner:2004ApJS..154....1W}, which revealed a large number of protostars with luminosities well below theoretical predictions. For example, Fig. 14 of \citet{Evans:2009ApJS..181..321E} shows the luminosity distribution of Class 0/I sources, with 59\% having luminosities below 1.6 $L_\odot$. 

A proposed solution is non-steady (episodic) accretion, as suggested by \citet{Kenyon:1990AJ.....99..869K}, \citet{Kenyon:1994AJ....108..251K}, and \citet{Kenyon:1995ApJS..101..117K}. In this scenario, material first accumulates in the circumstellar disk and is then accreted onto the protostar in short-lived, high-accretion bursts, such as FU Orionis-type events, with accretion rates of $\dot{M} > 10^{-5}$ \sm\ yr$^{-1}$. Several mechanisms have been proposed to explain episodic accretion, including close encounters in binary systems \citep{Bonnell:1992ApJ...401L..31B} or in young clusters \citep{Pfalzner:2008A&A...487L..45P}, thermal ionization instability \citep{Lin:1985prpl.conf..981L, Hartmann:1985ApJ...299..462H, Bell:1994ApJ...427..987B}, gravitational instability \citep{Vorobyov:2005ApJ...633L.137V, Vorobyov:2006ApJ...650..956V},  a combination of magneto-rotational instability (MRI) and gravitational instability \citep{Zhu:2009ApJ...694.1045Z}, magnetic walls
that form in some models \citep{Tassis:2005ApJ...618..783T}, and turbulence in the infalling envelope \citep{Offner:2008ApJ...686.1174O}.

Expanding on the idea of episodic accretion, \citet{Baraffe:2009ApJ...702L..27B} and \citet{Baraffe:2010A&A...521A..44B} conducted a detailed study of the impact of episodic accretion on the evolution of low-mass stars and brown dwarfs, providing a comprehensive framework for understanding how such accretion events can significantly alter the internal structure and evolutionary paths of these objects, in particular for Li depletion. In all their simulations, they found that episodic accretion leads to objects with smaller radii and higher temperatures compared to those formed through continuous accretion (see Figures 1 and 2 in \citealp{Baraffe:2010A&A...521A..44B}). For stars that end with masses greater than 0.35 \sm, the higher central temperatures lead to the development of a radiative core. The more compact and hotter these objects are, the higher the temperature at the base of the convective envelope, which in turn results in more significant Li depletion. For example, in a 1 \sm\ star formed via episodic accretion with an initial mass of 10 $M_{\rm{Jup}}$, the temperature at the base of the convective zone reaches $\sim 7 \times 10^{6}$ K, resulting in complete Li depletion at ages $<1$ Myr (see Fig. 4 in \citealp{Baraffe:2010A&A...521A..44B}). In \citet{Baraffe:2017A&A...597A..19B}, the authors extended their previous work by coupling a stellar evolution code with the two-dimensional (2D) hydrodynamics code from \citet{Vorobyov:2010ApJ...719.1896V, Vorobyov:2013A&A...557A..35V}. Their results confirmed earlier findings that episodic accretion produces more compact and hotter stellar structures, leading to enhanced Li depletion at ZAMS.

More recently, three unexpected low-Li stars have been identified: one in the thick disk (HIP 38908) and two in the thin disk (HIP 10725 and HIP 8522). HIP 38908 and HIP 10725 have masses of 1.02 \sm\ and 0.96 \sm\ and [Fe/H] values of $-0.317$ dex and $-0.170$ dex, respectively. HIP 8522 is the only solar twin among them, with [Fe/H] $\sim 0.01$ dex and $M \sim 1.01$ \sm. Interestingly, these stars show even stronger \li\ depletion than those previously reported below the Spite Plateau. With abundances below \li\ $<1$ dex, HIP 10725 and HIP 38908 were reported as FBSSs by \citet{Schirbel:2015AA...584A.116S} and \citet{Rathsam:2025AA...693A..26R}, respectively. Their low \li\ and \ber\ abundances have been linked to mass transfer from an AGB companion or stellar mergers. The solar twin HIP 8522 \citep{Yana_Galarza_2025} is the third and most puzzling case. Despite its young age ($<1$ Gyr), its \li\ is unusually low, with an upper limit of $\sim0.8$ dex, far below the expected $\sim3$ dex for such a young star. The mechanism behind its Li depletion is unknown, with both FBSS via stellar merger and early episodic accretion remaining viable explanations. 

In this paper, we report the discovery of six additional low-Li stars that together to HIP 8522, appear to form a distinct population, further testing the limits of current mixing models. These stars challenge both standard and non-standard stellar evolution models and offer valuable insight into mechanisms beyond typical `extra mixing' processes. As in the case of HIP 8522, we explore BSSs and early episodic accretion as potential channels for \li\ depletion, but also test the viability of a planet engulfment scenario.

In Section \ref{sec:obs}, we describe the sample selection and observations. Section \ref{sec:stellaparam} presents the determination of stellar parameters, while Section \ref{sec:chem} covers the determination of chemical composition and Section \ref{sec:ages} discusses the inferred age and mass. Section \ref{sec:li} presents the spectral synthesis of Li. Section \ref{sec:companions} describes the spectral energy distribution, kinematics, and radial velocity analysis for the detection of unresolved companions. 
In Section \ref{sec:disc}, we discuss the possible channels of Li depletion, and in Section \ref{sec:conclusions} we provide a summary and conclusions. 

%%%%%%%%%%%%%%%%%%%%%%%%%%%%%%%%%%%%%%%%%%%
\section{Sample Selection and Observations}
\label{sec:obs}
Our sample originates from the solar twin candidates reported in \cite{Yana_Galarza:2021MNRAS.504.1873Y}. In brief, these candidates were selected from \textit{Gaia} data \citep{GaiaDR3:2023A&A...674A...1G} by applying known color constraints for solar twins (see Table 1 of \citealp{Yana_Galarza:2021MNRAS.504.1873Y}). Most candidates were observed with the Robert G. Tull Coud\'e Spectrograph (TS23; \citealp{Tull:1995PASP..107..251T}), mounted on the 2.7-m Harlan J. Smith Telescope at McDonald Observatory. Using  this instrument, we detected our first low-Li star, HIP 8522 \citep{Yana_Galarza_2025}. Two additional stars (HD 221103, HD 236254), also showing significant Li depletion, were identified with the TS23. Similarly to HIP 8522, both stars were observe with the High Dispersion Spectrograph \citep[HDS;][]{Noguchi:2002PASJ...54..855N} at the National Astronomical Observatory of Japan. 

In addition, we detected significant Li depletion in four other stars, whose data were acquired with the High Accuracy Radial velocity Planet Searcher \cite[HARPS;][]{Mayor:2003Msngr.114...20M} and made publicly available in the European Southern Observatory (ESO) Science Archive\footnote{\url{https://archive.eso.org/wdb/wdb/adp/phase3_spectral/form?phase3_collection=HARPS}}. Because HARPS spectra do not cover wavelengths beyond 7000~\AA --- where oxygen is essential for our chemical analysis --- we obtained additional data using the Magellan Inamori Kyocera Echelle \cite[MIKE;][]{Bernstein:2003SPIE.4841.1694B, Shectman:2003SPIE.4837..910S} spectrograph, which covers wavelengths up to $\sim10,000$ \AA. In the following subsections, we present detailed information about the different instruments we used and their respective configurations. 

\subsection{Las Campanas Observatory}
We collected spectra for HIP 91700, HIP 93858, HIP 116937, and HIP 8522 with the Magellan Inamori Kyocera Echelle (MIKE; \citealp{Bernstein:2003SPIE.4841.1694B, Shectman:2003SPIE.4837..910S}) spectrograph on the 6.5-m Magellan Clay Telescope at Las Campanas Observatory (LCO). Data were taken on 2025 June 27. We used the $0.''35$ slit with standard blue and red grating azimuths, yielding spectra between 3200 and $10,000$~\AA. This narrow slit provides a spectral resolution ($R = \lambda/\Delta \lambda$) of $83,000$ and $65,000$ in the blue and red arms, respectively.

We collected calibration data (quartz and milky flat fields, and ThAr lamp frames) in the afternoon before the night of observation. We also took ThAr lamp frames every 1–2 hours to ensure an optimal wavelength solution. The MIKE spectra were reduced using the latest version of the CarPy\footnote{\url{https://code.obs.carnegiescience.edu/pipelines/carnegie-python-distribution}} software package \citep{Kelson:2000ApJ...531..159K, Kelson:2003PASP..115..688K, Kelson:2014ApJ...783..110K}.

\begin{deluxetable*}{lcccccccc} 
\tablecaption{Basics of the instruments and observation log. \label{tab:instruments}} 
\tablewidth{0pt} 
\tablehead{
\colhead{\textit{Gaia} DR3} &
\colhead{Other ID} & 
\colhead{R.A.} & 
\colhead{Decl.} & 
\colhead{$G$} & 
\colhead{RUWE} &
\colhead{Instrument} & 
\colhead{$R$} & 
\colhead{S/N}  \\ 
\colhead{} &
\colhead{} & 
\colhead{(deg)} & 
\colhead{(deg)} &
\colhead{(mag)} &
\colhead{} &
\colhead{} & 
\colhead{$\lambda/\Delta \lambda$} & 
\colhead{at $6500$\,\AA}
}
\startdata
3550148743432500992 & HIP 53087  & 162.912 & $-22.071$ & 7.68 & 0.972 & HARPS/MIKE & 115,000/83,000 & 300/350 \\
6735442725215002112 & HIP 91700  & 280.509 & $-34.466$ & 7.77 & 1.017 & HARPS/MIKE & 115,000/83,000 & 330/360 \\
6718894388002453120 & HIP 93858  & 286.717 & $-37.812$ & 5.98 & 0.947 & HARPS/MIKE & 115,000/83,000 & 320/370 \\
6498197630933903872 & HIP 116937 & 355.554 & $-53.423$ & 7.82 & 0.978 & HARPS/MIKE & 115,000/83,000 & 310/390 \\
2869747927140286208 & HD 221103  & 352.234 & $+30.150$ & 8.75 & 0.980 & HDS/TS23   & 165,000/60,000 & 450 \\
1993887478540418432 & HD 236254  & 359.331 & $+55.143$ & 8.67 & 0.935 & HDS/TS23   & 165,000/60,000 & 480 \\
\hline
\multicolumn{9}{c}{\textbf{Low-Li stars known in the literature}} \\
\hline
297548019938221056$^{a}$ & HIP 8522    & 27.479  & $+25.746$ & 8.40 & 0.986 & HDS/MIKE & 165,000/83,000 & 400 \\
5291028181119851776$^{b}$ & HIP 38908  & 119.450 & $-60.302$ & 5.44 & 0.921 & HARPS & 115,000 & 350 \\
4938221494201968512$^{c}$ & HIP 10725  & 34.510  & $-50.818$ & 8.34 & 2.496 & UVES & 110,000 & 300 \\
\enddata 
\tablenotetext{a}{Reported by \citet{Yana_Galarza_2025}.}
\tablenotetext{b}{Reported by \citet{Rathsam:2025AA...693A..26R}.}
\tablenotetext{c}{Reported by \citet{Schirbel:2015AA...584A.116S}, but not analyzed in this study.}
\end{deluxetable*}

\subsection{McDonald Observatory}
 HD 221103 and HD 236254 were observed for the first time at high-resolution ($R = 60,000$) using the Robert G. Tull Coudé Spectrograph (hereafter TS23; \citealp{Tull:1995PASP..107..251T}), mounted on the 2.7-m Harlan J. Smith Telescope at McDonald Observatory, on 2020 August 28, and 2020 September 22, respectively. Both stars were previously identified as solar twins in the Inti catalog \citep{Yana_Galarza:2021MNRAS.504.1873Y}. Since there is no official pipeline to reduce TS23 spectra, we developed a script in \textsc{PYRAF} \citep{PyRAF:2012ascl.soft07011S} to perform the standard procedures: bias subtraction, flat fielding, order extraction, and wavelength calibration. The resulting spectra were then combined using the \textsc{scombine} task in \textsc{IRAF\footnote{\textsc{IRAF} is distributed by the National Optical Astronomy Observatory, operated by the Association of Universities for Research in Astronomy, Inc., under a cooperative agreement with the National Science Foundation.}}. The TS23 spectral coverage spans approximately 3700–9900~\AA.

\subsection{National Astronomical Observatory of Japan}
We collected spectra for two stars with the HDS \citep{Noguchi:2002PASJ...54..855N}, Subaru program (S22B-TE010-G), under the Gemini time exchange program (PI: Jhon Yana Galarza, BR\_2022B\_010), on the 8.2-m Subaru Telescope of the National Astronomical Observatory of Japan (NAOJ), located at the Maunakea summit. Data were taken on 2023 January 11. The observations were made using the image slicer IS \#3, with a slit of $0.2''\times×3$, providing a maximum spectral resolution of $R = 165,000$. We used the standard setup Yc (StdYc), which delivers a spectral coverage ranging from approximately 4400 \AA\ to 7050 \AA.  

We reduced the HDS data using the Image Reduction and Analysis Facility (\textsc{IRAF}) framework\footnote{\url{https://www.naoj.org/Observing/Instruments/HDS/hdsql-e.html}}, performing flatfield and bias corrections, spectral order extractions, wavelength calibration, and removing the fringe pattern from the red side of the spectra. 

\subsection{La Silla Observatory}
We found publicly available HARPS \citep{Mayor:2003Msngr.114...20M} spectra for HIP 53087, HIP 91700, HIP 93858, and HIP 116937 from 26 ESO programs. The reduced spectra were downloaded from the ESO archive. HARPS is a high-resolution spectrograph ($R \sim 115,000$) covering a spectral range from 3782 to 6913~\AA. It is mounted on the ESO 3.6-m telescope at La Silla Observatory. The observations of our sample were taken between 2003 and 2022 and were automatically reduced by the HARPS pipeline after each exposure.

\subsection{Solar Spectra and Data Treatment}
The determination of stellar parameters and chemical abundances follows a strictly differential analysis between the Sun and the target stars. To perform this, we obtained a solar spectrum using each instrument. For MIKE, we observed the solar spectrum via light reflected from the asteroid Vesta. For HDS and HARPS, we used Ganymede and Vesta, respectively. For TS23, we observed the light reflected from the Moon. On average the solar spectra have a signal-to-noise ratio (S/N) of 400 per pixel at $\sim$6500 \AA. 

The MIKE, HDS, and TS23 spectra were corrected by their radial and barycentric velocities using \textsc{iSpec}\footnote{\url{https://www.blancocuaresma.com/s/iSpec}} \citep{Blanco:2014A&A...569A.111B, Blanco:2019MNRAS.486.2075B}. For the HARPS instrument, the radial velocities were provided by the HARPS pipeline. The rest frame spectra were individually normalized in pixel space with the \textsc{continuum} task in \textsc{IRAF}, by fitting the spectra with low degree spline functions (mostly third degree). We co-added the spectra after normalization with the \textsc{IRAF scombine} task using average values. The S/N and the spectral resolution $R$ achieved with each instrument are listed in Table \ref{tab:instruments}.

\begin{deluxetable*}{lcccccccr} 
\tablecaption{Spectroscopic and photometric stellar parameters for our sample. $^{a}$Photometric temperature. $^{b}$Trigonometric surface gravity. \label{tab:fundamental parameters}} 
\tablewidth{0pt}
\setlength{\tabcolsep}{4pt}
\tablehead{
\colhead{Designation} &
\colhead{\teff} & 
\colhead{\logg} & 
\colhead{\feh} & 
\colhead{\vmic} & 
\colhead{Age} &
\colhead{Mass} &  
\colhead{\teff$^{a}$} & 
\colhead{\logg$^{b}$} \\ 
\colhead{} &
\colhead{(K)} & 
\colhead{(dex)} & 
\colhead{(dex)} &
\colhead{(km s$^{-1}$)} &
\colhead{(Gyr)} &
\colhead{(\sm)} & 
\colhead{(K)} &
\colhead{(dex)}
}
\startdata
HIP 53087  & $5633\pm10$ & $4.432\pm0.017$ & $+0.095\pm0.010$ & $0.87\pm0.02$ & $4.6\pm0.7$ & $0.98\pm0.01$ & $5621\pm30$ & $4.480 \pm0.023$  \\
HIP 91700  & $5533\pm10$ & $4.443\pm0.016$ & $-0.008\pm0.010$ & $0.82\pm0.01$ & $5.3\pm0.7$ & $0.93\pm0.01$ & $5556\pm32$ & $4.520\pm0.023$ \\
HIP 93858  & $5671\pm10$ & $4.437\pm0.013$ & $+0.135\pm0.010$ & $0.93\pm0.01$ & $3.8\pm0.7$ & $1.01\pm0.01$ & $5634\pm38$& $4.479\pm0.023$ \\
HIP 116937 & $5659\pm10$ & $4.462\pm0.016$ & $+0.037\pm0.010$ & $0.85\pm0.01$ & $3.1\pm0.6$ & $0.99\pm0.01$ & $5653\pm34$ & $4.507\pm0.022$ \\
HD 221103 & $5884\pm10$  & $4.435\pm0.020$ & $+0.119\pm0.010$ & $1.10\pm0.02$ & $2.7\pm0.5$ & $1.08\pm0.01$ & $5874\pm34$ & $4.428\pm0.023$  \\
HD 236254 & $5632\pm10$  & $4.440\pm0.020$ & $+0.023\pm0.010$ & $0.90\pm0.02$ & $4.9\pm0.6$ & $0.96\pm0.01$ &  $5649\pm30$& $4.483\pm0.023$  \\
HIP 8522$^{c}$ & $5729\pm7$ & $4.532\pm0.016$ & $+0.005\pm0.010$ & $1.08 \pm 0.02$  & $<1$  & $1.01\pm0.01$ &  $5707\pm36$& $4.537\pm0.020$  \\
HIP 38908$^{d}$ & $5962\pm10$ & $4.480\pm0.022$ & $-0.317\pm0.010$ & $1.18\pm0.02$ & $4.8\pm0.6$ & $1.02\pm0.01$ &  $5986\pm50$& $4.403\pm0.023$  \\
HIP 10725$^{e}$ & $5777\pm16$ & $4.450\pm0.050$ & $-0.170\pm0.010$ & $0.97\pm0.04$ & $5.4\pm0.8$ & $0.96\pm0.01$ &  $5760\pm28$& $4.467\pm$ 0.023 \\
\enddata 
\tablenotetext{c}{Parameters from \citet{Yana_Galarza_2025}.}
\tablenotetext{d}{Parameters reported in \citet{Shejeelammal:2024AA...690A.107S}, but updated in this work.}
\tablenotetext{e}{Parameters from \citet{Schirbel:2015AA...584A.116S}.}
\end{deluxetable*}

%%%%%%%%%%%%%%%%%%%%%%%%%%%%
\section{Stellar Parameters}
\label{sec:stellaparam}
The stellar parameters, namely effective temperature (\teff), surface gravity (\logg), metallicity (\feh), and microturbulent velocity (\vmic), were estimated through differential spectroscopic analysis relative to the Sun, and their errors were derived from the uncertainty of the spectroscopic equilibrium. The method is based on measurements of the equivalent widths (EWs) of \ion{Fe}{1} and \ion{Fe}{2} atomic transitions by choosing pseudo-continuum regions of 6 \AA. These measurements are obtained by fitting Gaussians to the line profiles using the \textsc{KAPTEYN} kmpfit package \citep{KapteynPackage}. We adopted the line list from \citet{Melendez:2014ApJ...791...14M}. The measurement is performed entirely manually line by line to achieve maximum precision. We used only iron lines with EW $<$ 130 m\AA\ to prevent saturation effects in the determination of stellar parameters.

The \ion{Fe}{1} and \ion{Fe}{2} abundances were computed using the Qoyllur-quipu ($\mathrm{q}^{2}$) Python code \citep{Ramirez:2014A&A...572A..48R}\footnote{\url{https://github.com/astroChasqui/q2_tutorial}}, which is configured to employ the Kurucz's \textsc{ODFNEW} model atmospheres \citep{Castelli:2003IAUS..210P.A20C} and the 2019 version of the local thermodynamic equilibrium (LTE) radiative transfer code \textsc{MOOG} \citep{Sneden:1973PhDT.......180S} to reach the spectroscopic equilibrium. The iron abundance is computed using the curve-of-growth method through the \textsc{abfind} driver of \textsc{moog}. The stellar parameter uncertainties were estimated following \citet{Epstein:2010ApJ...709..447E} and \citet{Bensby:2014A&A...562A..71B}, and include both observational errors and parameter degeneracies. The same approach was used to estimate the abundance uncertainties, with further details provided in those papers. Table \ref{tab:fundamental parameters} lists the stellar parameters inferred from the MIKE, HDS and HARPS spectra. 

While \citet{Shejeelammal:2024AA...690A.107S} reported stellar parameters for HIP 38908, we determined them independently to avoid systematics in our chemical analysis. We adopted the stellar parameters of HIP 8522 and HIP 10725 from \citet{Yana_Galarza_2025} and \citet{Schirbel:2015AA...584A.116S}, respectively. The latter is not included in our analysis, as it is a well-established FBSS. The stellar parameters of the solar twins HD 221103 and HD 236254 were previously reported by \citet{Yana_Galarza:2021MNRAS.504.1873Y}. Using new data from the HDS instrument, we derived more precise parameters, which are consistent with the earlier values.

We also applied independent methods to verify the consistency of our spectroscopic stellar parameters. We calculated trigonometric surface gravity values following Equation (3) of \citet{Yana_Galarza:2021MNRAS.504.1873Y}, which incorporates \textit{Gaia} DR3 parallaxes, Johnson $V$ magnitudes \citep{Kharchenko:2001KFNT...17..409K}, bolometric corrections from \citet{Melendez:2006ApJ...641L.133M}, and stellar masses obtained through isochrone fitting (see Section \ref{sec:ages}). Table \ref{tab:fundamental parameters} shows no significant discrepancies between the trigonometric and spectroscopic \logg\ values for any set of stellar parameters. We also estimated photometric effective temperatures using the InfraRed Flux Method (IRFM) as described by \citet{Casagrande:2021MNRAS.507.2684C} through the Color-\teff\ routine \textsc{colte}\footnote{\url{https://github.com/casaluca/colte}}, which employs \textit{Gaia} photometry. We found an average difference of 19 K between photometric and spectroscopic \teff\ using \textit{Gaia} data, which supports the excitation equilibrium. These additional analyses confirm the consistency of the determination of spectroscopic stellar parameters.

\begin{figure*}
    \centering
    \begin{tabular}{ccc}
        \includegraphics[width=0.32\textwidth]{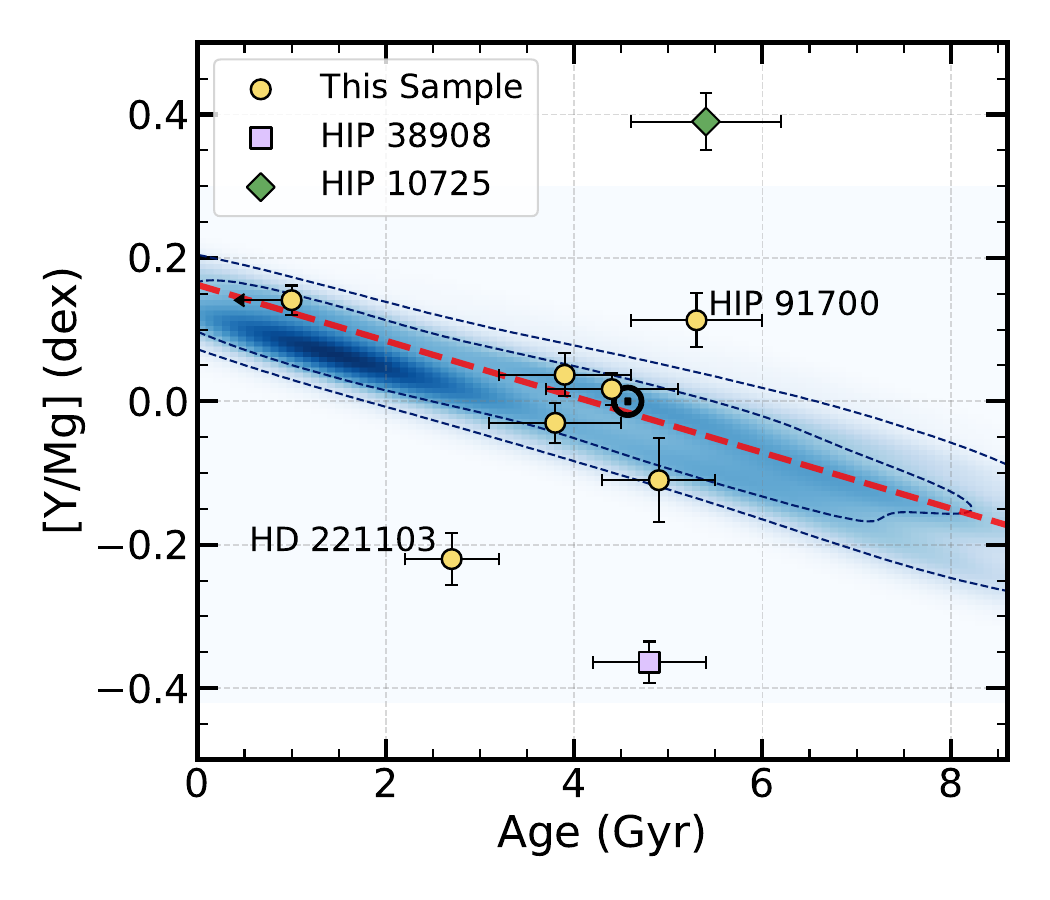} &
        \includegraphics[width=0.32\textwidth]{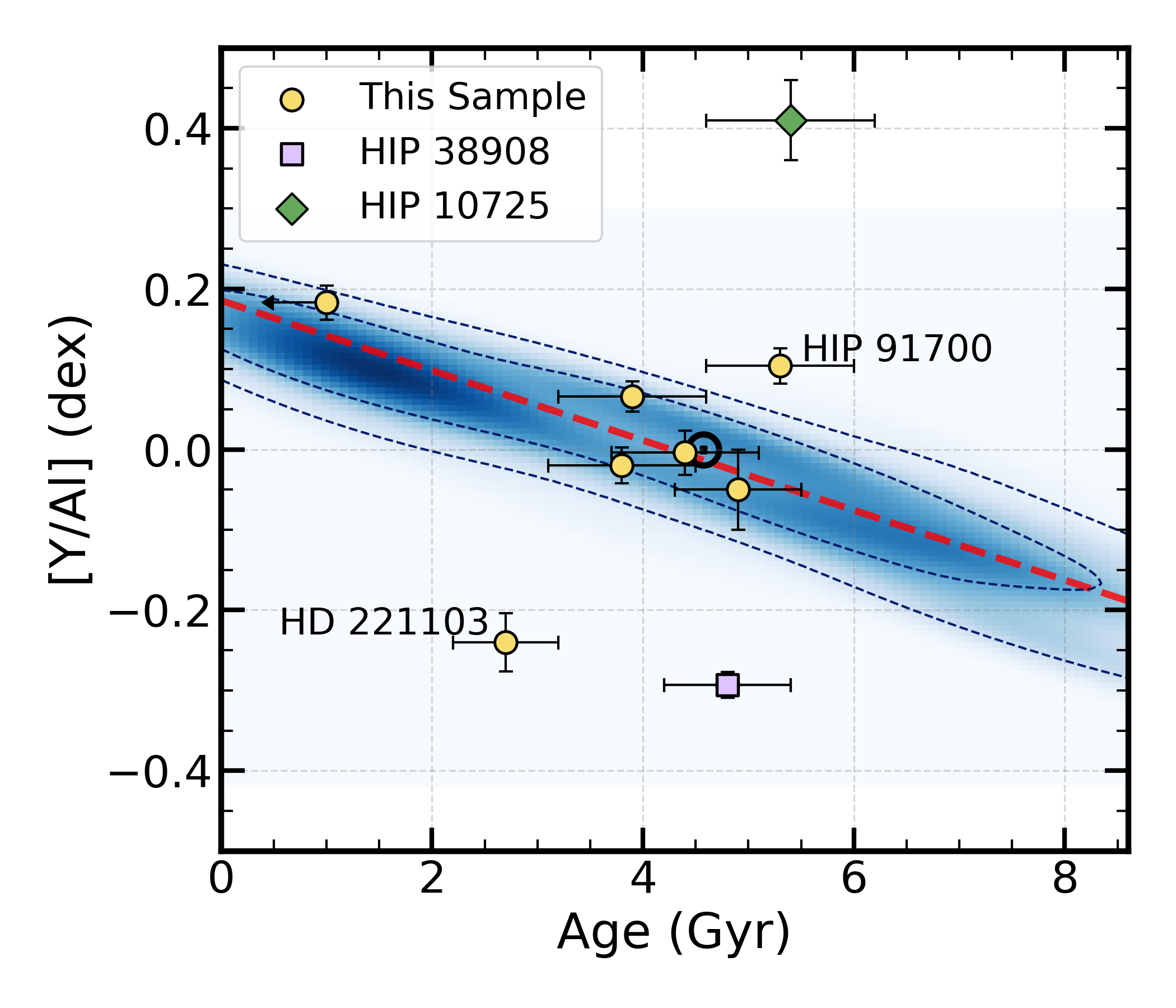} &
        \includegraphics[width=0.315\textwidth]{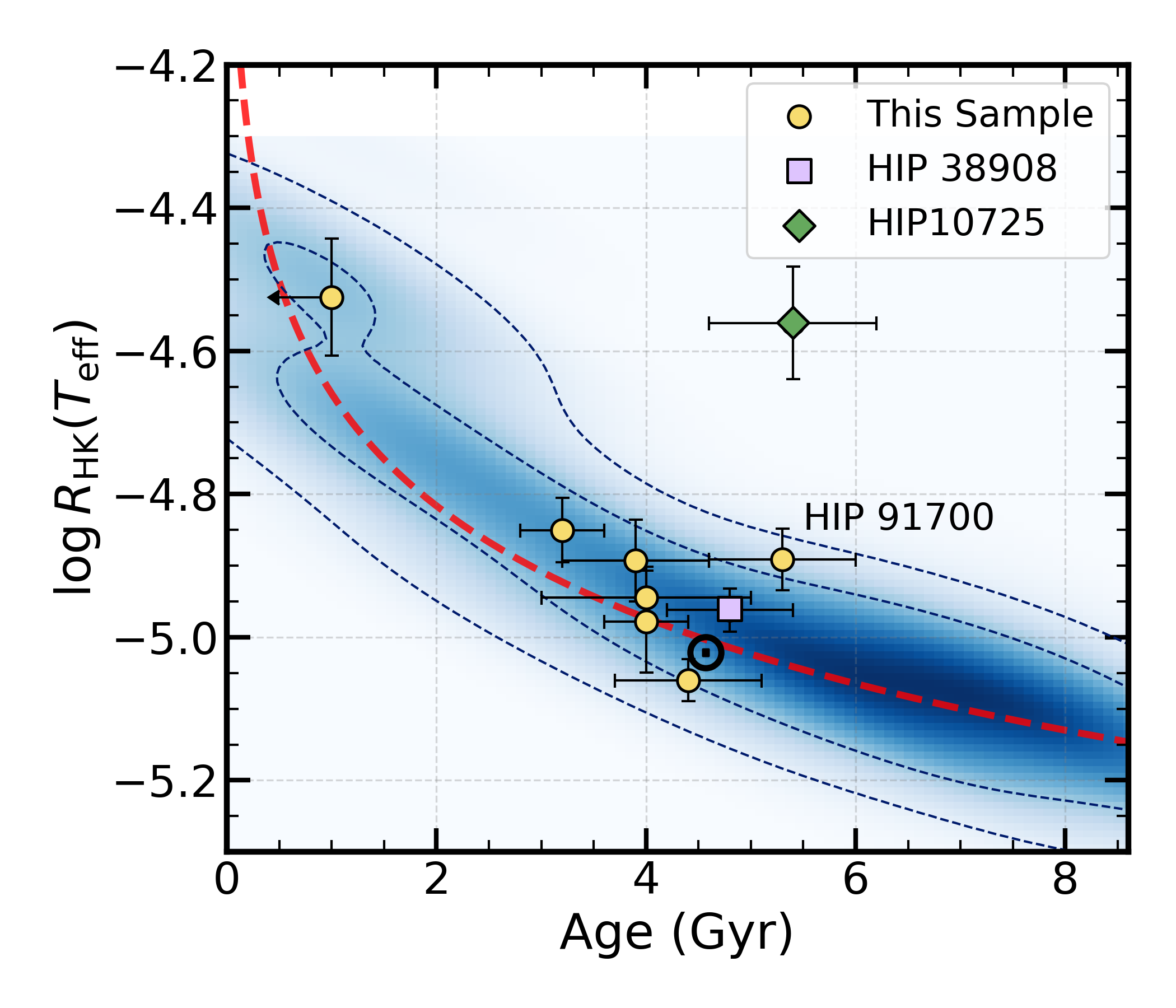}
    \end{tabular}
    \caption{[Y/Mg] (left panel) and [Y/Al] (middle panel) as a function of isochronal age. The red dashed lines show the linear fits applied to the solar-type stars from \citet{Yana_Galarza:2021MNRAS.504.1873Y}, as given in Equations (1) and (2). The right panel shows the activity proxy $\log{R_{\rm{HK}}(T_{\rm{eff}})}$ versus isochronal age. The blue density regions indicate the distribution of stars from \citet{Yana_Galarza:2021MNRAS.504.1873Y} and \citet{Diego:2018A&A...619A..73L}. The contours represent the 1$\sigma$ and 2$\sigma$ distributions of the solar twins. The yellow circles represent stars from our sample, while the purple square (HIP 38908) and green diamond (HIP 10725) symbols show field blue straggler stars previously reported by \citet{Schirbel:2015AA...584A.116S} and \citet{Rathsam:2025AA...693A..26R}, respectively. The Sun is represented by its standard symbol in black.}
    \label{fig:chemical_clocks}
\end{figure*}

%%%%%%%%%%%%%%%%%%%%%%%%%%%%%
\section{Chemical Abundances}
\label{sec:chem}
The chemical abundances were estimated by measuring EWs using the same procedure to that used for \ion{Fe}{1} and \ion{Fe}{2} via the curve-of-growth method, following the differential technique between the stars and the Sun implemented in $\mathrm{q}^{2}$. We used HARPS, which provides the highest-resolution spectra for HIP 53087, HIP 91700, HIP 93858, HIP 116937, and HIP 38908, to determine their chemical compositions. We also used MIKE to improve and confirm measurements when only a few absorption lines were available. Additionally, MIKE offers broader wavelength coverage than HARPS (see Section \ref{sec:obs}), which adds significant value to our analysis, particularly for volatile elements. For HD 221103 and HD 236254, we used HDS, and oxygen and potassium were obtained from TS23 spectra.

In total, we obtained abundances for 32 species: \ion{C}{1}, \ion{O}{1}, \ion{Na}{1}, \ion{Mg}{1}, \ion{Al}{1}, \ion{Si}{1}, \ion{S}{1}, \ion{K}{1}, \ion{Ca}{1}, \ion{Sc}{1}, \ion{Sc}{2}, \ion{Ti}{1}, \ion{Ti}{2}, \ion{V}{1}, \ion{Cr}{1}, \ion{Cr}{2}, \ion{Mn}{1}, \ion{Fe}{1}, \ion{Fe}{2}, \ion{Co}{1}, \ion{Ni}{1}, \ion{Cu}{1}, \ion{Zn}{1}, \ion{Sr}{1}, \ion{Y}{2}, \ion{Ba}{2}, \ion{La}{2}, \ion{Ce}{2}, \ion{Nd}{2}, \ion{Sm}{2}, \ion{Eu}{2}, and \ion{Li}{1}. Oxygen abundances were inferred from the high-excitation \ion{O}{1} $\lambda 777$ nm triplet using TS23/MIKE spectra and corrected for NLTE employing the grids of \citet{Ramirez:2007A&A...465..271R}. We accounted for hyperfine structure and isotopic splitting for \ion{Sc}{1}, \ion{Sc}{2}, \ion{V}{1}, \ion{Mn}{1}, \ion{Co}{1}, \ion{Cu}{1}, \ion{Y}{2}, \ion{Ba}{2}, and \ion{Eu}{2} from \cite{McWilliam:1998AJ....115.1640M,Prochaska2000a,Prochaska2000b,klose2002,Cohen:2003ApJ...588.1082C,Blackwell-Whitehead2005a,Blackwell-Whitehead2005b,Lawler2014}, and from the Kurucz\footnote{\url{http://kurucz.harvard.edu/linelists.html}} line lists. The chemical abundances of our sample relative to the Sun are summarized in Table \ref{tab:chemical_abundances}.

%%%%%%%%%%%%%%%%%%%%%%%%%%%%%%%%%%%%
\section{Isochronal Age and Mass}
\label{sec:ages}
We also used the $\mathrm{q}^{2}$ package to estimate the ages and masses of our sample. Briefly, this method is Bayesian and uses the Yonsei–Yale stellar evolution isochrones \citep{Yi:2001ApJS..136..417Y, Demarque:2004ApJS..155..667D}, along with the estimated stellar parameters, \textit{Gaia} DR3 $G$ magnitude, and parallax, to compute the probability distribution functions of age and mass. More details can be found in \citet{Ramirez:2014A&A...572A..48R} and  \citet{Yana_Galarza:2021MNRAS.504.1873Y}. The results are shown in Table~\ref{tab:fundamental parameters}. For HIP 10725, we applied our methodology to obtain a more precise isochronal age of $5.4 \pm 0.8$ Gyr, consistent with the $5.2 \pm 2.0$ Gyr reported by \citet{Schirbel:2015AA...584A.116S}.

To test the consistency of our age determinations, we compared our isochronal ages against the so-called chemical clocks, specifically the activity-age relation and the [Y/Mg] and [Y/Al] abundance-age correlations. These are well-established age indicators for solar-mass and solar-metallicity stars in the literature using high-resolution spectroscopy \citep[e.g.,][]{daSilva:2012A&A...542A..84D, Nissen:2017A&A...608A.112N, Spina:2018MNRAS.474.2580S, Diego:2018A&A...619A..73L}. For the [Y/Mg] and [Y/Al] relations, we used our own calibration based on the Inti sample \citep{Yana_Galarza:2021MNRAS.504.1873Y} for solar twins, whose stellar parameters and chemical compositions were measured following the same procedure as in this study. We applied a linear fit to the data using the KAPTEYN package, accounting for errors in both axes. The resulting relations are:
\begin{align}
\mathrm{[Y/Mg]} &= 0.162(\pm0.008) - 0.039(\pm0.002) \times \mathrm{age}, \\
\mathrm{[Y/Al]} &= 0.185(\pm0.008) - 0.043(\pm0.001) \times \mathrm{age}.
\end{align}

These correlations are consistent with those reported in previous studies \citep[e.g.,][]{Nissen:2017A&A...608A.112N, Spina:2018MNRAS.474.2580S}. By inverting the equations, we derived ages that agree with those from isochrone fitting (see Table \ref{tab:clocks}), except for HIP 91700 and HIP 38908, as seen in the left and middle panel of Fig. \ref{fig:chemical_clocks}. 

We also computed activity indices ($S_{\rm{HK}}$) by measuring the fluxes of the \ion{Ca}{2} H\&K emission lines fluxes at 3933.664 \AA\ and 3968.470 \AA, respectively. The activity levels, $\log{R^{'}_{\rm{HK}}(T_{\rm{eff}})}$, were then estimated following \citet{Diego:2018A&A...619A..73L}, which accounts for photospheric contamination (see right panel of Fig. \ref{fig:chemical_clocks}). Activity-age estimates were derived using the solar-metallicity relation of \citet[][Eq. 11]{Diego:2018A&A...619A..73L}, while for HIP 38908 and HIP 10725 we applied the metallicity-corrected calibration of \citet[Eq. 2][]{Gaby:2025ApJ...983L..31C}. HIP 38908 shows an anomalously low activity level compared to other stars of similar metallicity and isochronal age, appearing as a solar twin in the age-activity diagram only when metallicity is not considered. We found consistent ages for most stars in the sample when compared to the isochronal ages. However, we found discrepancies for HIP 91700, HD 221103, HIP 10725, and HIP 38908, the latter two being a BSSs. These cases are discussed in detail in Section~\ref{sec:disc}. The isochronal ages and those estimated from the chemical clocks are listed in Table \ref{tab:clocks}. 

\begin{table*}
\centering
\caption{Comparison of isochronal ages with those from chemical clocks, including activity-based ages.}
\begin{tabular}{lccccccc}
\hline
\hline
 \bf{ID}    & [Y/Mg]  & [Y/Al]  & $\log{R^{'}_{\rm{HK}}(T_{\rm{eff}})}$  &    Isochronal   & [Y/Mg]-age      & [Y/Al]-age     & $\log{R^{'}_{\rm{HK}}(T_{\rm{eff}})}$-age  \\
            &  (dex)  & (dex)   &    &    (Gyr)        &  (Gyr)          & (Gyr)          &  (Gyr)                                     \\
 \hline
 HIP 53087  & $0.03\pm0.02$   & $0.01\pm0.02$  & $-5.06\pm0.03$  &  $4.6\pm0.7$    &  $3.4\pm0.5$  & $4.0\pm0.5$  & $5.9\pm0.8$  \\
 HIP 91700  & $0.08\pm0.04$   & $0.11\pm0.02$  & $-4.89\pm0.04$  &  $5.3\pm0.7$    &  $2.1\pm1.0$  & $1.8\pm0.6$  & $2.8\pm0.5$  \\
 HIP 93858  & $0.00\pm0.02$   & $0.01\pm0.02$  & $-4.99\pm0.04$  &  $3.8\pm0.7$    &  $4.1\pm0.6$  & $4.1\pm0.5$  & $4.2\pm0.8$  \\
 HIP 116937 & $0.05\pm0.02$   & $0.07\pm0.02$  & $-4.89\pm0.06$  &  $3.1\pm0.6$    &  $2.9\pm0.5$  & $2.7\pm0.4$  & $2.8\pm0.7$  \\
 HD 221103  & $-0.12\pm0.02$  & $-0.16\pm0.02$ & $-4.85\pm0.05$  &  $2.7\pm0.5$    &  $7.3\pm0.6$  & $8.0\pm0.7$  & $2.3\pm0.5$ \\ 
 HD 236254  & $0.02\pm0.02$   & $0.04\pm0.04$  & $-4.99\pm0.07$  &  $4.9\pm0.6$    &  $3.7\pm0.7$  & $3.4\pm0.9$  & $4.6\pm1.1$ \\ 
 HIP 8522   & $0.14\pm0.02$   & $0.18\pm0.02$  & $-4.53\pm0.08$  &     $<1$        &  $<0.5$         & $<0.5$     & $0.6\pm0.2$ \\ 
 HIP 38908  & $-0.36\pm0.03$  & $-0.29\pm0.02$ & $-4.96\pm0.03$  &  $4.8\pm0.6$    &  $13.5\pm1.0$ & $11.1\pm0.7$ & $6.4\pm0.9^a$ \\ 
 HIP 10725$^b$  & $0.39\pm0.04$  & $0.41\pm0.05$ & $-4.56\pm0.08$  &  $5.4\pm0.8$    & $<0.1$  & $<0.1$ & $8.9\pm0.2^a$ \\ 
\hline
\end{tabular}
\\[1ex]
\textbf{Notes.} $^a$ Inferred using activity-age correlation from \citet{Gaby:2025ApJ...983L..31C}. $^b$ Abundances and activity levels from \citet{Schirbel:2015AA...584A.116S}. 
\label{tab:clocks}
\end{table*}

\begin{table*}
\centering
\caption{Broadening velocities, lithium abundances, and kinematics.}
\label{tab:broadening_lithium}
\begin{tabular}{lccccccccc}
\hline
ID             & $v_{\text{macro}}$ & $v \sin i$          & A(Li)$_{\rm LTE}$ & A(Li)$_{\rm 3DNLTE}$  & $z_{\rm{max}}$ & U             & V             & W             \\
               &  (km s$^{-1}$)     & (km s$^{-1}$)       &  (dex)            &  (dex)                & (kpc)          & (km s$^{-1}$) & (km s$^{-1}$) & (km s$^{-1}$) \\
\hline   
HIP 53087      & $2.71 \pm 0.16$    & $2.40 \pm 0.26$     & $< 0.1$           & $< 0.1$               & 0.13           & $-44.6$       & $-15.3$       & $-16.3$       \\
HIP 91700      & $2.36 \pm 0.16$    & $2.10 \pm 0.17$     & $0.49 \pm 0.15$   & $0.43 \pm 0.15$       & 0.09           & $-11.9$       & $-9.9 $       & $-0.7$        \\
HIP 93858      & $2.83 \pm 0.16$    & $1.60 \pm 0.34$     & $0.58 \pm 0.11$   & $0.52 \pm 0.11$       & 0.12           & $+57.7$       & $-33.4$       & $-15.8$       \\
HIP 116937     & $2.75 \pm 0.16$    & $2.00 \pm 0.10$     & $0.48 \pm 0.13$   & $0.42 \pm 0.13$       & 0.28           & $-22.5$       & $-17.6$       & $-26.2$       \\
HD 221103      & $3.57 \pm 0.16$    & $1.00 \pm 0.25$     & $< 0.40$          & $< 0.40$              & 0.07           & $+12.1$       & $-28.7$       & $-12.1$       \\
HD 236254      & $2.78 \pm 0.16$    & $1.20 \pm 0.27$     & $0.42 \pm 0.14$   & $0.36 \pm 0.14$       & 0.27           & $+8.5 $       & $+8.7 $       & $-24.3$       \\
HIP 8522$^a$   & $2.94 \pm 0.16$    & $2.98 \pm 0.16$     & $< 0.40$          & $< 0.40$              & 0.03           & $+11.1$       & $+1.7 $       & $-6.5$        \\
HIP 38908      & $3.83 \pm 0.16$    & $2.00 \pm 0.27$     & $< 0.23$          & $< 0.23$              & 0.74           & $+11.8$       & $-24.4$       & $+33.9$       \\
HIP 10725$^b$  & $3.60$             & $3.30 \pm 0.10$     & $< 0.90$          & $< 0.90$              & 0.44           & $-37.6$       & $-30.9$       & $-35.3$       \\
Sun (HDS)$^a$  & $3.30 \pm 0.15$    & $1.90 \pm 0.80$$^c$ & $1.05 \pm 0.03$   & $0.99 \pm 0.03$       &                &               &               &               \\
Sun (HARPS)    & $3.30 \pm 0.17$    & $1.90 \pm 0.80$$^c$ & $1.03 \pm 0.04$   & $0.97 \pm 0.04$       &                &               &               &               \\
\hline
\end{tabular}
\\[1ex]
\textbf{Notes.} $^a$ Adopted from \citet{Yana_Galarza_2025}. 
$^b$ Parameters from \citet{Schirbel:2015AA...584A.116S}. No uncertainties are given for $v \sin i$. 
$^c$ Adopted from \citet{Saar:1997MNRAS.284..803S}.
\label{tab:li_abundance}
\end{table*}

%%%%%%%%%%%%%%%%%%%%%%%%%%%%%%%
\section{Lithium Determination}
\label{sec:li}
We determined the absolute lithium abundance (\li) through spectral synthesis of the Li feature at $\sim6707.8$ \AA, using the 2019 version of the radiative transfer code MOOG \citep{Sneden:1973PhDT.......180S} and Kurucz model atmospheres under the assumption of Local Thermodynamic Equilibrium (LTE) \citep{Castelli:2003IAUS..210P.A20C}. The methodology follows the procedure described in detail by \citet{Yana_Galarza:2016A&A...589A..17Y} and \citet{ Yana_Galarza_2025}. Before synthesizing the Li line, we computed the broadening velocities, with macroturbulence and rotational broadening being the most important. Other components, such as atomic and Doppler broadening, are calculated internally by MOOG. The macroturbulence velocity ($v_{\rm{macro}}$) was derived using the calibration from \citet{Leo:2016A&A...592A.156D}. To determine the projected rotational velocity (\vsini), we synthesized four \ion{Fe}{1} lines (6027.050 \AA, 6151.618 \AA, 6165.360 \AA, and 6705.102 \AA) and one \ion{Ni}{1} line (6767.772 \AA). Once the broadening parameters were determined, we performed the Li synthesis using the line list from \citet{Carlos:2019MNRAS.tmp..667C}. The Li abundance was obtained by fitting the observed spectrum to the synthetic one through $\chi^2$ minimization. Finally, we applied 3D non-LTE (NLTE) corrections using the grid from \citet{Wang:2021MNRAS.500.2159W}\footnote{\url{https://github.com/ellawang44/Breidablik}}.

\begin{figure}
    \centering
    \begin{tabular}{c}
        \includegraphics[width=\columnwidth]{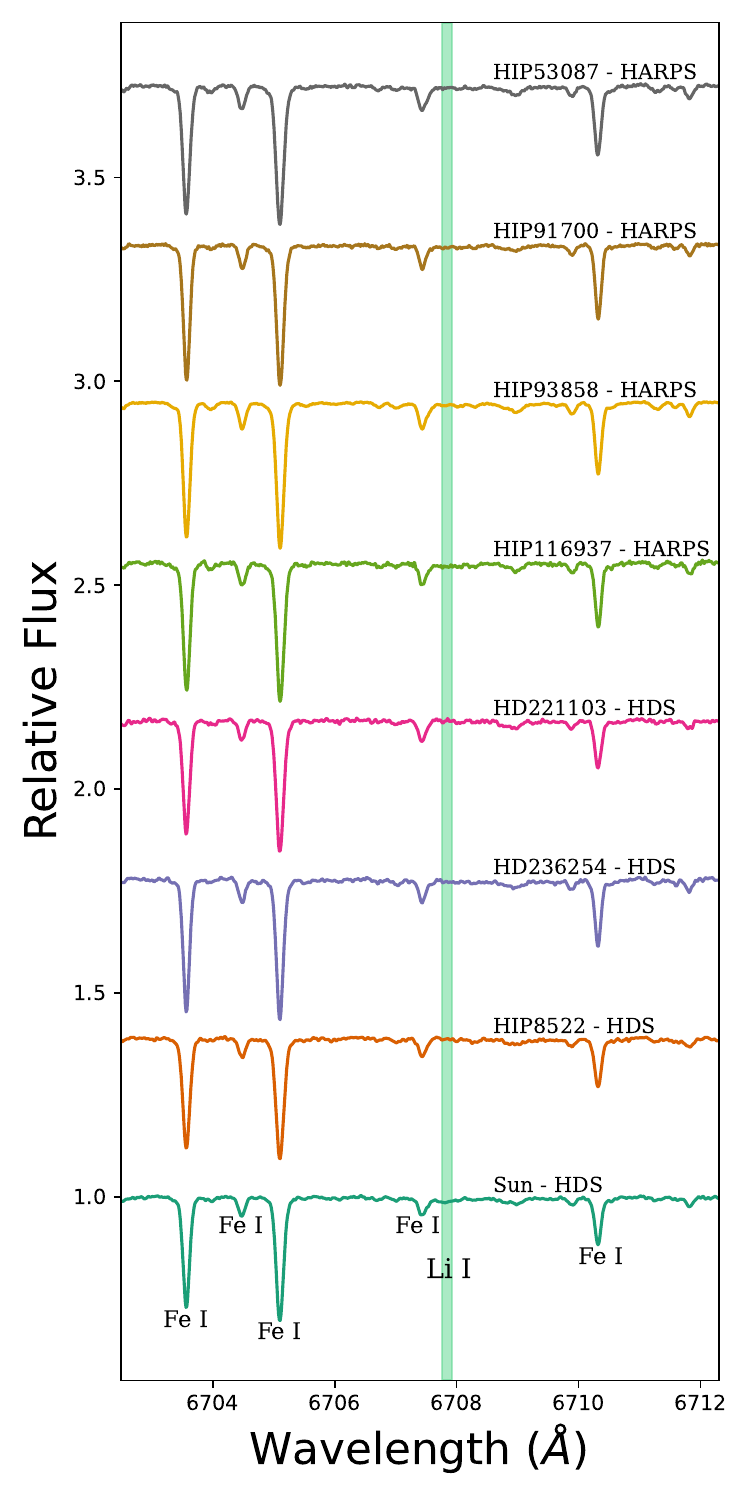}
    \end{tabular}
    \caption{HARPS ($R = 115,000$) and HDS ($R = 165,000$) spectra of our sample of low-Li stars. The Li line region is highlighted in green. The average S/N around the Li line is $\sim$300 in each spectrum. This figure includes only stars discovered through our Inti program for the search of new solar twins \citep{Yana_Galarza:2021MNRAS.504.1873Y}.}
    \label{fig:Li}
\end{figure}

As we used two different high-resolution instruments (HDS and HARPS), it is important to check for possible systematic effects, especially in the macroturbulence velocity, which is sensitive to spectral resolution, as discussed in \citet{Yana_Galarza_2025}. We first determined the macroturbulence velocity of the Sun using HARPS spectra by fixing the projected rotational velocity \vsini$_{\odot}$ to 1.9 km s$^{-1}$ \citep{Bruning:1984ApJ...281..830B, Saar:1997MNRAS.284..803S}. Through spectral synthesis of iron and nickel lines, we obtained a macroturbulence velocity of 3.3 km s$^{-1}$, consistent with the values reported by \citet[][see their Table 2]{Doyle:2014MNRAS.444.3592D}, based on solar spectra with resolutions between $R = 76,000$ and $R = 300,000$. Using both broadening velocities, we determined the solar lithium abundance from the synthesis of the Li I 6707.8 \AA\ line, obtaining a value of \li$_{\rm{LTE}, \odot} = 1.03 \pm 0.05$ dex. This agrees with the Li abundance previously reported using HDS spectra by \citet{Yana_Galarza_2025}, which was \li$_{\rm{LTE}, \odot} = 1.05 \pm 0.01$ dex. Since there is no significant difference between both Li solar abundances, we adopted the weighted average for our analysis and then applied the 3D NLTE correction, yielding \li$_{\rm{3DLTE}, \odot} = 0.98 \pm 0.02$ dex. 

Using the same methodology as described above for the Sun, we estimated $v_{\text{macro}}$, \vsini, and \li\ for all our sample, which are listed in Table \ref{tab:li_abundance}. Fig. \ref{fig:Li} shows the high-resolution HARPS and HDS spectra of the Li I resonance doublet at 6708 \AA\ for the low-Li stars from our sample. We found \li\ $\lesssim 0.5$ dex, which is unexpected for stars with solar-mass, solar-metallicity, and ages lower than $\sim5$ Gyr. Additionally, we recomputed the \li\ of HIP 38908 and obtained an upper limit of 0.23 dex, consistent with the value reported by \citet{Rathsam2023}. 

\begin{figure}
    \centering
    \begin{tabular}{c}
        \includegraphics[width=\columnwidth]{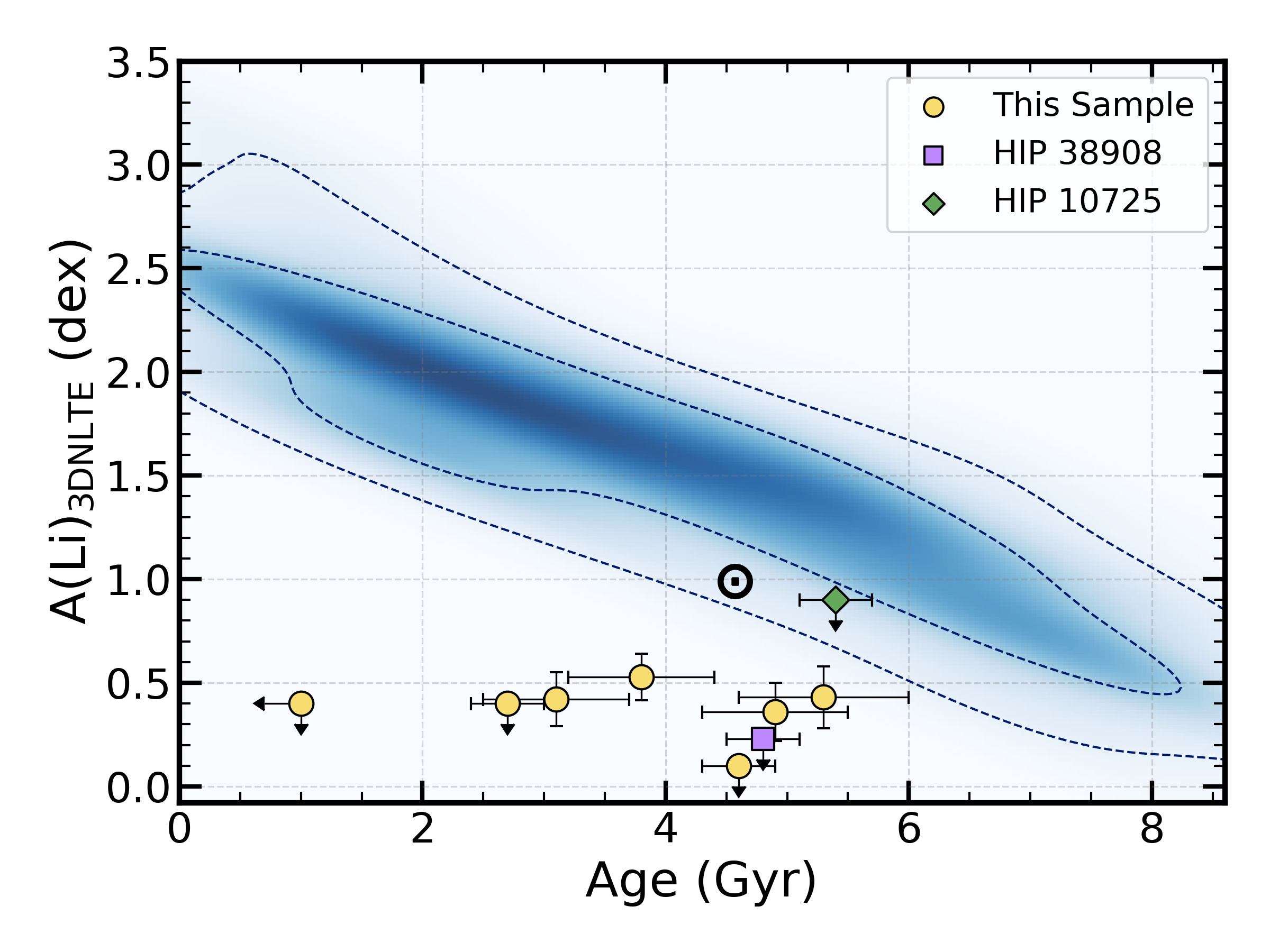}
    \end{tabular}
    \caption{3D NLTE Li abundance as a function of age. In blue we show the density region of solar twins from the Inti Survey, with contours of $1\sigma$ and $2\sigma$. The yellow circles represent the seven recently detected low-Li solar twins and the purple square and green diamond represent the reported field blue stragglers. The Sun is represented in black by its standard symbol.}
    \label{fig:Li_age}
\end{figure}

Fig. \ref{fig:Li_age} shows the 3D NLTE Li abundance as a function of isochronal age and reveals that our stars do not follow the typical \li\ depletion trend observed in solar twins. Instead, they appear to suffer from depletion effects unlike the remainder of the sample, having already depleted most of their \li\ at a younger age. Even when using the ages obtained from chemical clock correlations, the stars remain anomalous compared to solar twins, except for HD 221103. The implications of our findings will be discussed in detail in Section \ref{sec:disc}. The previously reported FBSSs, represented by a purple square and a green diamond, are also shown in Fig.  \ref{fig:Li_age}. However, HIP 10725 is the only star that shows no significant Li depletion, as it falls within the $2\sigma$ distribution of the solar twins.

%%%%%%%%%%%%%%%%%%%%%%%%%%%%%%%%%%%%%
\section{Resolving Hidden Companions}
\label{sec:companions}
Given the significant \li\ depletion detected in our sample, this section explores the possibility of an unresolved companion through spectral energy distribution (SED) fitting and radial velocity (RV) analysis. Additionally, we include an analysis of their Galactic disk membership.

\subsection{Spectral Energy Density Fitting}
We performed $\chi^2$ fitting of the SED for the six stars in our sample. The SED fits for HIP 8522, HIP 38908, and HIP 10725 are shown in \cite{Yana_Galarza_2025}, \citet{Rathsam:2025AA...693A..26R}, and \cite{Schirbel:2015AA...584A.116S}, respectively. The SED fit was carried out using archival photometry spanning from $\lambda = 0.355~\mu m$ to $\lambda = 22.08~\mu m$ within the Virtual Observatory SED Analyser (VOSA; \citealt{Bayo:2008A&A...492..277B}). The data includes observations from the Javalambre Physics of the Accelerating Universe Astrophysical Survey (J-PAS; \citealt{Benitez:2014arXiv1403.5237B}), the Javalambre-Photometric Local Universe Survey (J-PLUS; \citealt{Cenarro:2019A&A...622A.176C}), \emph{Gaia} DR3 \citep{GaiaDR3:2023A&A...674A...1G}, the Wide-field Infrared Survey Explorer (WISE; \citealt{Wright:2010AJ....140.1868W}), the 2-Micron All-Sky Survey (2MASS; \citealt{Ochsenbein:2000A&AS..143...23O, skrutskie2006}), the DEep Near-Infrared Survey (DENIS) of the southern sky \citep{Fouque:2000A&AS..141..313F, Epchtein:1999A&A...349..236E, Ochsenbein:2000A&AS..143...23O}, the Tycho-2 Catalogue \citep{Ochsenbein:2000A&AS..143...23O, Hog:2000A&A...357..367H}, the Str$\ddot{o}$mgren-Crawford uvby$\beta$ photometry \citep{Ochsenbein:2000A&AS..143...23O, Paunzen:2015A&A...580A..23P}, AKARI/IRC mid-infrared all-sky Survey \citep{Ochsenbein:2000A&AS..143...23O, Murakami:2007PASJ...59S.369M, Ishihara:2010A&A...514A...1I}, the AAVSO Photometric All Sky Survey \citep[APASS;][]{Ochsenbein:2000A&AS..143...23O, Evans:2002A&A...395..347E}, the Galaxy Evolution Explorer \citep[GALEX;][]{Ochsenbein:2000A&AS..143...23O, Bianchi:2017ApJS..230...24B}, and the Japan Aerospace Exploration Agency (JAXA) AKARI telescope \citep{Murakami:2007PASJ...59S.369M}. Fig. \ref{fig:SED} shows the SED fitting for our sample with the synthetic spectrum from the {\sc BT-Settl} model grid \citep{Allard:2011ASPC..448...91A} of theoretical spectra. The nominal values obtained are consistent within 100 K in \teff, and 0.1 dex in \feh\ and 0.1 dex in \logg\ with those derived from high-resolution spectroscopy.

To rule out the presence of a hidden companion, we performed a two-component model fit to the observed SED. The primary component was fixed using the best-fit parameters from the single-star model, representing the known star in the system. The secondary component was varied over a range of effective temperatures ($T_{\rm eff} = 400$–$2300$ K) and surface gravities ($\log g = 4.5$–$6.0$ dex) using the {\sc BT-Settl} model grid, consistent with potential ultra-cool stars or brown dwarf companions \citep{Reid:2000nlod.book.....R}. We also explored the white dwarf parameter space, varying $T_{\rm eff} = 5000$–$80,000$ K and $\log g = 6.5$–$9.0$ dex using the corresponding model grid \citep{Koester:2010MmSAI..81..921K}. In all cases, the two-component fits did not significantly improve upon the single-star model, and we found no compelling evidence for a binary companion---neither an ultra-cool object nor a degenerate s. This supports the conclusion that the stars in our sample are single. This is further reinforced by their low \textit{Gaia} renormalized unit weight error (RUWE $<$1.2) values (see Table~\ref{tab:instruments}).

%%%%%%%%%%%%%%%%%%%%%%%%%%%%%%%%%%%%%%%%%%%%%%%%
\subsection{No Stellar Companions, Only Planets}
We modeled the radial velocity time-series of HIP 53087 and HIP 93858 using the {\sc pyaneti} suite \citep{2019MNRAS.482.1017B, 2022MNRAS.509..866B}, incorporating both a Keplerian and Gaussian Process (GP) model to account for the companion's orbital motion and stellar activity, respectively. Radial velocity modulations induced by stochastic motions at the surface of a star (e.g., \citealt{2004sipp.book.....H, 2020dmde.book.....H}) can be modeled as ${\rm RV} = A_0 \, \gamma_{\rm MQP}(t_i, t_j) + A_1 \, \frac{\mathrm{d}}{\mathrm{d}t} \gamma_{\rm MQP}(t_i, t_j)$, where adopting a Mat{\'e}rn quasi-periodic covariance kernel, observations at two different epochs, $t_i$ and $t_j$, are correlated through the covariance function $\gamma_{\rm MQP}(t_i, t_j)$ (see \citealt{2022MNRAS.509..866B}): 

\begin{eqnarray}
    \gamma_{\rm MQP}(t_i, t_j) &=& A_0^2 \, k_{\mathrm{Mat\acute{e}rn}, \nu=1/2}(|t_i - t_j|; \lambda_e) \notag \\ &\times& \exp\left[-\frac{\sin^2\left(\pi \frac{t_i - t_j}{P_{\rm GP}}\right)}{2 \lambda_p^2}\right],
\end{eqnarray}

where the Mat{\'e}rn kernel of order $\nu=1/2$ (the exponential kernel) is defined as

\begin{equation}
    k_{\mathrm{Mat\acute{e}rn}, \nu=1/2}(\tau; \lambda_e) = \exp\left(-\frac{\tau}{\lambda_e}\right),
\end{equation}

with $\tau = |t_i - t_j|$ denoting the time lag between observations. Here, $A_0$ represents the amplitude of stellar activity-induced variations, $A_1$ corresponds to the amplitude of covariance involving the time derivatives of the process (capturing correlated slopes), $P_{\rm GP}$ denotes the characteristic recurrence timescale of active regions --- typically the stellar rotation period \citep{2015MNRAS.452.2269R, 2023ARA&A..61..329A} ---, $\lambda_p$ is a smoothing parameter regulating the complexity of the periodic harmonics, and $\lambda_e$ characterizes the evolutionary timescale of active regions \citep[e.g.,][]{2014MNRAS.443.2517H, 2020MNRAS.495L..61L, 2021AJ....162..160N}.

We adopted Uniform priors for the orbital parameters: (1) orbital period $P_b\sim\mathcal{U}(220, 230)$ days, (2) time of mid-transit $T_0\sim\mathcal{U}(5588, 5596)$ BJD, (3) eccentricity coupling with argument of periastron $e\cos\omega\sim\mathcal{U}(-1.0, 1.0)$ and $e\sin\omega\sim\mathcal{U}(-1.0, 1.0)$, and (4) radial velocity semi-amplitude $K\sim\mathcal{U}(0.0, 0.05)$ km s$^{-1}$ for HIP 53087, whilst for HIP 93858 we used: (1) orbital period $P_b\sim\mathcal{U}(759, 779)$ days, (2) time of mid-transit $T_0\sim\mathcal{U}(7608, 7808)$ BJD, (3) eccentricity components $e\cos\omega\sim\mathcal{U}(-1.0, 1.0)$ and $e\sin\omega\sim\mathcal{U}(-1.0, 1.0)$, and (4) radial velocity semi-amplitude $K\sim\mathcal{U}(0.0, 0.30)$ km s$^{-1}$.

Moreover, we employed a GP regression using a Mat{\'e}rn quasi-periodic kernel with hyper-parameters uniformly distributed following: (5) amplitude components $A_1\sim\mathcal{U}(0.0, 0.6)$ and $A_2\sim\mathcal{U}(0.0, 0.6)$, (6) long-term evolution timescale $\lambda_e\sim\mathcal{U}(1, 200)$ days, (7) harmonic complexity parameter $\lambda_p\sim\mathcal{U}(0.2, 0.7)$, and (8) characteristic GP timescale $P_{GP}\sim\mathcal{U}(1.0, 50.0)$ days for both stars. 

Markov Chain Monte Carlo (MCMC) sampling used 1500 independent chains, and to assess model convergence, we employed a Gelman-Rubin statistics test \citep{Gelman1992}, and applying a burn-in phase of 1500 iterations and a thinning factor of 10, yielded a final set of 450,000 independent samples per hyper-parameter. 

\begin{figure}
    \centering
    \includegraphics[width = \linewidth]{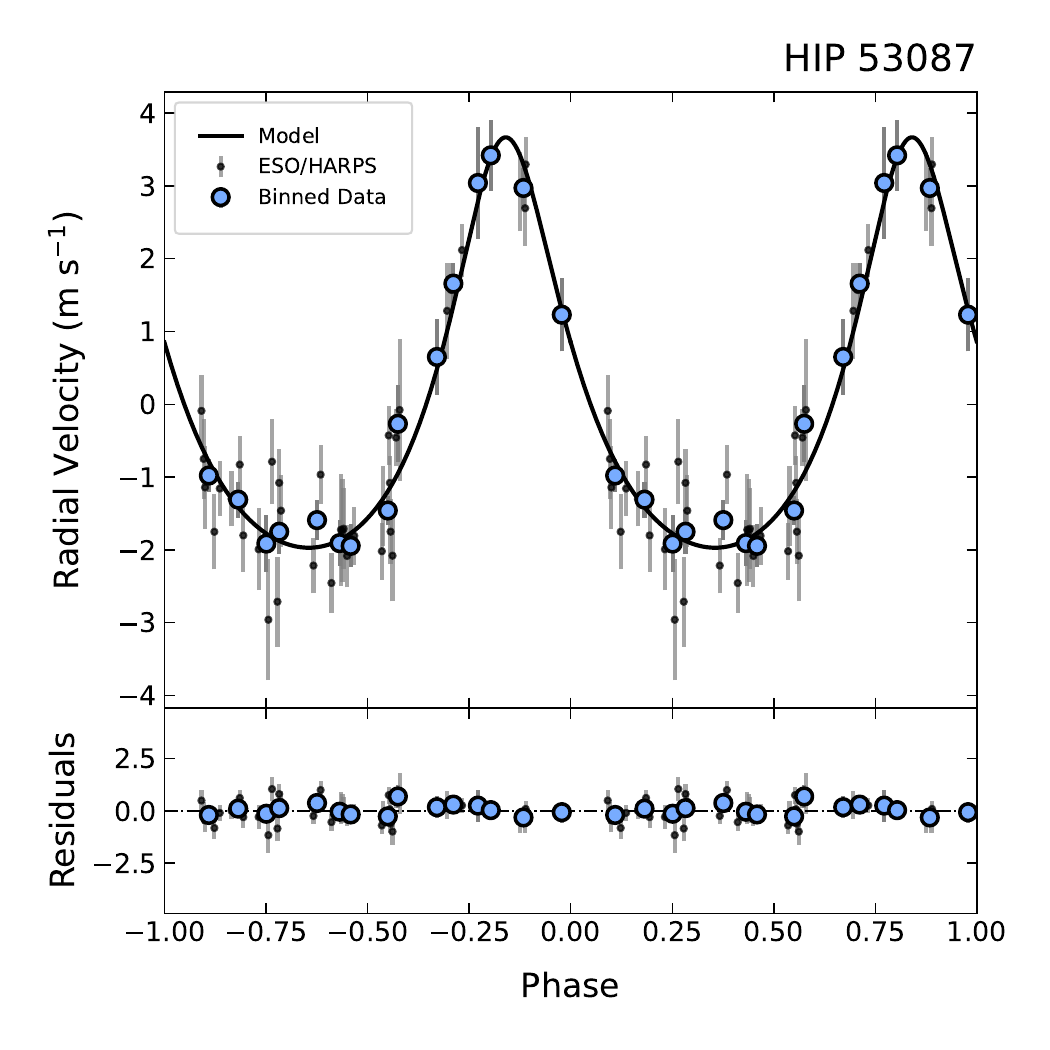}
    \includegraphics[width = \linewidth]{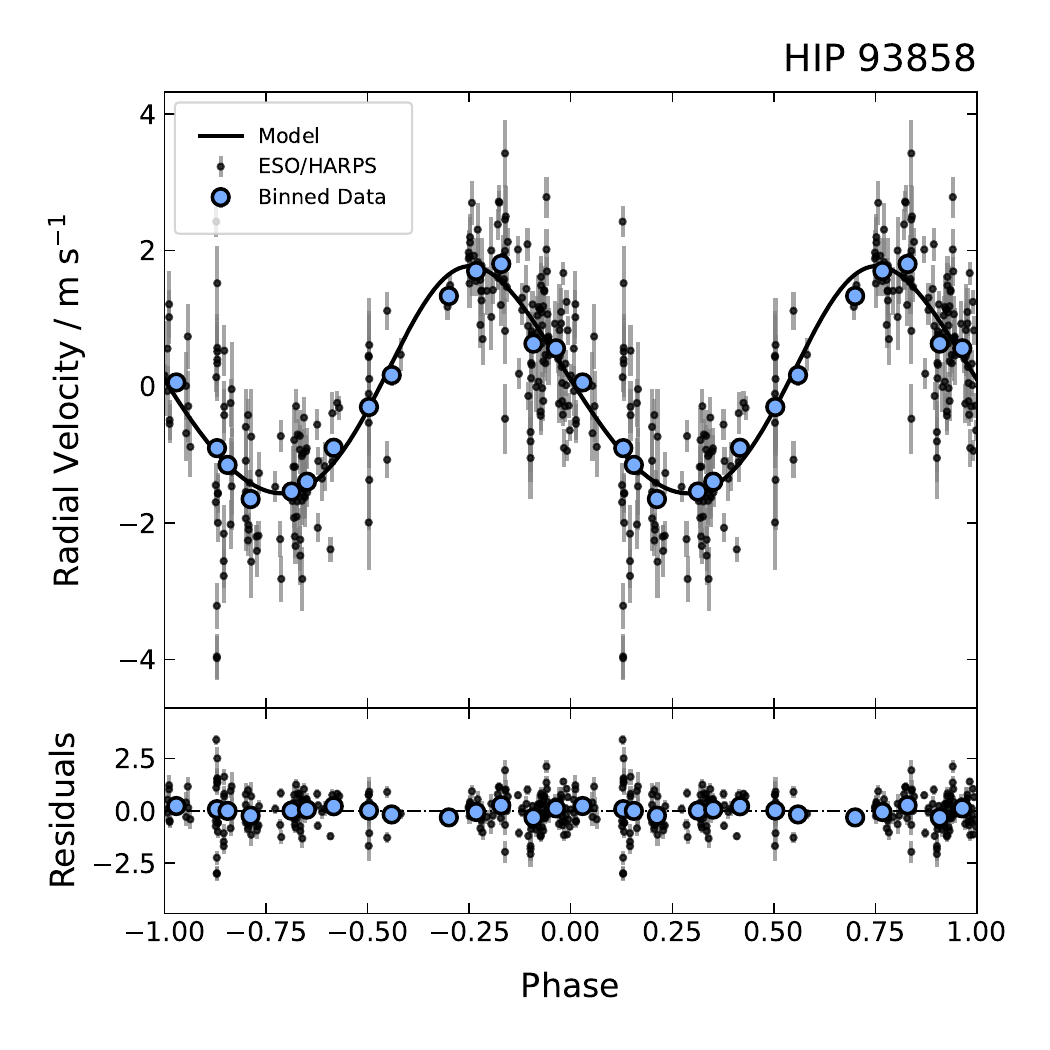}
    \caption{Phase folded HARPS radial velocity modulations of HIP 53087 (top panel) and HIP 93858 (bottom panel). The best-fit models are shown in black mirrored across the interval to cover the range [-1, 1]. The blue circles represent the binned data.}
    \label{fig:exoplanetsHIP}
\end{figure}

Fig. \ref{fig:exoplanetsHIP} and Fig. \ref{fig:phase_rv} display the radial velocity modulations for HIP 53087 (top panel) and HIP 93858 (botom panel), accounting for both the Keplerian and stellar activity models. Their companions' derived physical and orbital parameters are provided in Table \ref{tab:pyaneti_results}.

\begin{table}[h]
    \caption{Keplerian and Gaussian Process derived parameters for HIP 53087 and HIP 93858.}
    \centering
    \renewcommand{\arraystretch}{1.2}
    \begin{tabular}{l r}
    \hline\hline
    \textbf{Parameter} & \textbf{Value} \\
    \hline
    \multicolumn{2}{l}{\it HIP 53087 $b$} \\
    Orbital period, $P$ (days) & $225.16_{-2.28}^{+2.49}$ \\
    Time of inferior conj., $T_{0;b}$ (BJD$-2450000$) & $3142.01_{-2.76}^{+2.71}$ \\
    $\sqrt{e}\cos\omega$ & $-0.023_{-0.512}^{+0.555}$ \\
    $\sqrt{e}\sin\omega$ & $0.230_{-0.282}^{+0.246}$ \\
    RV semi-amplitude, $K$ (m s$^{-1}$) & $2.82_{-1.20}^{+1.46}$ \\
    Combined jitter, $\sigma_{\rm RV}$ (m s$^{-1}$) & $0.71_{-0.27}^{+0.31}$ \\
    $A_0$ (m s$^{-1}$) & $3.32_{-0.60}^{+0.74}$ \\
    $A_1$ (m s$^{-1}$ d$^{-1}$) & $4.51_{-3.19}^{+5.12}$ \\
    Active region timescale, $\lambda_e$ (days) & $142.8_{-62.2}^{+42.6}$ \\
    Harmonic complexity, $\lambda_p$ & $0.640_{-0.082}^{+0.044}$ \\
    GP characteristic period$^\alpha$, $P_{\rm GP}$ (days) & $36.9_{-4.2}^{+8.8}$ \\
    Minimum mass, $M_b\sin i$ ($M_{\oplus}$) & $23.4_{-10.6}^{+11.1}$ \\
    Eccentricity, $e$ & $0.30_{-0.19}^{+0.42}$ \\
    Argument of periastron, $\omega$ (deg) & $-4.3_{-73.3}^{+84.6}$ \\
    Semi-major axis, $a_b$ (au) & $0.732_{-0.005}^{+0.005}$ \\
    \hline
    \multicolumn{2}{l}{\it HIP 93858 $b$} \\
    Orbital period, $P$ (days) & $766.59_{-4.41}^{+5.17}$ \\
    Time of inferior conj., $T_{0;b}$ (BJD$-2450000$) & $5255.39_{-36.12}^{+26.36}$ \\
    $\sqrt{e}\cos\omega$ & $-0.051_{-0.239}^{+0.200}$ \\
    $\sqrt{e}\sin\omega$ & $-0.063_{-0.264}^{+0.261}$ \\
    RV semi-amplitude, $K$ (m s$^{-1}$) & $1.67_{-0.35}^{+0.40}$ \\
    Combined jitter, $\sigma_{\rm RV}$ (m s$^{-1}$) & $0.99_{-0.08}^{+0.08}$ \\
    $A_0$ (m s$^{-1}$) & $1.59_{-0.27}^{+0.35}$ \\
    $A_1$ (m s$^{-1}$ d$^{-1}$) & $2.59_{-0.95}^{+1.18}$ \\
    Active region timescale, $\lambda_e$ (days) & $150.4_{-32.0}^{+24.2}$ \\
    Harmonic complexity, $\lambda_p$ & $0.228_{-0.018}^{+0.061}$ \\
    GP characteristic period$^\alpha$, $P_{\rm GP}$ (days) & $30.0_{-0.1}^{+0.1}$ \\
    Minimum mass, $M_b\sin i$ ($M_{\oplus}$) & $23.5_{-5.2}^{+4.9}$ \\
    Eccentricity, $e$ & $0.084_{-0.062}^{+0.205}$ \\
    Argument of periastron, $\omega$ (deg) & $-43.1_{-103.8}^{+180.9}$ \\
    Semi-major axis, $a_b$ (au) & $1.656_{-0.007}^{+0.007}$ \\
    
    \hline\hline
    \end{tabular}
    \label{tab:pyaneti_results}
\end{table}

\subsection{Kinematics}
\begin{figure*}
    \centering
    \begin{tabular}{cc}
        \includegraphics[width=0.4\textwidth]{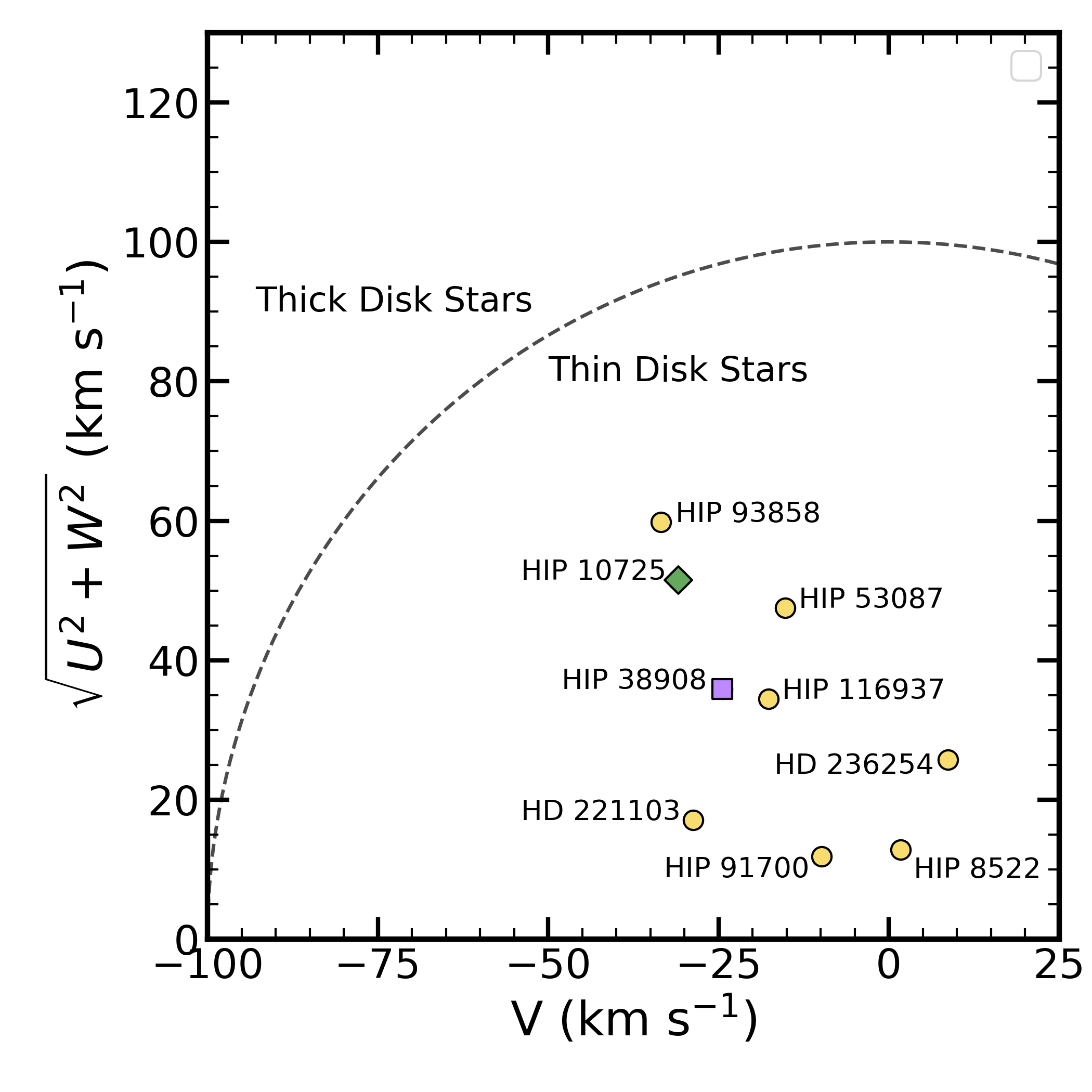} &
        \includegraphics[width=0.55\textwidth]{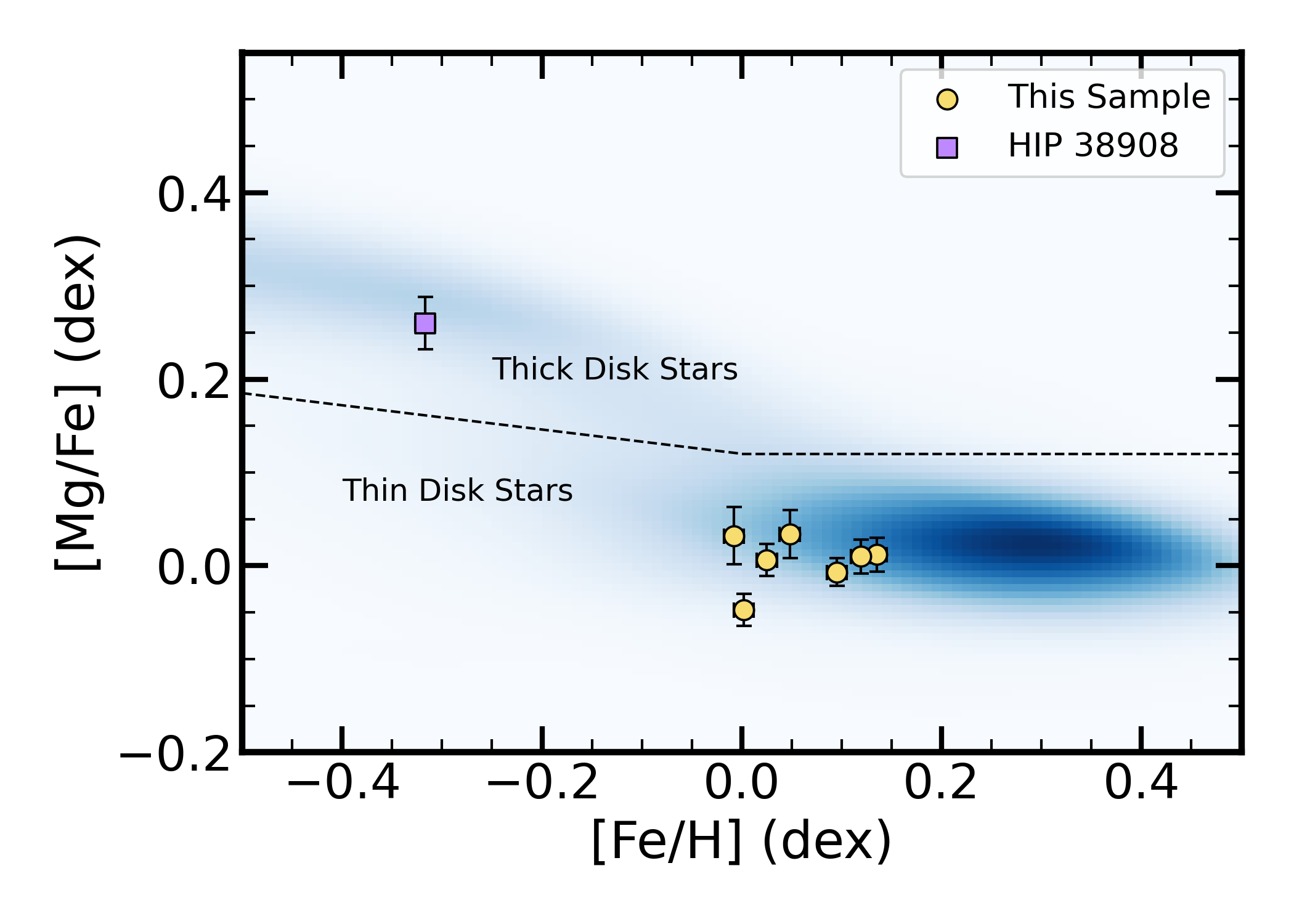}
    \end{tabular}
    \caption{\textbf{Left panel:} Toomre diagram for our sample of low-Li stars, including the two already reported in the literature. \textbf{Right panel:}  [Mg/Fe] as a function of [Fe/H]. Circles represent thin disk stars, and the square denotes a thick disk star. The density map shows APOGEE stars \citep{Majewski:2017AJ....154...94M}, and the dashed lines mark the separation between the thick (upper) and thin (lower) disks.}
    \label{fig:disk_membership}
\end{figure*}

We calculated the galactic orbits for our full sample, as well as for the previously reported FBSSs HIP 38908 and HIP 10725, using the \textsc{Gala} package\footnote{\url{https://github.com/adrn/gala}} \citep{gala, adrian_price_whelan_2020_4159870}, along with proper motions and radial velocities from \textit{Gaia} DR3 \citep{GaiaDR3:2023A&A...674A...1G}. We used the default {\sc MilkyWayPotential} model in \textsc{Gala}, which includes a spherical nucleus and bulge \citep{Hernquist:1990ApJ...356..359H}, a Miyamoto–Nagai disk \citep{Miyamoto:1975PASJ...27..533M, Bovy:2015ApJS..216...29B}, and a spherical Navarro-Frenk-White (NFW) dark matter halo \citep{Navarro:1996ApJ...462..563N}. We adopted the Sun's position and velocity as $x_{\odot} = (-8.3, 0, 0)$ kpc and $v_{\odot} = (-11.1, 244, 7.25)$ km s$^{-1}$ \citep{Schonrich:2010MNRAS.403.1829S, Schonrich:2012MNRAS.427..274S}. The galactic space velocities ($U, V, W$, see Table \ref{tab:li_abundance}) indicate that the low-Li stars are located right within the thin disk kinematic distribution on the Toomre diagram  ($V$ vs. $\sqrt{U^{2}+W^{2}}$; see left panel of Fig. \ref{fig:disk_membership}). 

This result is supported by their [Mg/Fe] ratios (see right panel of Fig. \ref{fig:disk_membership}), which serve as a good proxy for the [$\alpha$/Fe]\footnote{[$\alpha$/Fe] = 1/4 ([Mg/Fe] + [Si/Fe] + [Ca/Fe] + [Ti/Fe])} ratio. The only exception is HIP 38908, which, despite showing thin-disk-like kinematics, has a [Mg/Fe] ratio that suggests a thick disk origin. This was already reported by \citet{Rathsam:2025AA...693A..26R}, and we will discuss this in detail in the next section.

%%%%%%%%%%%%%%%%%%%%
\section{Lithium Depletion Scenarios}
\label{sec:disc}
HIP 8522 is an exceptional solar twin, as it is the only young star in our sample that shows significant \li\ depletion for its age. In \citet{Yana_Galarza_2025}, we investigated several scenarios to explain this depletion and found that only two were plausible: FBSS formation through binary mergers, and episodic accretion onto young stars. Here, we take advantage of our larger sample to perform a similarly detailed analysis and further investigate extreme \li\ depletion in solar-twins.

%%%%%%%%%%%%%%%%%%%%%%%%%%%%%%
\subsection{Planet Engulfment}
When chemical abundances of stars relative to the Sun are plotted against condensation temperature\footnote{Temperature at which 50\% of an element condenses from a gas with solar composition at a total pressure of $10^{-4}$ bar \citep{Lodders:2003ApJ...591.1220L}.} (\Tc), the slope---defined here for refractory elements\footnote{Rock-forming elements with \Tc\ $> 900$ K.} and referred to as the \Tc-slope---may provide signatures of planet formation or engulfment. Negative \Tc-slopes are thought to reflect the depletion of refractory elements due to planet formation, with the missing refractories observed in the Sun attributed to the formation of the terrestrial planets \citep{Melendez:2009ApJ...704L..66M, Ramirez:2010A&A...521A..33R, Booth:2020MNRAS.493.5079B, Yana:2021MNRAS.502L.104Y, Huhn:2023A&A...676A..87H}. Positive \Tc-slopes are attributed to planet engulfment \citep[e.g.,][]{Melendez:2017A&A...597A..34M, Yana_Galarza:2021ApJ...922..129G, Behmard:2023MNRAS.521.2969B, Liu:2024Natur.627..501L, Flores:2024MNRAS.52710016F}, often accompanied by an increase in \li\ that can be rapidly erased through thermohaline convection after accretion. This mechanism was first proposed by \citet{Theado:2012ApJ...744..123T}, who concluded that planet engulfment could reduce \li\ to below its pre-engulfment level. Similarly, \citet{Sevilla:2022MNRAS.516.3354S}, using {\sc MESA} stellar evolution models \citep{paxton2011, paxton2013, paxton2015, paxton2018, paxton2019}, showed that the magnitude and duration of \li\ enrichment strongly depend on internal mixing processes, leading to a rapid post-engulfment depletion of \li.

\begin{figure*}
    \centering
    \begin{tabular}{cc}
        \includegraphics[width=0.48\textwidth]{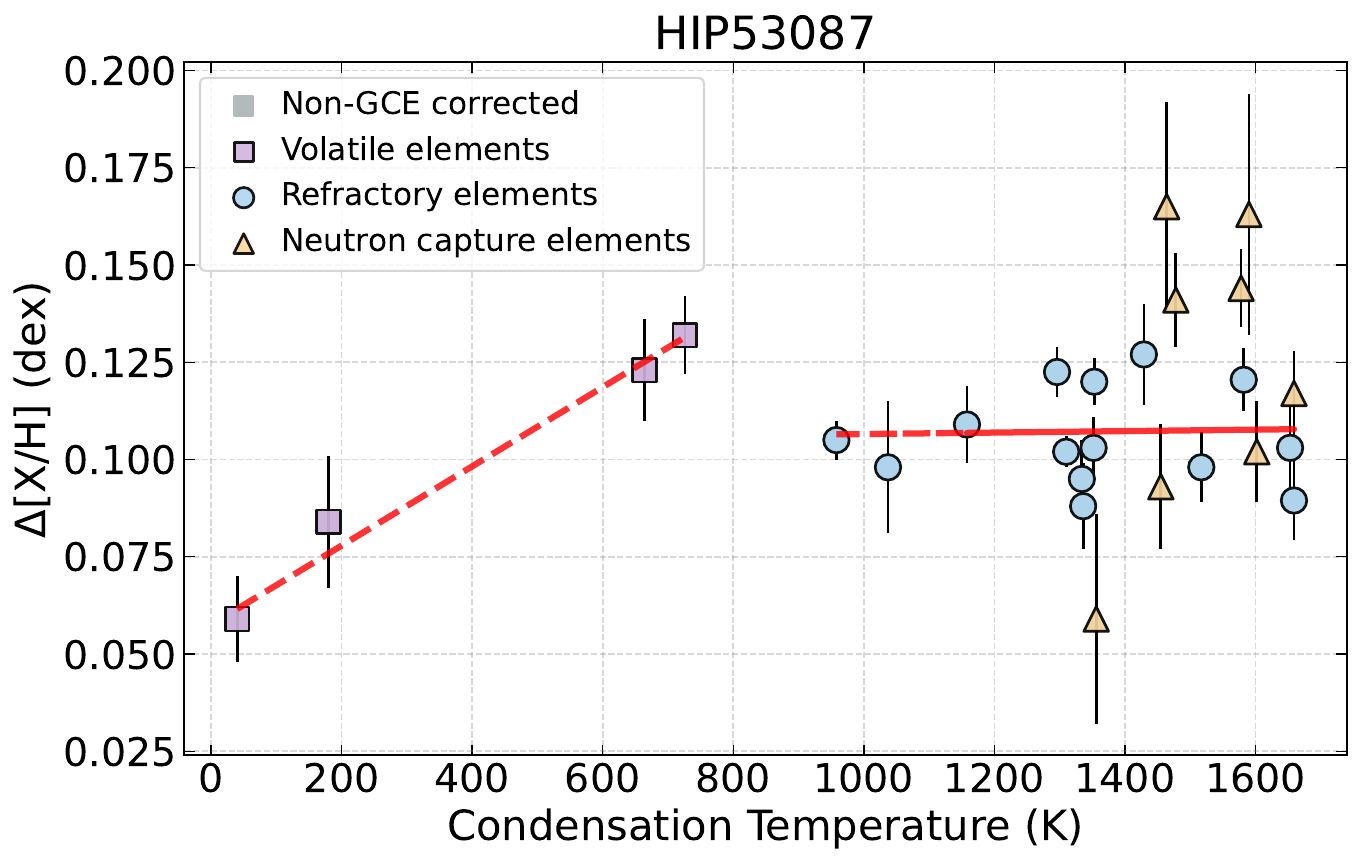} &
        \includegraphics[width=0.48\textwidth]{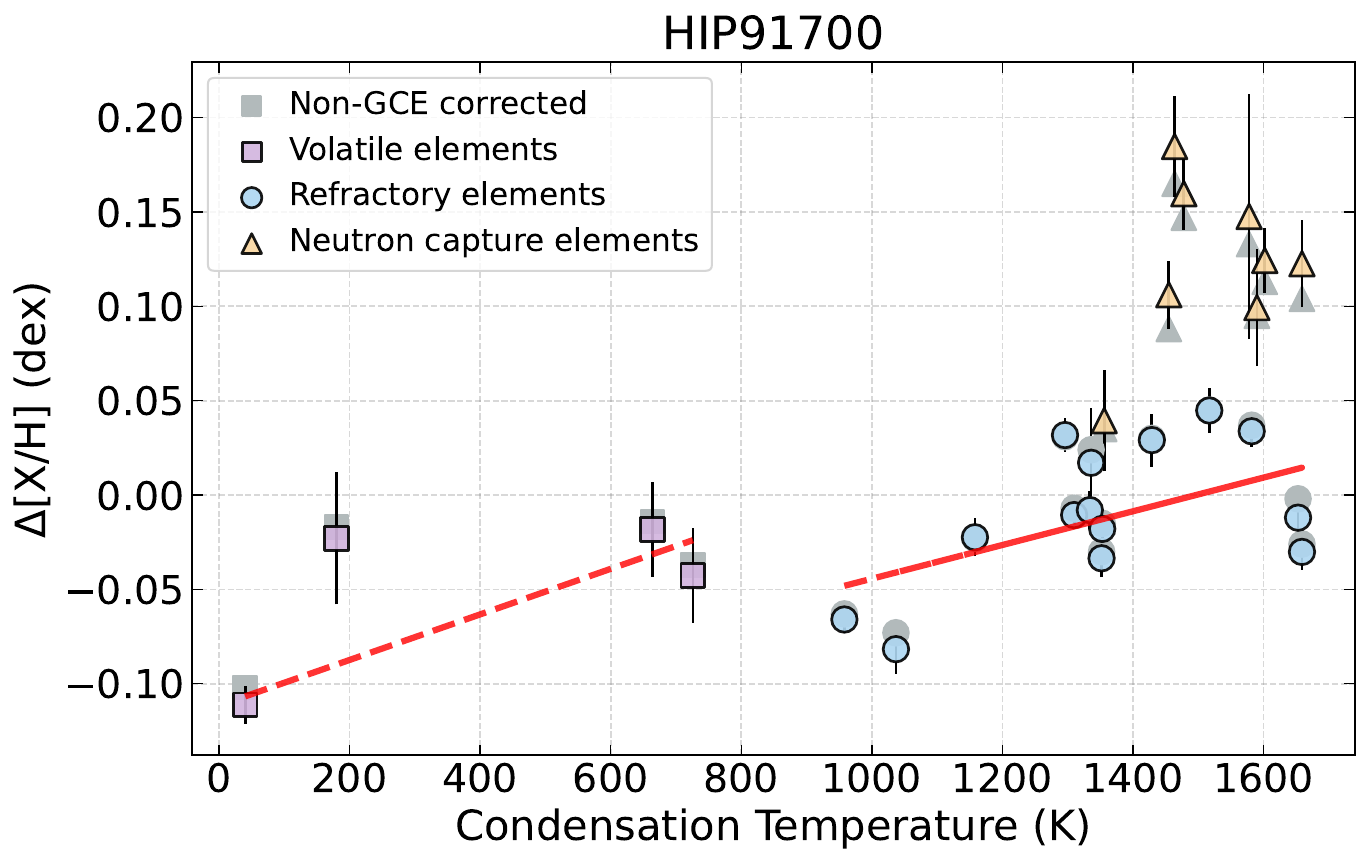} \\
        \includegraphics[width=0.48\textwidth]{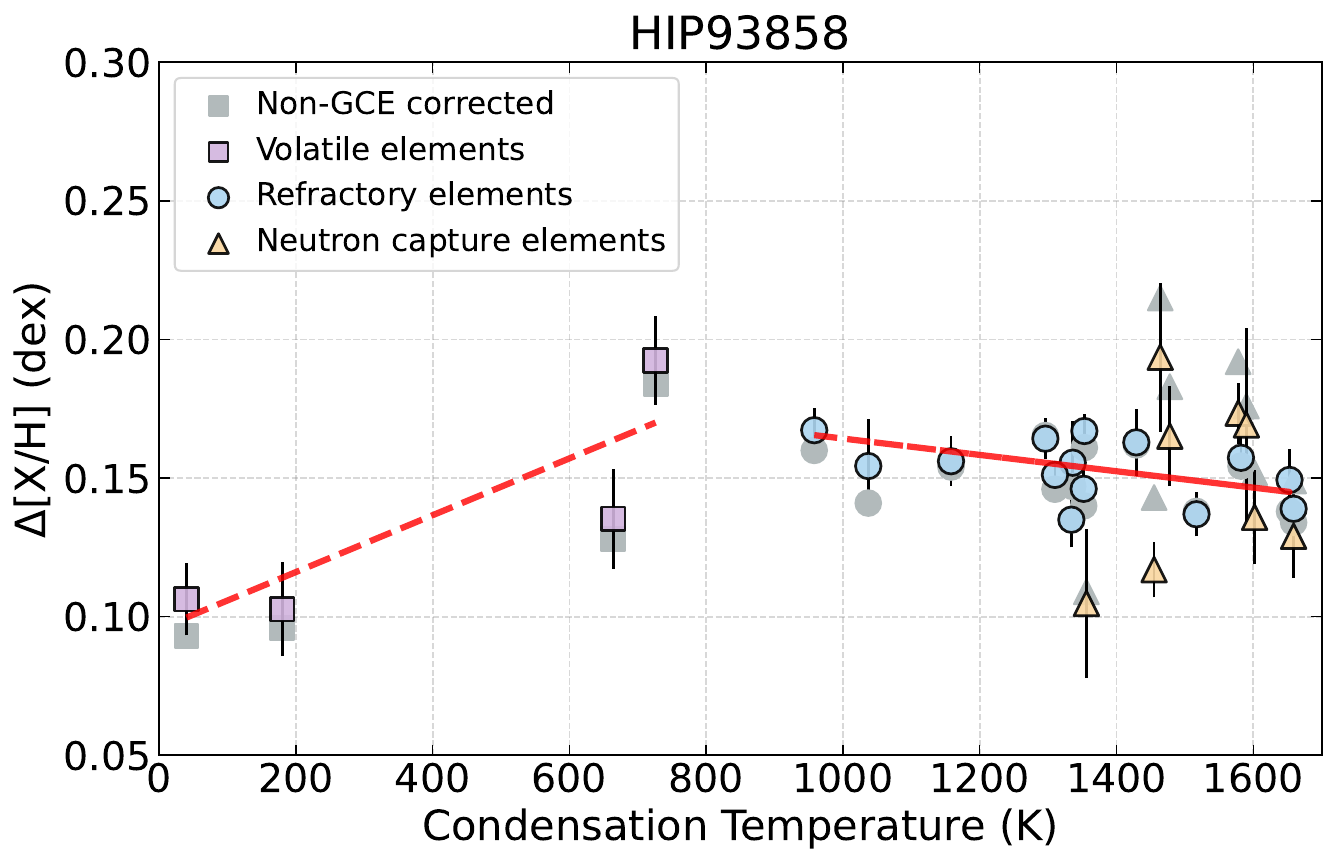} &
        \includegraphics[width=0.48\textwidth]{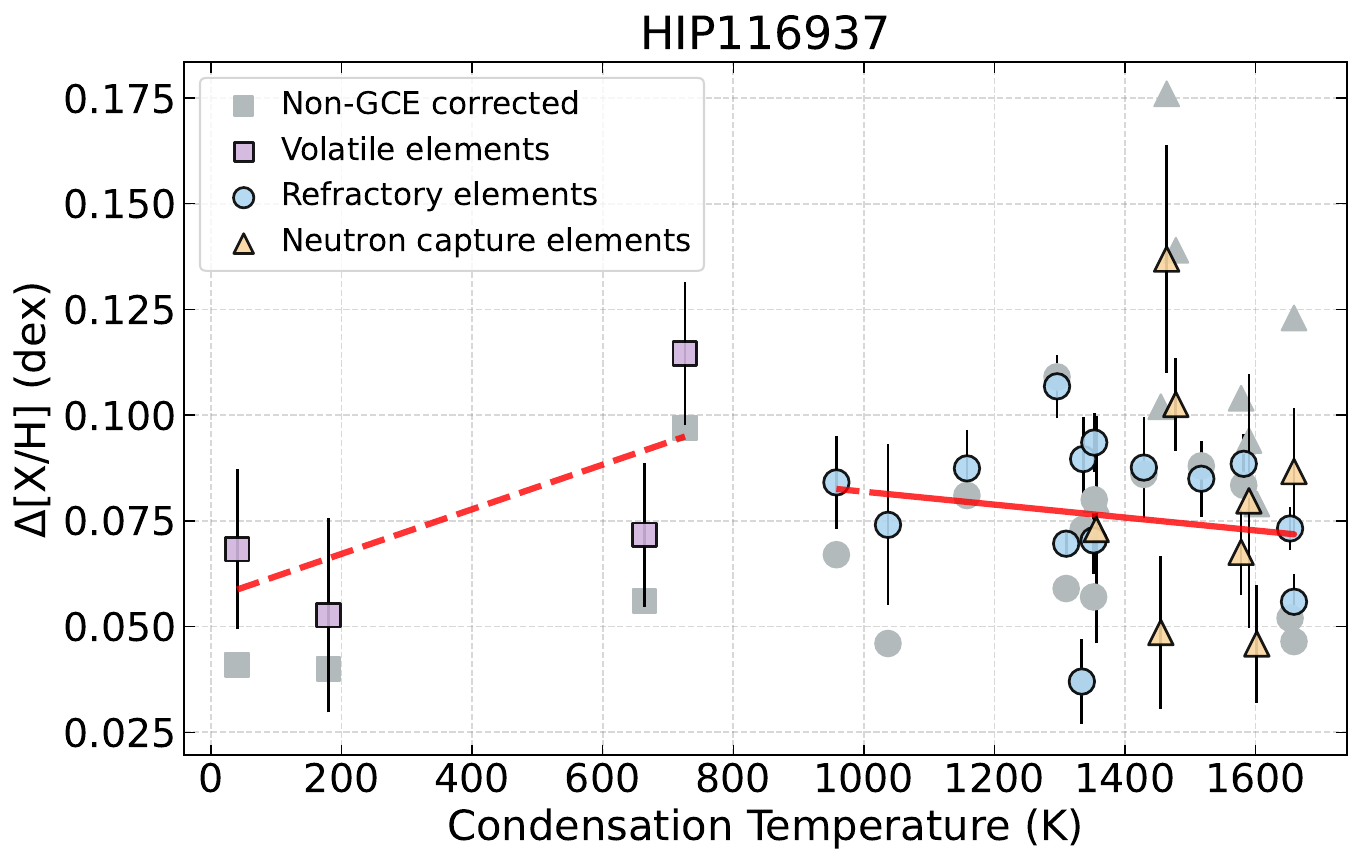} \\
        \includegraphics[width=0.48\textwidth]{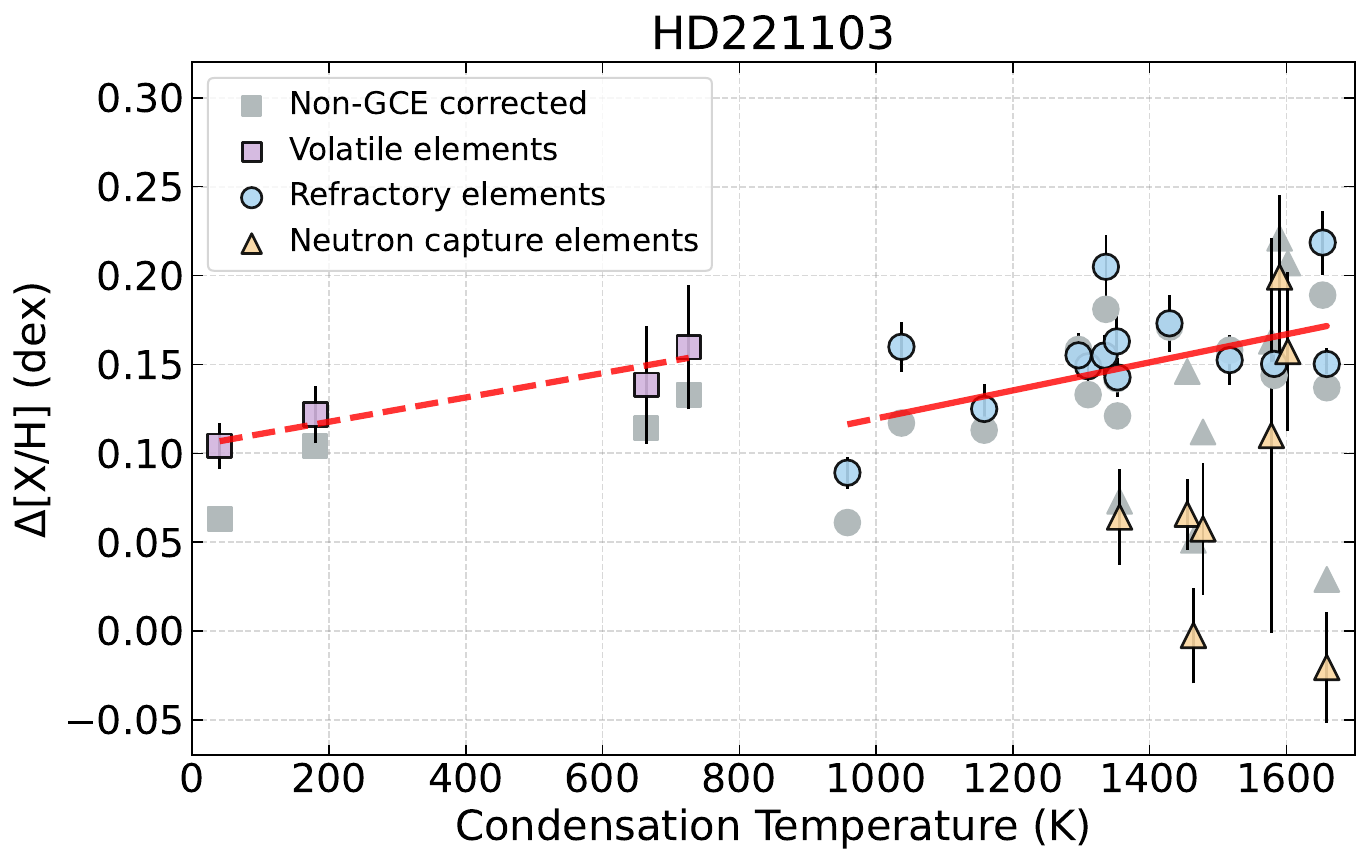} &
        \includegraphics[width=0.48\textwidth]{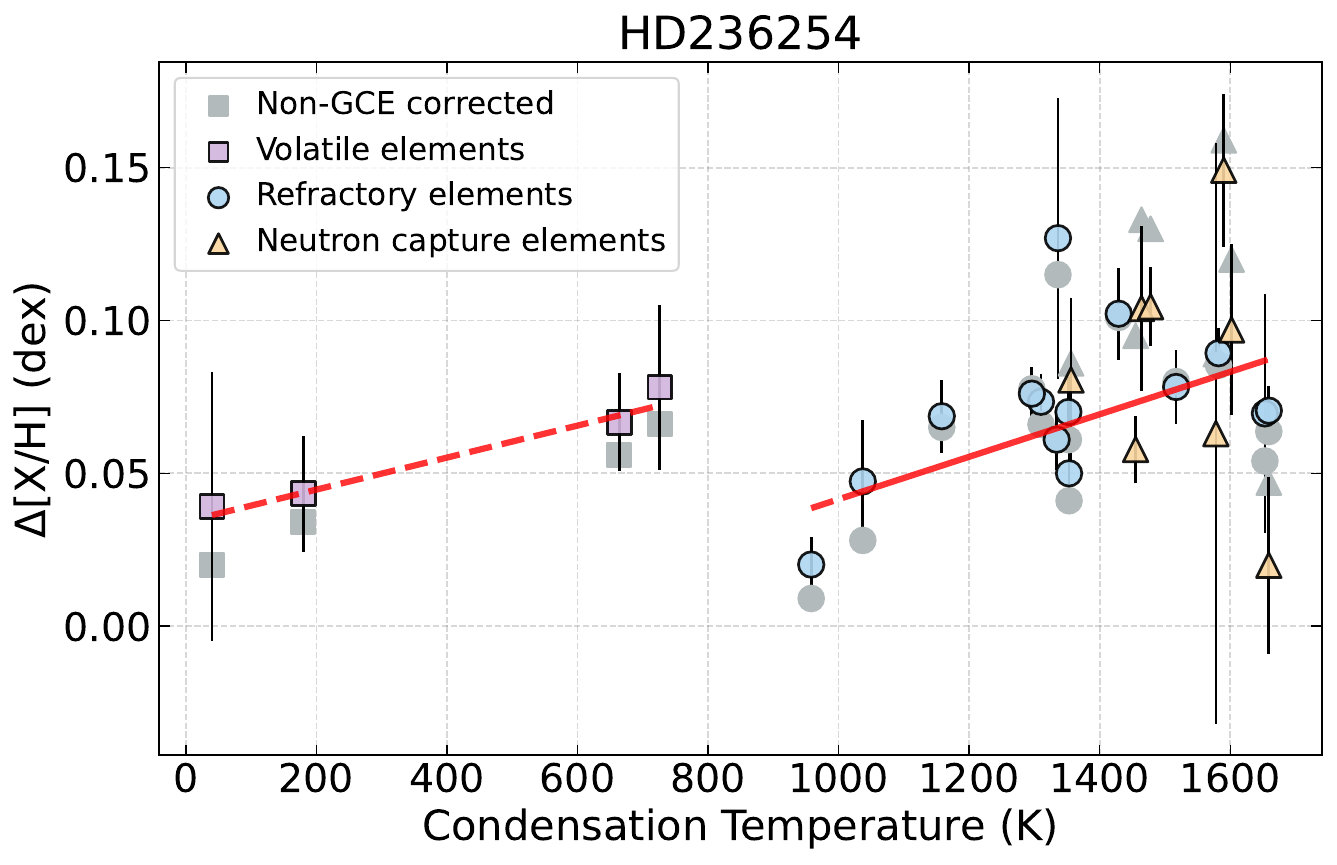} \\
        \includegraphics[width=0.48\textwidth]{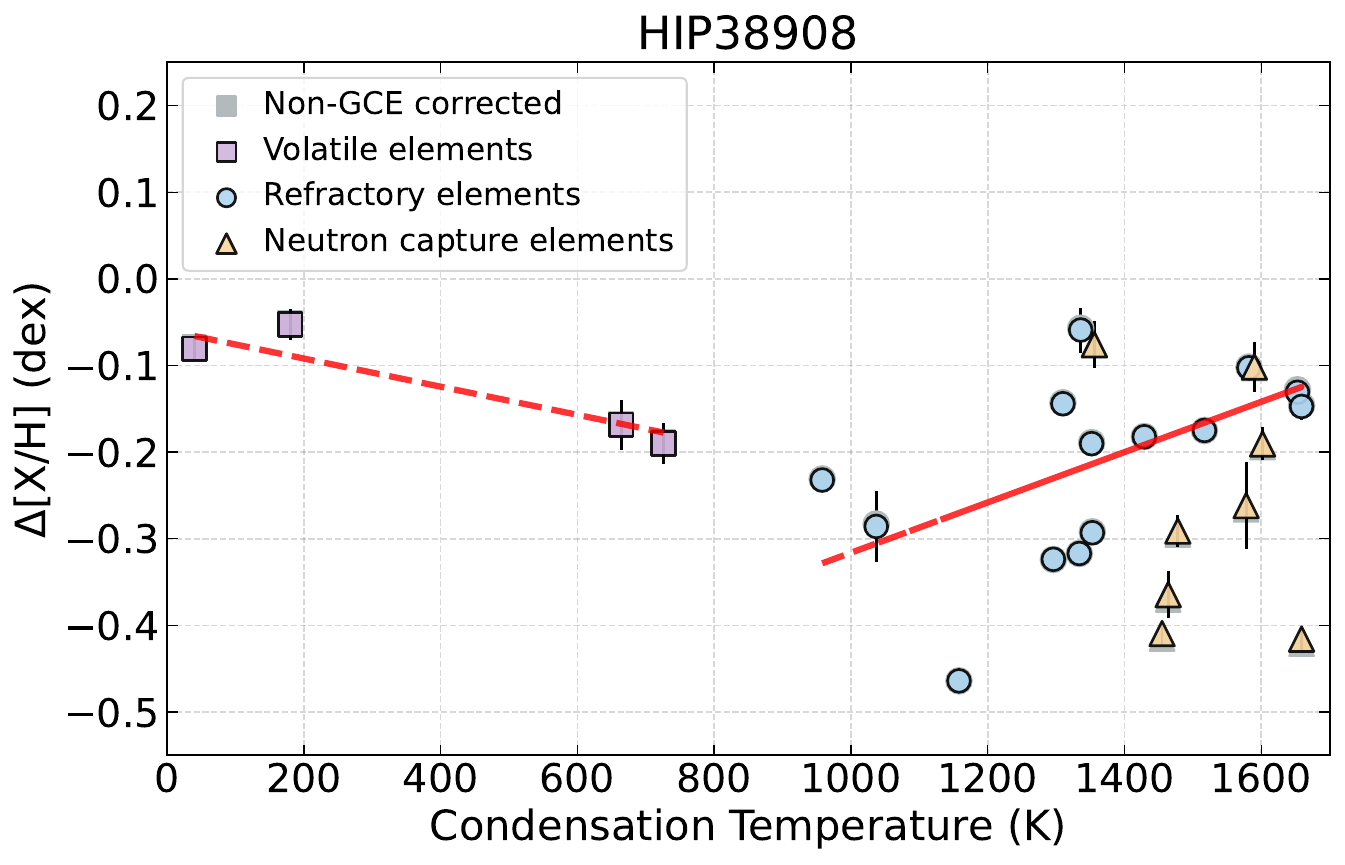} &
        \includegraphics[width=0.48\textwidth]{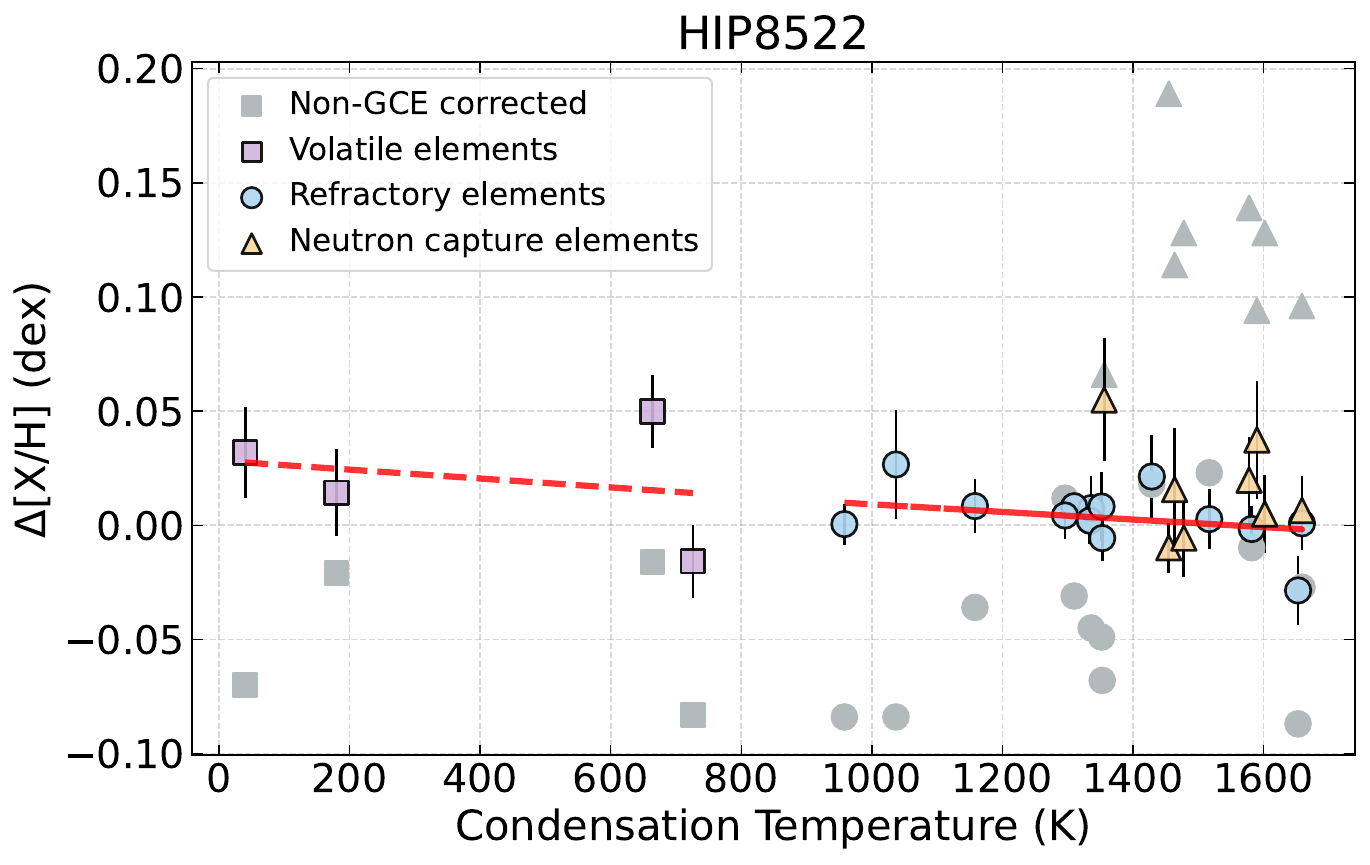}
    \end{tabular}
    \caption{GCE-corrected differential abundances (colored symbols) as a function of \Tc. The red dashed and solid lines show the fits for GCE-corrected volatile ($T_{c} < 900$ K) and refractory elements ($T_{c} > 900$ K), respectively. GCE corrections follow \citet{Cowley:2022MNRAS.512.3684C}, except for HIP 53087, which is solar-aged and requires no correction. Grey symbols indicate the observed (uncorrected) abundances.}
    \label{fig:chemical_tc}
\end{figure*}

To properly analyze the \Tc-slope, the chemical abundances must be corrected for Galactic Chemical Evolution (GCE) effects. These corrections are based on correlations between elemental abundances, stellar age, and metallicity in solar twins \citep[e.g.,][]{Nissen:2015A&A...579A..52N, Spina:2018MNRAS.474.2580S}. In this study, we adopt the GCE corrections from \citet[][Eq. 3]{Cowley:2022MNRAS.512.3684C}, which are based on non-linear relations that better reproduce the observed trends for elements such as Na and Cu. For further details on the effects of adopting linear versus non-linear relations, we refer the reader to \citet{Yana_Galarza_2025}. 

The GCE corrections have little impact on the \Tc-slope in our sample, as most stars have similar metallicities and ages close to solar. Fig. \ref{fig:chemical_tc} shows that HIP 91700, HD 221103, HD 236254, and HIP 38908 exhibit positive \Tc-slopes even after applying the GCE corrections, with p-values $< 0.05$ (see Table \ref{tab:tc_significance}), indicating that the slopes are statistically significant. This positive slope suggests that these stars may have engulfed planets. Although they do not currently show \li\ enhancement---a typical signature of planet engulfment---they may have accreted planets at an earlier stage in their evolution, allowing mechanisms such as thermohaline convection to deplete most of the \li\ by the time they are observed. Following the same approach used for HIP 8522 \citep{Yana_Galarza_2025}, we carried out planet engulfment simulations using two key elements: Li and Fe. Iron is representative of refractory elements and is likely among the most sensitive tracers of planet engulfment after Li. Be could also be a sensitive tracer, possibly even more so than Fe, but no dedicated study on this has yet been conducted.

\begin{table}
    \centering
    \caption{Tc-slopes with their uncertainties and associated p-values.}
    \renewcommand{\arraystretch}{1.2}
    \begin{tabular}{lcr}
    \hline\hline
    \textbf{Designation} & \textbf{Tc-Slope ($\times10^{-5}$ dex K$^{-1}$)}  & \textbf{p-value}\\
    \hline\hline
    HIP 53087  & $0.19\pm1.57$   & $0.979$ \\
    HIP 91700  & $8.93\pm3.53$   & $0.029$ \\
    HIP 93858  & $-2.95\pm1.42$  & $0.104$ \\
    HIP 116937 & $-1.53\pm2.38$  & $0.579$ \\
    HD 221103  & $10.77\pm2.46$  & $0.010$ \\
    HD 236254  & $7.24\pm2.13$   & $0.019$ \\
    HIP 8522   & $-1.67\pm1.25$  & $0.059$ \\
    HIP 38908  & $29.04\pm12.85$ & $0.043$ \\
    \hline\hline
    \end{tabular}
    \label{tab:tc_significance}
\end{table}

Assuming that Li is rapidly destroyed within a few Myr after engulfment through mechanisms such as atomic diffusion, convective overshooting, and thermohaline mixing, we followed the prescription and {\sc MESA} {\tt inlists} given in \citet{Sevilla:2022MNRAS.516.3354S} to model planet engulfment and reproduce the \li\ and \fe\ observed in HIP 91700, HD 221103, HD 236254, and HIP 38908. In brief, the model runs in three stages. The first stage evolves the star up to its ZAMS phase with a metallicity value of $Z = 0.017$ dex, with convection as the only mixing process, yielding a \li\ $\sim$ 3 dex at the ZAMS, which is consistent with the observed \li\ in our sample of young solar twins (see Fig.  \ref{fig:Li_age}). During the second stage a planetary engulfment is simulated through accretion of bulk Earth composition material \citep{MCDONOUGH1995223} at a rate of $3\times 10^{-4} M_{\oplus}$ yr$^{-1}$ at ZAMS. The final stage evolves the star until 10 Gyr. Phases 2 and 3 include mixing mechanisms such as convective overshooting, thermohaline mixing, and atomic diffusion. None of the stages include rotation. For more details, we refer to \citet{Sevilla:2022MNRAS.516.3354S}.

\begin{figure*}
\centering
\begin{tabular}{cc}
\includegraphics[width=0.4\textwidth]{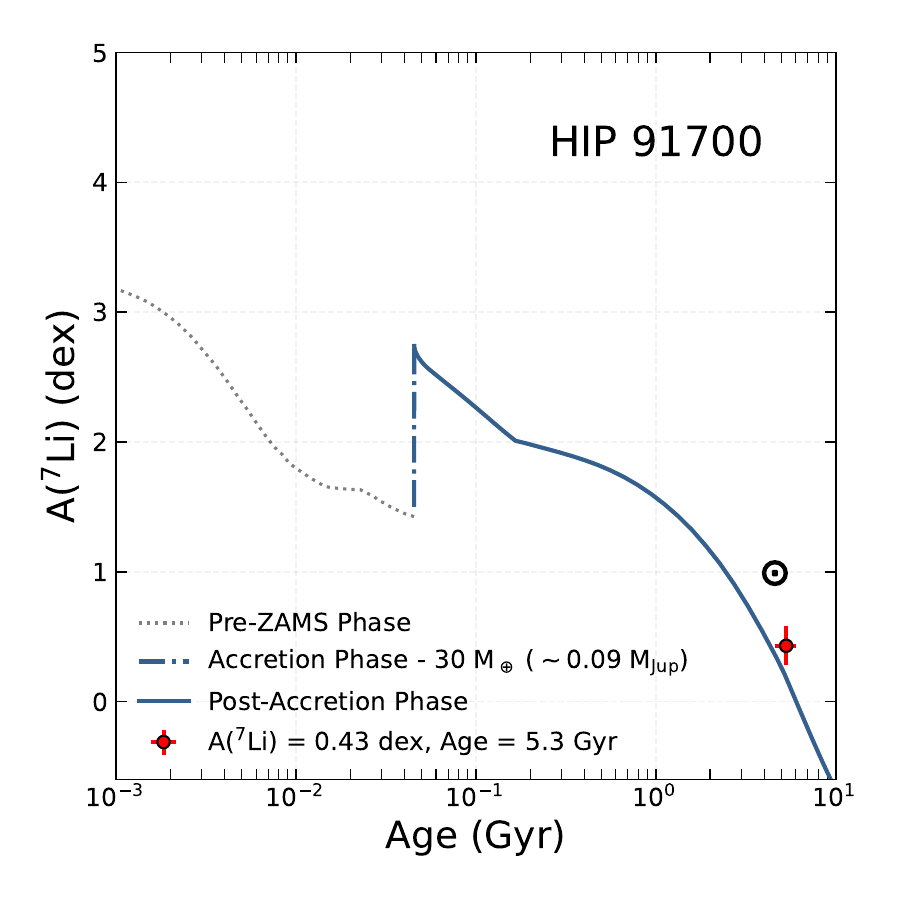} &
\includegraphics[width=0.4\textwidth]{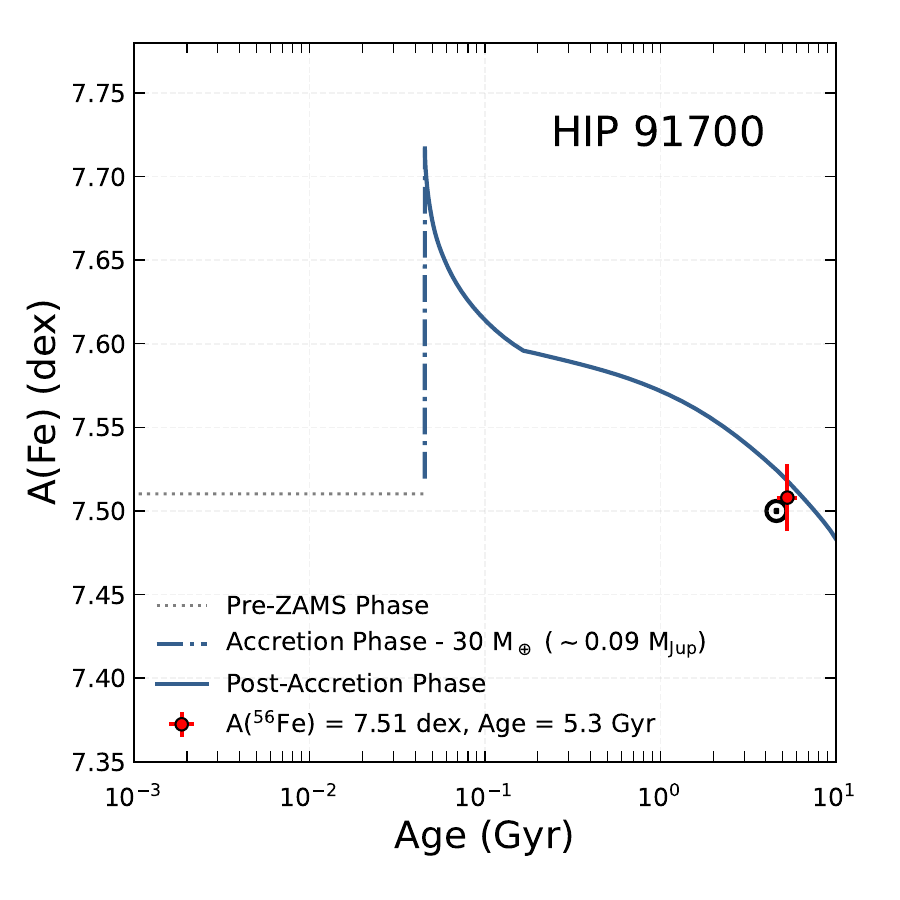} \\
\includegraphics[width=0.4\textwidth]{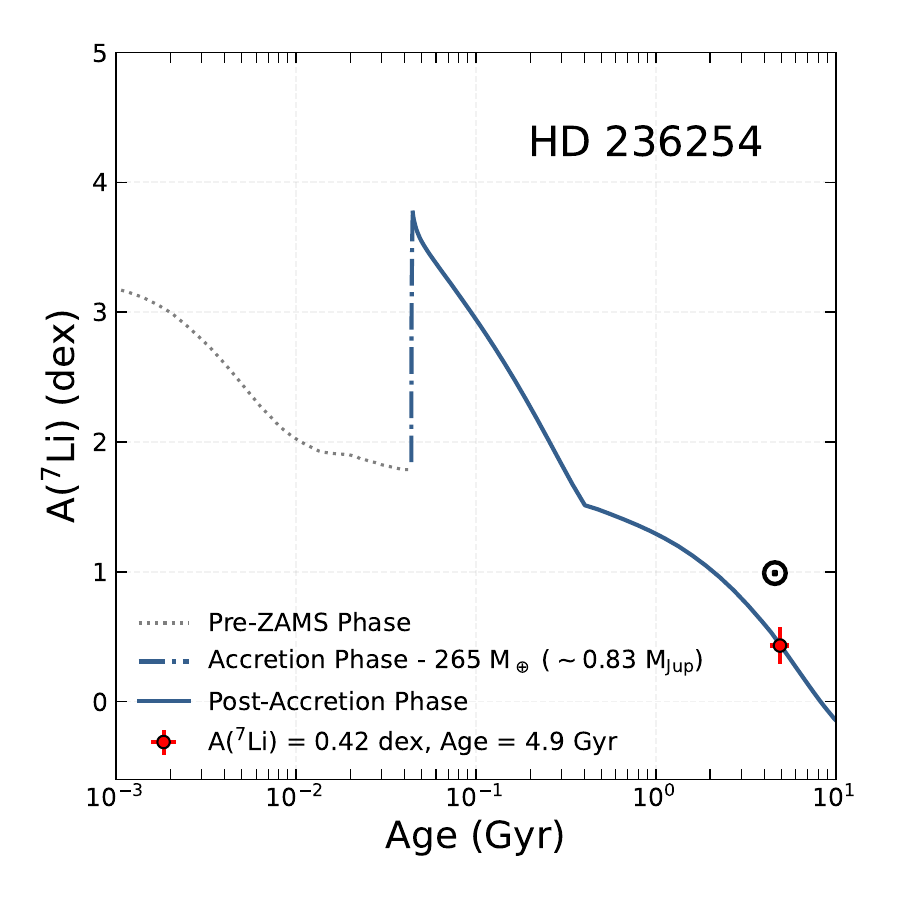} &
\includegraphics[width=0.4\textwidth]{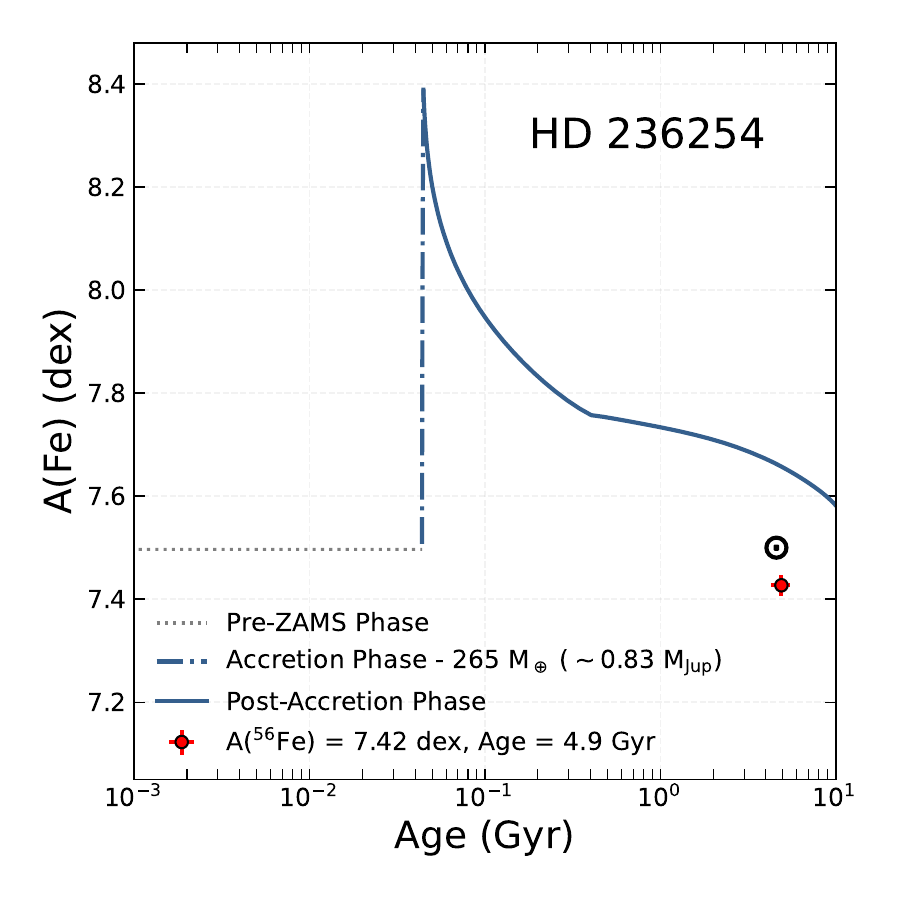}\\
\includegraphics[width=0.4\textwidth]{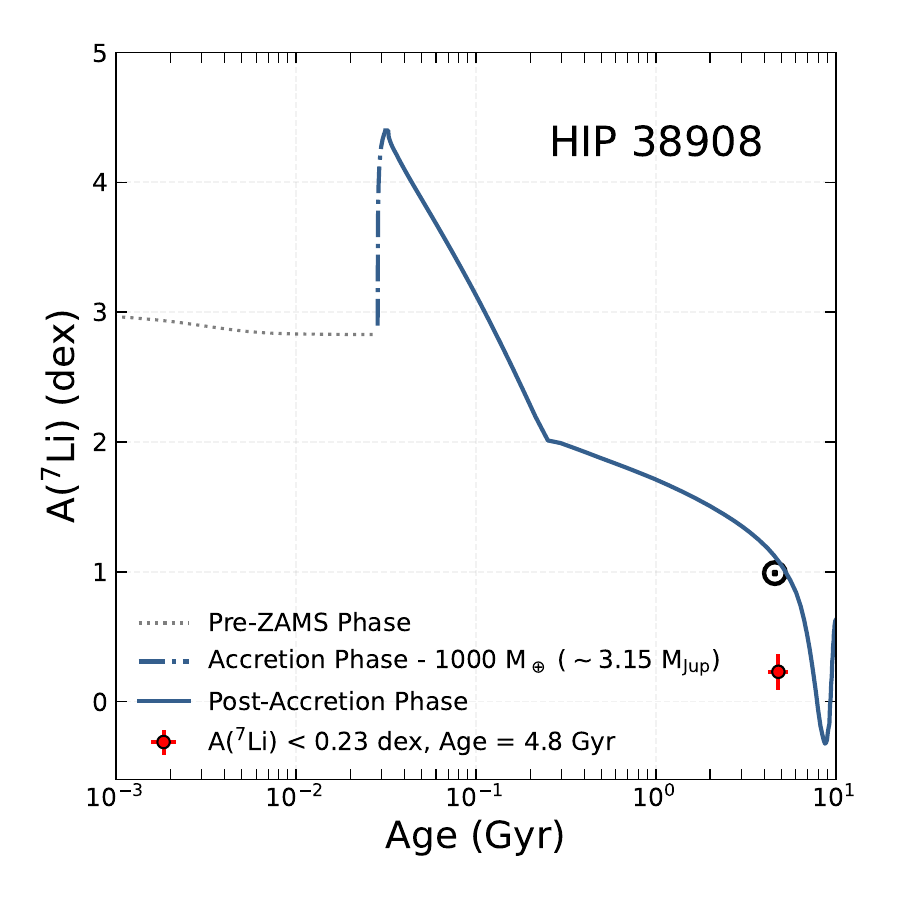} &
\includegraphics[width=0.4\textwidth]{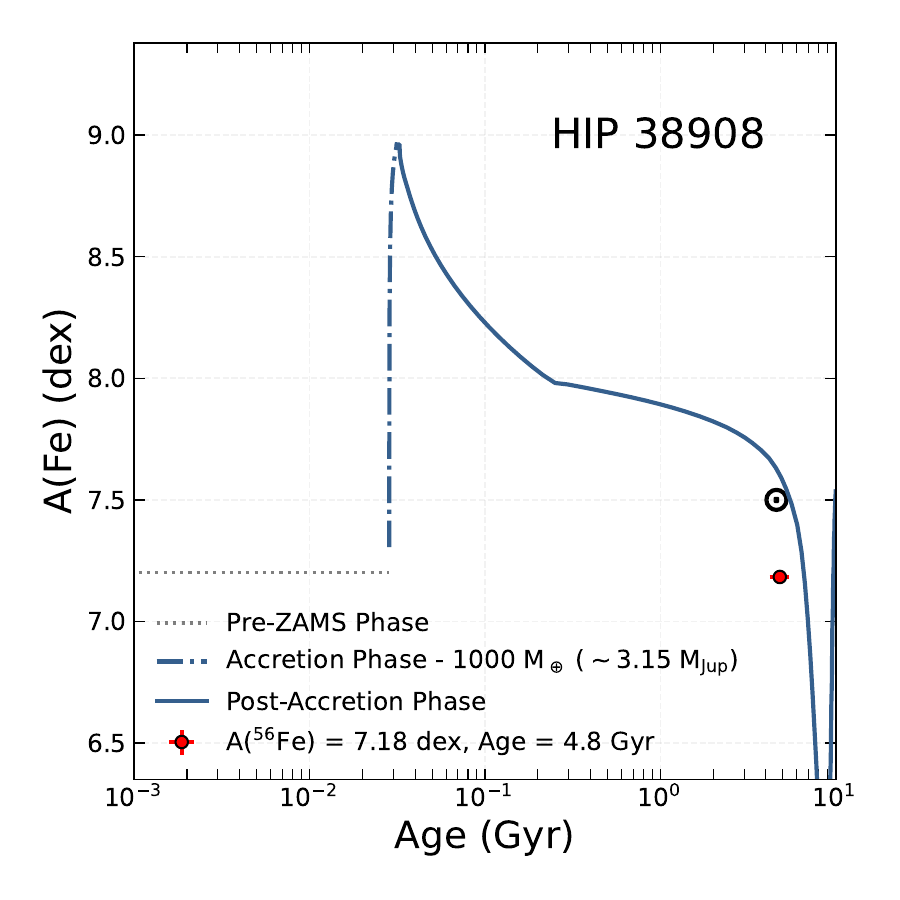}
\end{tabular}
\caption{Lithium and iron abundance for HIP 91700 (upper panels), HD 236254 (middle panels), and HIP 38908 (bottom panels) following a $\sim 30~M_{\oplus}$, $\sim 265~M_{\oplus}$, and $\sim 1000~M_{\oplus}$ engulfment model of bulk Earth composition as a function of time, respectively. The three stages of the simulation are shown as dashed grey lines (Pre-ZAMS), blue dashed-dotted lines (planetary engulfment at ZAMS), and a solid blue line (post-accretion). The stars are represented by the red circles, while the Sun is represented in black by its standard symbol.}
\label{fig:engulfment}
\end{figure*}

Fig. \ref{fig:engulfment} shows the evolution of \li\ (left panel) over time before, during, and after the accretion of $\sim$30 $M_{\oplus}$, $\sim$265 $M_{\oplus}$, and $\sim$1000 $M_{\oplus}$ of bulk Earth composition material for HIP 91700, HD 236254, and HIP 38908, respectively. We also simulated the evolution of A(Fe) (right panel), a representative refractory element, since planet engulfment should enhance refractory elements as well. Unlike Li, Fe is not a fragile element, so its enhancement should still be visible. Our simulation shows that the engulfment of $\sim$30 $M_{\oplus}$ might explain the low \li\ and \fe\ in HIP 91700, making it a strong candidate for planet engulfment. In the case of HD 236254, the engulfment of $\sim$265 $M_{\oplus}$ explains the \li\ depletion but not the \fe, suggesting the refractory elements are not accounted for. For HIP 38908 and HD 221103, we find that an engulfment of more than $\sim$1000 $M_{\oplus}$ is needed to explain both \li\ and \fe\ depletion, but this amount exceeds the total mass of all planets in the solar system, which is unrealistic. A similar conclusion was reported by \citet{Yana_Galarza_2025} HIP 8522. We conclude that HIP 91700 might have engulfed one or more planets, leading to the rapid destruction of \li.

%%%%%%%%%%%%%%%%%%%%%%%%%%%%%%%%%
\subsection{Blue Straggler Stars}
As discussed earlier, blue stragglers (BSSs) are stars that appear younger and more massive than the main-sequence turnoff stars in their population. They are more easily identified in dense environments such as open and globular clusters. Field blue stragglers (FBSSs) are rarer as they are difficult to identify. They are the result of binary evolution, usually through mass transfer or stellar mergers. These formation channels can leave unique chemical signatures of their past, including significant \li\ and \ber\ depletion.

BSSs are, in most cases, metal-poor stars, found in the thick disk or halo of the Galaxy. Some have near-solar masses and are found in the thin disk. One example is HIP 10725, reported by \citet{Schirbel:2015AA...584A.116S}, a solar analog with unusually low lithium (\li$<$0.9 dex) and beryllium (\ber$<$0.2 dex) abundances for its age ($\sim$5.2 Gyr). The authors concluded that this FBSS formed through mass accretion from a companion, based on three main arguments: (1) HIP 10725 shows a strong enhancement in $s$-process elements, likely due to pollution from an AGB star; (2) it exhibits radial velocity variations of up to 70 km s$^{-1}$, suggesting the presence of a companion, probably a white dwarf; and (3) it has a high projected rotational velocity (\vsini\ = 3.3 km s$^{-1}$) and elevated chromospheric activity ($R_{\rm HK} = -4.51$), both inconsistent with its isochronal age ($\sim5.2^{+1.9}_{-2.1}$ Gyr), indicating that HIP 10725 has been rejuvenated.

More recently, HIP 38908 was reported as a FBSS by \citet{Rathsam:2025AA...693A..26R}. The authors concluded that it likely formed through the merger of two Population II stars, based on: (1) the lack of a stellar companion in the radial velocity data, (2) a discrepancy between chemical clock ages, from [Y/Mg] and activity-age relations, and the isochronal age, and (3) the upper limits on Li ($<0.25$ dex) and Be ($<0.21$ dex) abundances. As shown in Table \ref{tab:fundamental parameters}, our inferred stellar parameters, isochronal age, and mass agree well with those reported by \citet{Rathsam:2025AA...693A..26R}, including \li\ (see Table \ref{tab:li_abundance}). Furthermore, our chemical abundance analysis shows no enrichment in neutron-capture elements (see Fig.  \ref{fig:chemical_tc}), which would be expected from a mass-transfer scenario. Kinematically, HIP 38908 is consistent with the thin disk (Fig. \ref{fig:disk_membership}), but its alpha-enhancement ([Mg/Fe] $= 0.30$ dex) and moderate metallicity ([Fe/H] $= –0.317$ dex) place it at the metal-rich end of the thick disk population. Its orbital parameters (e.g., $z_{\rm max}$ = 0.74 kpc) are also typical of thick-disk stars \citep{Li:2018ApJ...860...53L}, further supporting this classification. However, the final classification between thin and thick disk is always determined by its chemistry. Our analysis further corroborate the thick disk membership and the FBSS nature of HIP 38908. This system was likely formed through the merger of two stars, a formation pathway that appears common among FBSSs in the thick disk. We refer the reader to \citet{Preston:2015ASSL..413...65P} for a detailed discussion of FBSSs in the thick disk and halo populations.

Unlike the FBSSs HIP 10725 and HIP 38908, our previously reported anomalous star HIP 8522 is younger and shows activity, rotation, and chemical abundances (excluding Li) typical of young solar twins ($<1$ Gyr). \citet{Yana_Galarza_2025} rule out the possibility that HIP 8522 is an FBSS formed through binary mass transfer, as RV measurements reveal no companion, and the star lacks the carbon and oxygen depletion commonly associated with this channel \citep[see subsection 3.5 of][]{Wang:2024arXiv241010314W}. The stellar collision channel is also unlikely, as it tends to occur in dense environments such as GCs and OCs \citep[e.g.,][]{Hills:1976ApL....17...87H, Lombardi:2002ApJ...568..939L, Sills:2010MNRAS.407..277S}. However, formation via binary mergers remains a possibility, especially in less crowded environments where this mechanism dominates. Three-dimensional magnetohydrodynamic (3D MHD) simulations \citep{Schneider:2019Natur.574..211S} suggest that BSSs formed via mergers may not retain clear signatures of their origin, and can resemble typical main-sequence single stars in surface properties and composition \citep{Glebbeek:2013MNRAS.434.3497G}, except for Li, which is strongly depleted and may remain as the only signature of the merger event. 

This seems to be the case for HIP 8522. MHD simulations of binary mergers support this hypothesis \citet{Schneider:2019Natur.574..211S}, however the models are limited to massive stars ($\sim8 M_\odot$), and blue straggler simulations in the solar-mass regime exist only for mass transfer in OCs and GCs, which is an unlikely formation channel for our stars. It is also important to note that HIP 8522 shows no discrepancies between isochronal ages and chemical clocks, and its activity, rotation, and chemical profile (aside from Li) are consistent with other young solar twins. If the current age reflects the time since a merger, we would expect to see additional signs—such as abnormal rotation, activity, or composition, which are not observed. This lowers the likelihood that HIP 8522 is an FBSS, though the scenario cannot be completely ruled out.

Following the same approach as for the previously reported stars, we considered the possibility that our targets are FBSSs. First, we ruled out the binary mass transfer scenario, as no stellar companions were detected in our sample. The second channel, FBSSs formed via collisions of binary stars, is also unlikely since this mechanism predominantly occurs in dense stellar environments. Thus, the only remaining plausible explanation for FBSSs in our sample, as in the case of HIP 8522, is stellar mergers. Although we cannot directly confirm this scenario, several arguments may provide indirect evidence. For instance, based on HIP 10725 and HIP 38908, it is known that FBSSs typically exhibit discrepancies between isochronal ages and chemical clocks. In our sample, only HIP 91700 and HD 221103 show such discrepancies. Both stars have similar ages, masses, and \li, but HD 221103 is slightly more metal-rich, with [Fe/H] enhanced by 0.03 dex.

If we assume the ages derived from [Y/Mg]- and [Y/Al]-age correlations represent the true ages, as in \citet{Rathsam:2025AA...693A..26R} for HIP 38908, then HD 221103 emerges as the best FBSS candidate. In this scenario, its true age would be around 7 Gyr, but the isochronal and activity ages (both $\sim4$ Gyr) would suggest the merger event occurred 4 Gyr ago. The chemical clock ages are consistent with its low projected rotational velocity, \vsini\ $\sim$ 1.0 km s$^{-1}$ (Table \ref{tab:clocks}), which is typical of old stars (see Fig.  5 of \citealp{Leo:2016A&A...592A.156D}). Fig. \ref{fig:GCE-solartwins} compares HD 221103 to the GCE trend of solar twins from the Inti sample, showing that other elements, such as \ion{Sr}{1}, \ion{Ba}{2}, \ion{Ce}{2}, \ion{Mg}{1}, \ion{Al}{1}, and \ion{Y}{2}, are also anomalous.

HIP 91700 presents the opposite case. Its age inferred from chemical clocks and stellar activity is lower than its isochronal age. If we instead assume the older age is correct, as in \citet{Schirbel:2015AA...584A.116S} with HIP 10725, the chemical clock ages may simply reflect the consequences of the merger event. In this scenario, the isochronal age is supported by its \vsini\ $\sim$ 2.1 km s$^{-1}$, typical of solar-age stars---considering that the Sun has \vsini$_{\odot} = 1.9$ km s$^{-1}$. Both stars could be good FBSS candidates, but confirmation is challenging. The lack of simulations for FBSSs in solar-mass stars and their effects on chemical composition complicates determining which inferred age should be considered the true one. Fig. \ref{fig:GCE-solartwins} shows that \ion{Ca}{1}, \ion{Cr}{1}, \ion{Cr}{2}, \ion{Sr}{1}, \ion{Y}{2}, and \ion{Ce}{2} are anomalous in HIP 91700 compared to GCE trend of solar twins.

With the aim of searching for an additional signature of a binary merger, we compared the elements that show anomalies in HD 221103 and HIP 91700, finding that only \ion{Sr}{1}, \ion{Y}{2}, and \ion{Ce}{2} are common to both. However, more thin disk stars that are candidates for FBSSs need to be analyzed to determine whether these elements are reliable signatures of a binary merger. Figure~\ref{fig:GCE-solartwins} also shows that HIP 38908 does not follow the GCE trends of solar twins, except for \ion{Cr}{1}, \ion{Cr}{2}, \ion{Ni}{1}, \ion{Cu}{1}, \ion{Sr}{1}, \ion{La}{2}, \ion{Ce}{2}, and \ion{Nd}{2}, which is expected since it is a thick disk star. We conclude that there is no clear evidence pointing to which of these elements may be considered potential signatures of a binary merger.

On the other hand, HIP 53087, HIP 93858, HIP 116937, and HD 236254 are particularly intriguing. These stars do not show significant discrepancies between isochronal ages and chemical clocks as FBSSs generally do, and their \vsini\ values and activity levels are consistent with those of solar twins. As shown in Fig.  \ref{fig:GCE-solartwins}, they follow the GCE trends of solar twins. These stars could be FBSSs formed via stellar mergers that occurred long enough ago that any traces of their dramatic origins have been erased. As a result, they now appear typical compared to other solar twins, with Li being the only remaining evidence of their merger history.

%%%%%%%%%%%%%%%%%%%%%%%%%%%%%%%
\subsection{Episodic Accretion}
As discussed in the introduction, episodic accretion can significantly affect the evolution of low-mass stars and brown dwarfs at early ages. Compared to steady accretion, episodic accretion can produce more compact and hotter objects with faster \li\ depletion. The simulations of \citet{Baraffe:2009ApJ...702L..27B, Baraffe:2010A&A...521A..44B} show that, during high accretion episodes ($>10^{-4}$ \sm\ yr$^{-1}$), the temperature in the convective zone can increase enough to cause rapid Li destruction, with complete depletion occurring in less than 1 Myr for a 1 \sm\ star. Later work \citep{Baraffe:2017A&A...597A..19B}, coupling stellar evolution with 2D hydrodynamics, confirmed these findings and further emphasized that episodic accretion, under the assumption of a cold or hybrid scenario (see details in \citet{Baraffe:2017A&A...597A..19B}), yields hotter, smaller stars with enhanced \li\ depletion.

\citet{Baraffe:2017A&A...597A..19B} found that, among the 60 models computed under various assumptions and parameters and covering the mass range from $\sim0.05$ \sm\-$1.2$ \sm, only one---assuming either cold or hybrid accretion scenario with a final mass of 0.735 \sm---showed very strong \li\ depletion at 3 Myr (see their Figures 4 and 5). While violent accretion bursts can lead to strong \li\ depletion, their results show no clear correlation between burst intensity and the level of Li depletion across all models (see their Fig. 6). Unexpectedly, some accreting models in the 0.5–1.1 \sm\ mass range retain higher surface Li than their non-accreting counterparts. Overall, the study concludes that there is no simple link between burst strength and \li\ depletion.

\begin{figure}
    \centering
    \begin{tabular}{c}
        \includegraphics[width=\columnwidth]{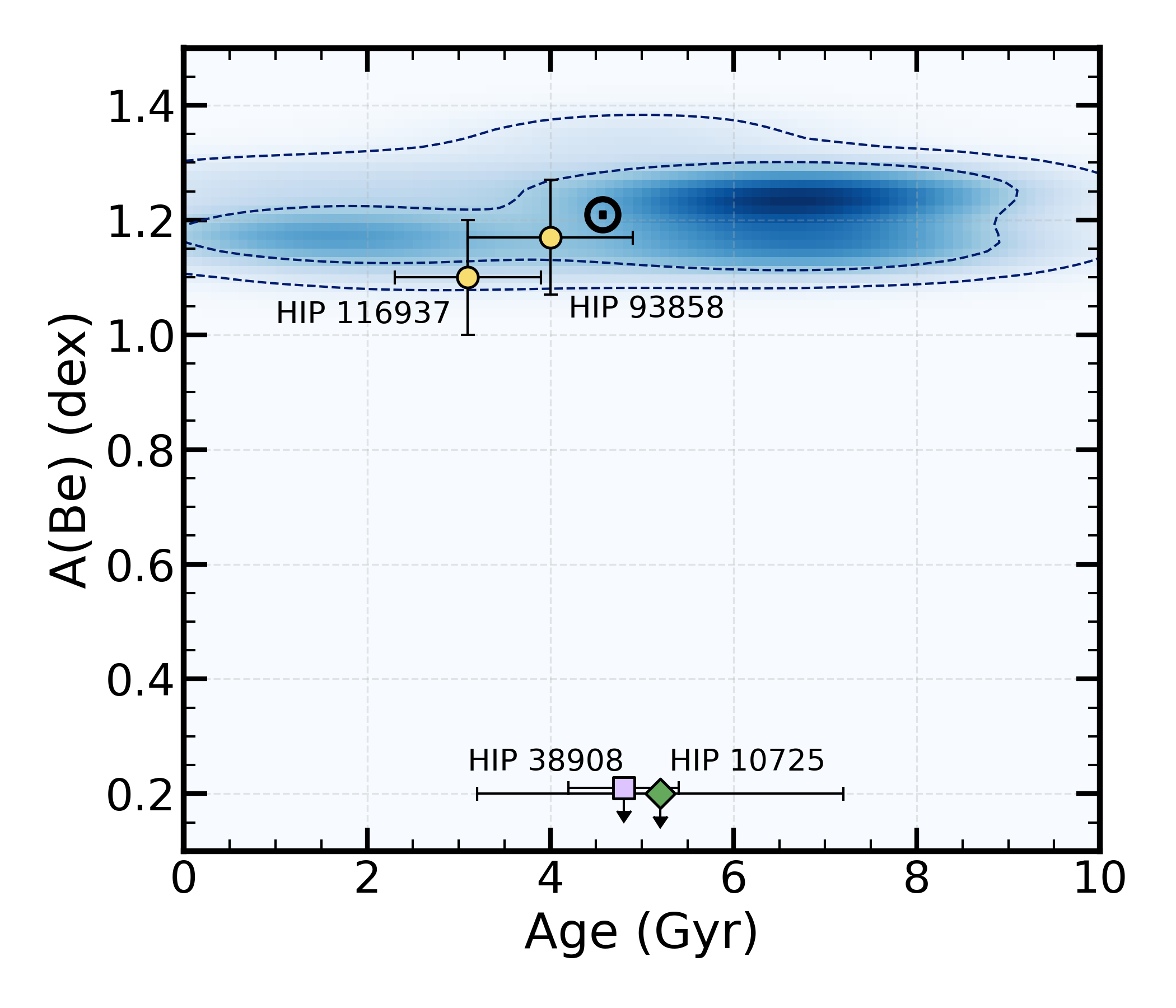}
    \end{tabular}
    \caption{Beryllium abundances as a function of isochronal age. Data are taken from \citet{Reggiani:2025ApJ...984..108R} and \citet{Boesgaard:2023ApJ...943...40B}. The yellow circles represent stars from our sample (HIP 116937, HIP 93858), while the square and diamond symbols show the previously reported field blue straggler stars HIP 38909 and HIP 10725, respectively. The Sun is represented by its standard symbol in black.}
    \label{fig:Be_age}
\end{figure}

Both models of episodic accretion suggest that Li depletion is possible, particularly for stars with final masses around one solar mass, matching the mass range of our sample. In the models from \citet{Baraffe:2017A&A...597A..19B}, one case (0.735 \sm) predicts total \li\ depletion and another (0.926 \sm) predicts partial depletion (\li\ $\sim 0.4$ dex), both under a cold accretion scenario (see left panel of their Fig. 4 and 5). These two cases represent approximately $\sim$3\% of the 60 accretion models explored. In our study of 150 stars, we find that seven, including HIP 8522, show significant \li\ depletion ($\sim 0.4$ dex), roughly $\sim$5\% of the sample. This fraction is consistent with the predictions of episodic accretion models, assuming this is the only mechanism responsible for Li depletion. Note that the hybrid scenario also yields one case for very strong Li depletion, but as pointed by \citet{Baraffe:2017A&A...597A..19B}, hybrid accretion tends to moderate the effect of Li depletion and produces less strongly Li depleted models. 

To further support episodic accretion, we explored the Be evolution of solar twins recently reported by \citet{Reggiani:2025ApJ...984..108R}. Fig. \ref{fig:Be_age} shows Be abundance as a function of age for solar twins, where the trend differs from that of Li (see Fig.  \ref{fig:Li_age}), as Be does not decrease as a function of time. Be abundances for HIP 116937 and HIP 93858, as measured by \citet{Reggiani:2025ApJ...984..108R}, follow the general trend observed in solar twins. We extended the models from \citet{Baraffe:2017A&A...597A..19B} to include Be evolution up to $\sim7$ Gyr and found that Be is not depleted (\ber\ $\sim$0.85 and $\sim$0.99 dex; see Table \ref{tab:extensionmods}) in the two cases (0.735 \sm\ and 0.926 \sm) that show significant \li\ depletion. This is consistent with our findings, at least for the stars with available Be measurements. 

Fig. \ref{fig:Be_age} also shows the Be abundances of two stars previously reported as FBSSs (HIP 38908 and HIP 10725), which clearly do not follow the Be evolution trend seen in solar twins. This would suggests that the stars in our sample are not FBSSs, but instead experienced episodic accretion events that depleted Li without affecting Be. However, Be measurements for the rest of the sample are needed to further test the episodic accretion scenario. Moreover, the lack of FBSS simulations involving mergers of solar-mass stars makes it difficult to assess whether such events could deplete Li without affecting Be.

\newcolumntype{Y}{>{\raggedright\arraybackslash}X}

\begin{table*}[]
\caption{Evidence for each Li-depletion scenario.}
\centering
\begin{tabularx}{\textwidth}{l Y Y Y Y}
\hline\hline
\textbf{Designation} & \textbf{Planetary Engulfment} & \textbf{BS - Binary Mass Transfer} & \textbf{BS - Binary Merger} & \textbf{Episodic Accretion} \\
\hline
HIP 53087            & No refractory enrichment                         & No companion; confirmed planet                                             & Possible       & Possible                    \\
HIP 91700            & Refractory enrichment; agreement with model      & No companion                                                               & Possible; anomalous composition                             & Possible   \\
HIP 93858            & No refractory enrichment                         & No companion; confirmed planet                                             & Possible & Possible; not Be-depleted           \\
HIP 116937           & No refractory enrichment                         & No companion; planet candidate                                             & Possible                             & Possible; not Be-depleted   \\
HD 221103            & No refractory enrichment                         & No companion                                                               & Possible; anomalous composition                             & Possible                    \\
HD 236254            & Refractory enrichment; poor agreement with model & No companion                                                               & Possible                             & Possible                    \\
HIP 8522             & No refractory enrichment                         & No companion                                                               & Possible                             & Possible                    \\
HIP 38908            & Refractory enrichment; poor agreement with model & No companion                                                               & Reported; anomalous composition      & Unlikely; Be-depleted       \\
HIP 10725            & No refractory enrichment                         & Reported; confirmed binary with anomalous stellar activity and composition & Unlikely; confirmed binary           & Unlikely; Be-depleted       \\
\hline\hline
\end{tabularx}
\label{tab:conclusion}
\end{table*}

%%%%%%%%%%%%%%%%%%%%%%%%%%%%%%%%%
\section{Summary and Conclusions}
\label{sec:conclusions}
In this study, we present a sample of six solar twins exhibiting significant \li\ depletion, which, together with our previously reported star HIP 8522, suggest a puzzling history of enhanced internal mixing or of accretion at very early stages of evolution. Using high-resolution, high S/N spectra from instruments such as HARPS, HDS, TS23, and MIKE (see Table \ref{tab:instruments}), we derived precise stellar parameters, chemical compositions, ages, masses, and Li abundances, the latter obtained via spectral synthesis of the $\sim 6707.8$ \AA\ line (see Tables \ref{tab:fundamental parameters}, and \ref{tab:clocks}, \ref{tab:li_abundance}). To explore the observed Li depletion, we considered several possible scenarios that could enhance such depletion, including blue straggler formation via stellar merger or mass transfer from an unresolved companion, planetary engulfment, and early episodic accretion.

We first explored the planet engulfment scenario, where Li initially enhanced by accretion is rapidly depleted, primarily through thermohaline convection. We detected enhancements in refractory elements as a function of condensation temperature for HIP 91700, HD 221103, HD 236254, and HIP 38908 (see Fig.  \ref{fig:chemical_tc}), which may be signatures of planet engulfment. Using \textsc{MESA}, we simulated the engulfment of planets with different masses to reproduce the observed \li\ and 
\fe\ abundances, with Fe as a representative element for refractory elements. Our results suggest that the engulfment of $\sim$30 M$\oplus$ can explain both \li\ and \fe\ of HIP 91700. For HD 236254, the accretion of $\sim$265 M$\oplus$ reproduces the observed \li\ but not \fe, making this scenario unlikely. For HIP 38908 and HD 221103, planetary engulfment is completely ruled out, as it would require the accretion of more than 1000 M$_\oplus$ to match both its \li\ and \fe.

For the blue straggler scenario, we ruled out binary mass transfer, as no evidence of stellar companions was found in the radial velocity analysis and the SED fitting (see Fig.  \ref{fig:SED}). Instead, we detected two new exoplanets (see Fig.  \ref{fig:exoplanetsHIP}) with the HARPS spectrograph: HIP 53087 $b$ ($P = 225$ d, $e = 0.30$, $M \sin i = 23.4,M_{\oplus}$) and HIP 116937 $b$ ($P = 766$ d, $e = 0.084$, $M \sin i = 23.5,M_{\oplus}$). We also ruled out mass transfer by collisions, as such events typically occur in dense stellar environments like open or globular clusters. The only remaining possibility is binary merging.

Following the examples of previously identified field blue stragglers HIP 38908 and HD 10725, we investigated additional binary merger signatures by analyzing stellar ages using multiple methods, including isochronal ages, chemical clocks, and activity indicators, as inconsistencies among these age indicators are common in FBSSs. We find that only HIP 91700 and HD 221103 exhibit such discrepancies, identifying them as potential FBSS candidates. The remaining stars (HIP 53087, HIP 93858, HIP 116937, and HD 236254) do not show discrepancies between age indicators and, similar to our previously studied HIP 8522, may represent particularly interesting objects within our sample. 

At present, it remains difficult to determine whether these stars are FBSSs, as no simulations currently exist for one-solar-mass. Furthermore, the fact that two of these stars host planets raises the question of whether planetary systems can survive such energetic events as binary stellar mergers. If they are indeed FBSSs, they would be the first known one-solar-mass FBSSs hosting exoplanets in the thin disk.

We also explored the early episodic accretion scenario proposed by \citet{Baraffe:2017A&A...597A..19B}, which predicts \li\ depletion in solar-mass stars that undergo episodic accretion bursts during their very early stages of evolution. In their simulations of episodic accretion for 60 objects with different masses, they found that about $\sim3\%$ experienced significant Li depletion and ended up with final masses similar to the Sun. The Inti sample \citep{Yana_Galarza:2021MNRAS.504.1873Y} currently includes around 150 solar twins with measured Li abundances. We found that seven of them, including HIP 8522, show extreme Li depletion, roughly $\sim5\%$ of the sample, consistent with the predictions of \citet{Baraffe:2017A&A...597A..19B}. We also have Be measurements for two of our stars (HIP 116937 and HIP 93858) from \citet{Reggiani:2025ApJ...984..108R}, and found that they follow the expected Be evolution of solar twins (see Fig.  \ref{fig:Be_age}). This agrees with the predictions of \citet{Baraffe:2017A&A...597A..19B}, which suggest that Be is not significantly affected by episodic accretion for the $\sim3\%$ of stars with Li depletion (see Table \ref{tab:extensionmods}). We conclude that early episodic accretion is the most likely scenario to explain the \li\ depletion in our sample. However, Be measurements for the remaining five stars are still needed to confirm this. Table \ref{tab:conclusion} summarizes the various scenarios that may explain the \li\ depletion observed in our sample.

The Li-depleted stars presented in this study represent a significant challenge to current non-standard models. Identification of more Li-depleted solar-twins, along with developments in simulations, especially of one-solar-mass FBSSs, will be key to testing different scenarios to explain the observed abundances. These stars offer valuable clues for improving our understanding of internal mixing processes and/or early stages of evolution in solar-mass stars.

%%%%%%%%%%%%%%%%%%%%%%%%%%
\section*{Acknowledgments}
I.W. acknowledges support from the Carnegie Astrophysics Summer Student Internship Program (CASSI) and Pomona College. J.Y.G. acknowledges support from a Carnegie Fellowship. H.R. acknowledges the support from NOIRLab, which is managed by the Association of Universities for Research in Astronomy (AURA) under a cooperative agreement with the National Science Foundation. T.F. acknowledges support from Yale Graduate School of Arts and Sciences. I.B. acknowledges the STFC Consolidated Grant ST/Y002164/1. D.L.O acknowledges the support from CNPq (PCI 301612/2024-2). N.W.C.L. gratefully acknowledges the generous support of a Fondecyt Regular grant 1230082, as well as support from Millenium Nucleus NCN2023\_002 (TITANs). R. L-V acknowledges support from CONAHCyT through a postdoctoral fellowship within the program ``Estancias Posdoctorales por M\'exico''. E.M. acknowledges funding from FAPEMIG under project number APQ-02493-22 and a research productivity grant number 309829/2022-4 awarded by the CNPq. G.C.S. thanks the FAPESP fellowships 2021/01303-3 and 2023/16319-8.

The observations were carried out within the framework of the Subaru-Keck/Subaru-Gemini time exchange program which is operated by the National Astronomical Observatory of Japan. We are honored and grateful for the opportunity of observing the Universe from Maunakea, which has the cultural, historical and natural significance in Hawaii. We are also deeply thankful to the Likan Antai of the Atacama Desert, who have observed the dark constellations of the Andean skies for millennia.

This work has made use of data from the European Space Agency (ESA) mission Gaia (\url{https://www.cosmos.esa.int/gaia}), processed by the Gaia Data Processing and Analysis Consortium (DPAC, \url{https://www.cosmos.esa.int/web/gaia/dpac/consortium}). The ID ESO programs used in this work are: 60.A-9709(G), 072.C-0488(E), 074.C-0364(A), 077.C-0364(E), 084.C-0229(A), 086.C-0230(A), 088.C-0011(A), 091.C-0936(A), 092.C-0579(A), 093.C-0062(A), 094.C-0797(A), 095.C-0040(A), 095.D-0026(A), 096.C-0053(A), 096.C-0499(A), 097.C-0021(A), 097.D-0150(A), 0100.C-0487(A), 105.2045.001, 105.2045.002, 109.2374.001, 183.C-0972(A), 190.D-0237(E), 192.C-0852(A), 196.C-1006(A), 198.C-0836(A). 

Funding for the DPAC has been provided by national institutions, in particular the institutions participating in the Gaia Multilateral Agreement. 

This publication makes use of VOSA, developed under the Spanish Virtual Observatory (\url{https://svo.cab.inta-csic.es}) project funded by MCIN/AEI/10.13039/501100011033/ through grant PID2020-112949GB-I00.
VOSA has been partially updated by using funding from the European Union's Horizon 2020 Research and Innovation Program, under Grant Agreement nº 776403 (EXOPLANETS-A).

\vspace{5mm}
\facilities{Magellan:Clay/MIKE, ESO:3.6m/HARPS, Smith:TS23, Subaru:HDS}

\software{
\textsc{numpy} \citep{van_der_Walt:2011CSE....13b..22V}, 
\textsc{matplotlib} \citep{Hunter:4160265}, 
\textsc{pandas} \citep{mckinney-proc-scipy-2010}, 
\textsc{astroquery} \citep{Ginsburg:2019AJ....157...98G}, 
\textsc{iraf} \citep{Tody:1986SPIE..627..733T}, 
\textsc{iSpec} \citep{Blanco:2014A&A...569A.111B, Blanco:2019MNRAS.486.2075B}, 
\textsc{Kapteyn} Package \citep{KapteynPackage}, 
\textsc{moog} \citep{Sneden:1973PhDT.......180S}, 
q$^{\textsc{2}}$ \citep{Ramirez:2014A&A...572A..48R}, 
\textsc{Gala} \citep{gala}}, 
\textsc{astrobase} \citep{2021zndo...4445344B}, 
\textsc{VOSA} \citep{2008A&A...492..277B}.

\appendix
\renewcommand{\thefigure}{A\arabic{figure}} % Redefines the figure numbering format to "A" followed by the Arabic numeral.
\renewcommand{\theHfigure}{A\arabic{figure}}
\renewcommand{\thetable}{A\arabic{table}}
\renewcommand{\theHtable}{A\arabic{table}}

\setcounter{figure}{0} % Resets the figure counter to 0, so the first figure will be A1.
\setcounter{table}{0}

\section{SED plots}

\begin{figure*}[htbp]
\centering
\begin{tabular}{cc}
\includegraphics[width=0.4\textwidth]{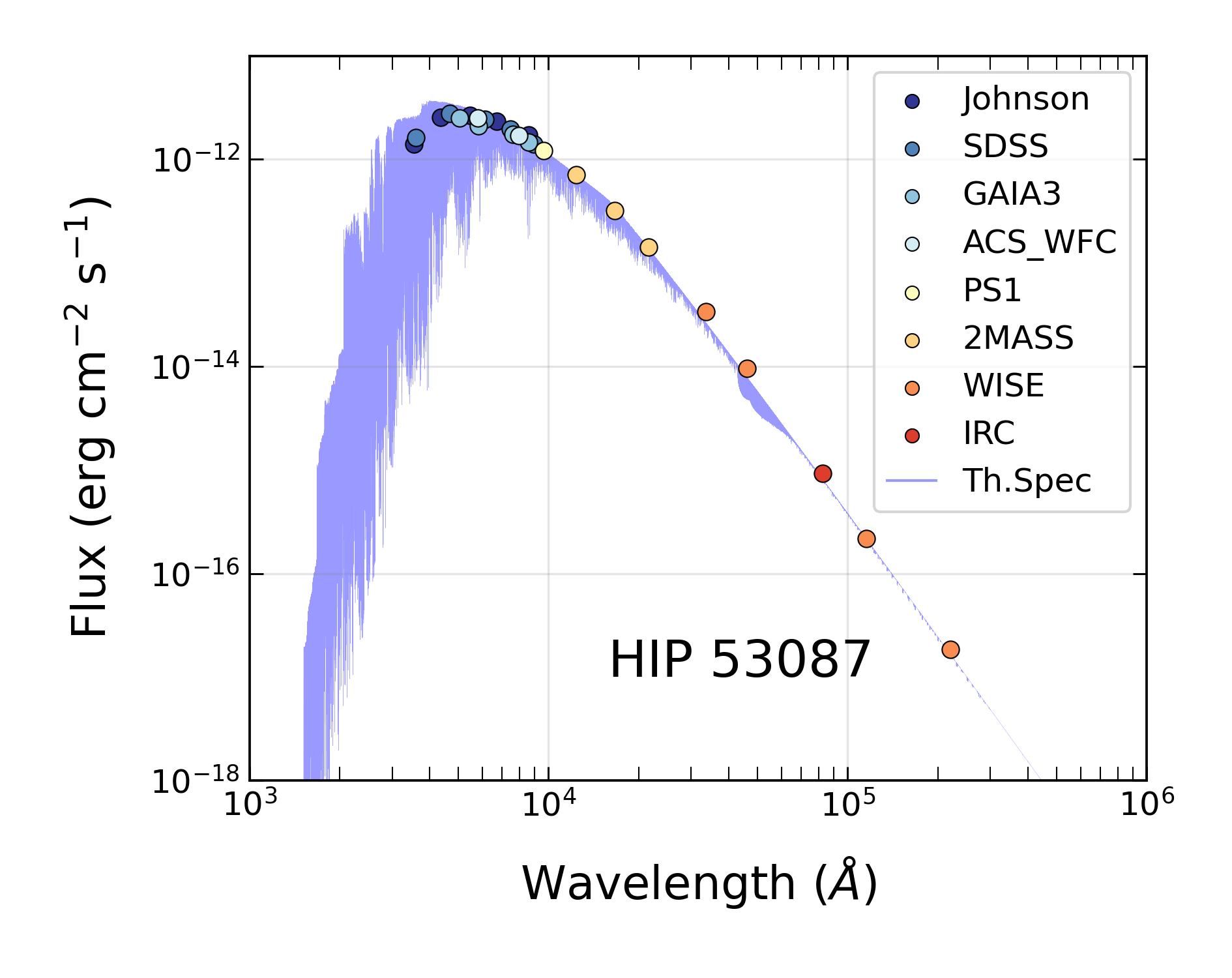} &
\includegraphics[width=0.4\textwidth]{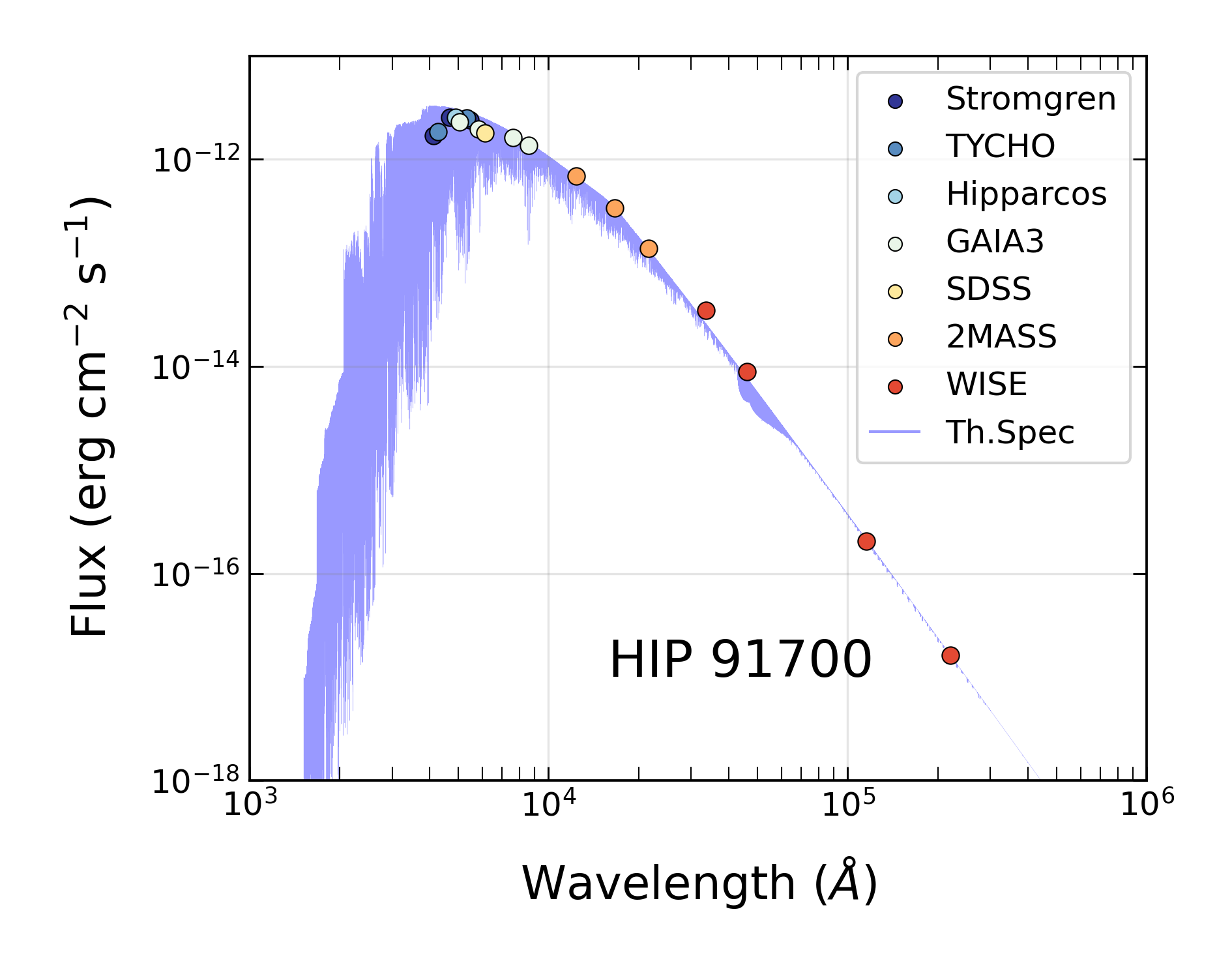} \\
\includegraphics[width=0.4\textwidth]{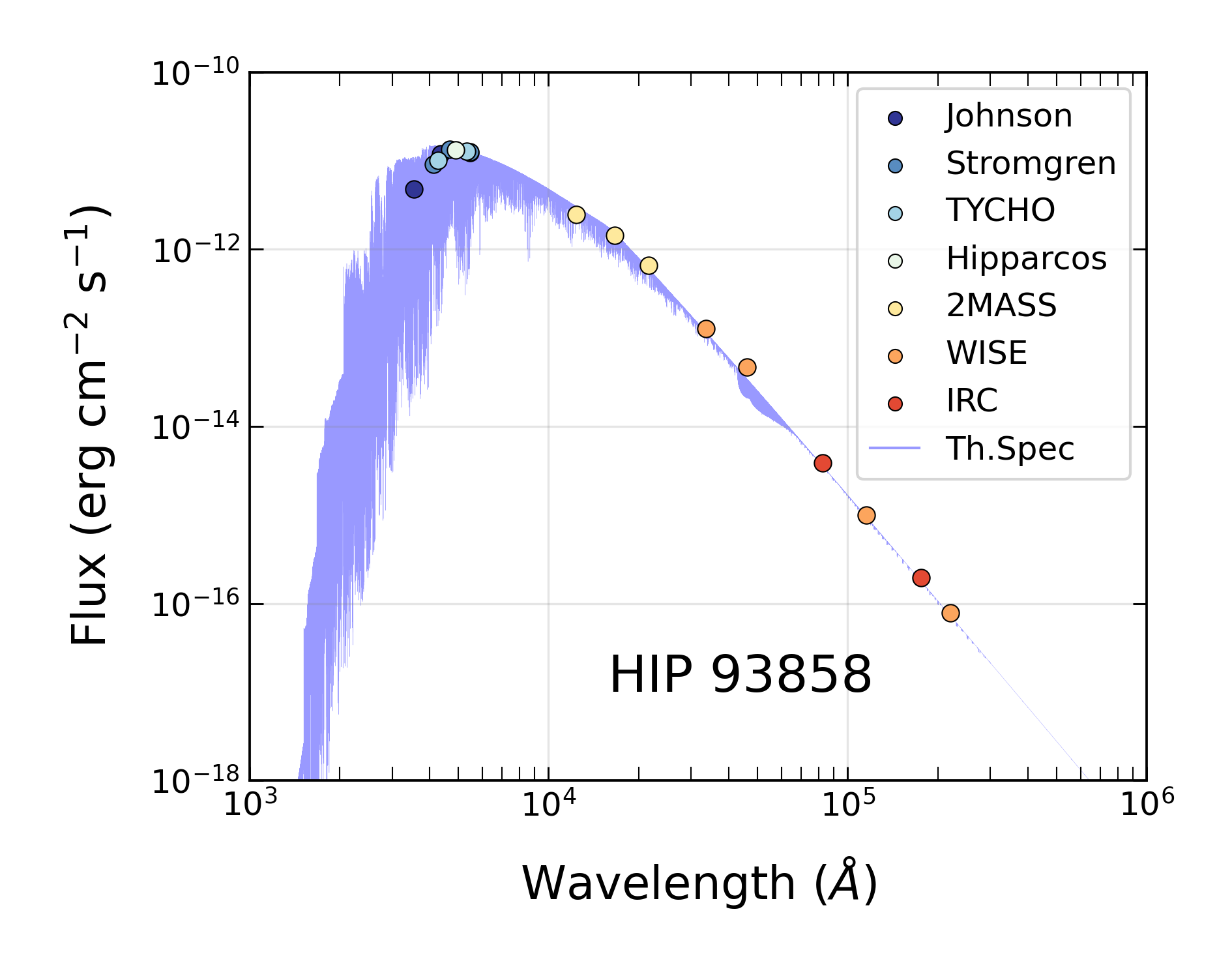} &
\includegraphics[width=0.4\textwidth]{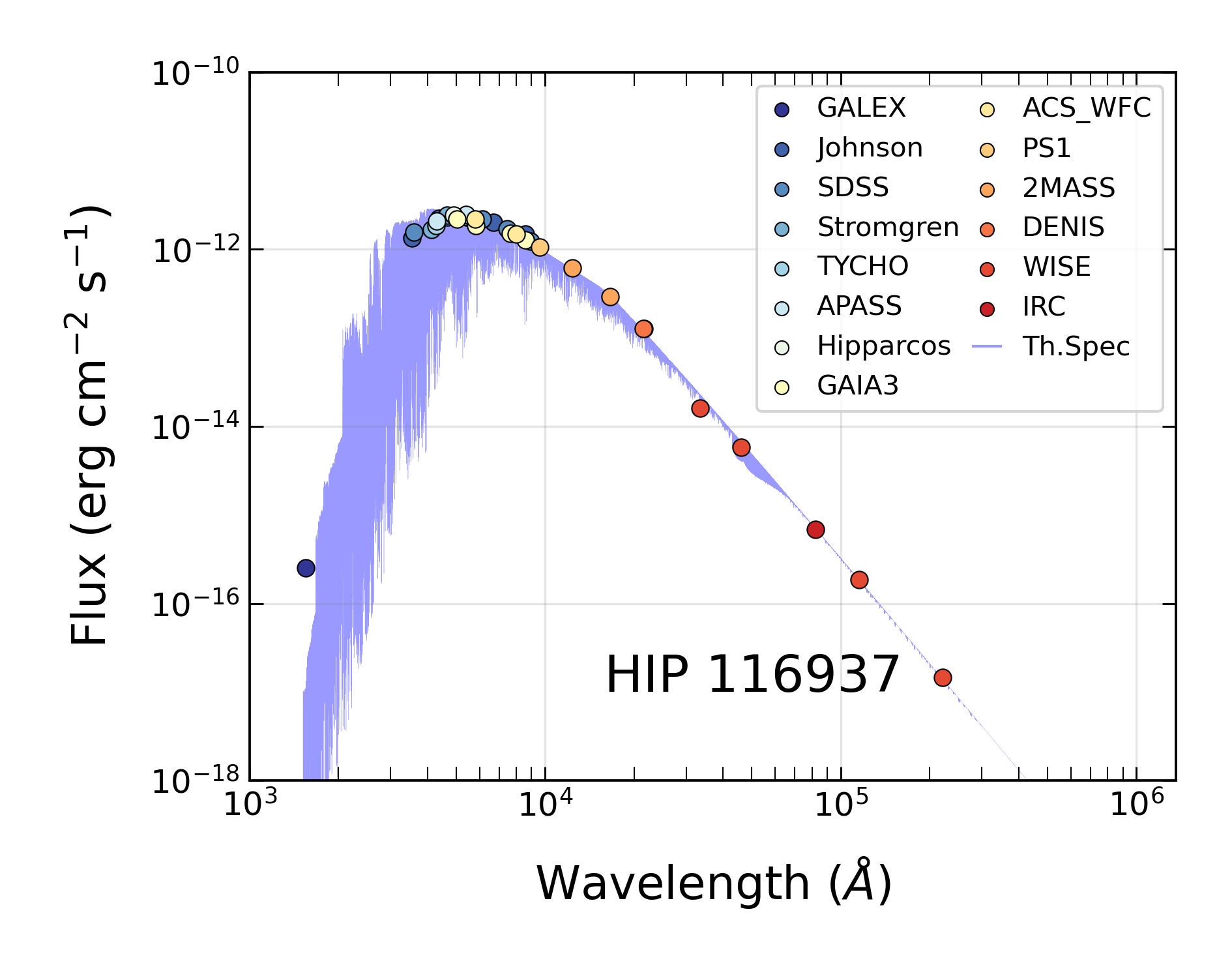} \\
\includegraphics[width=0.4\textwidth]{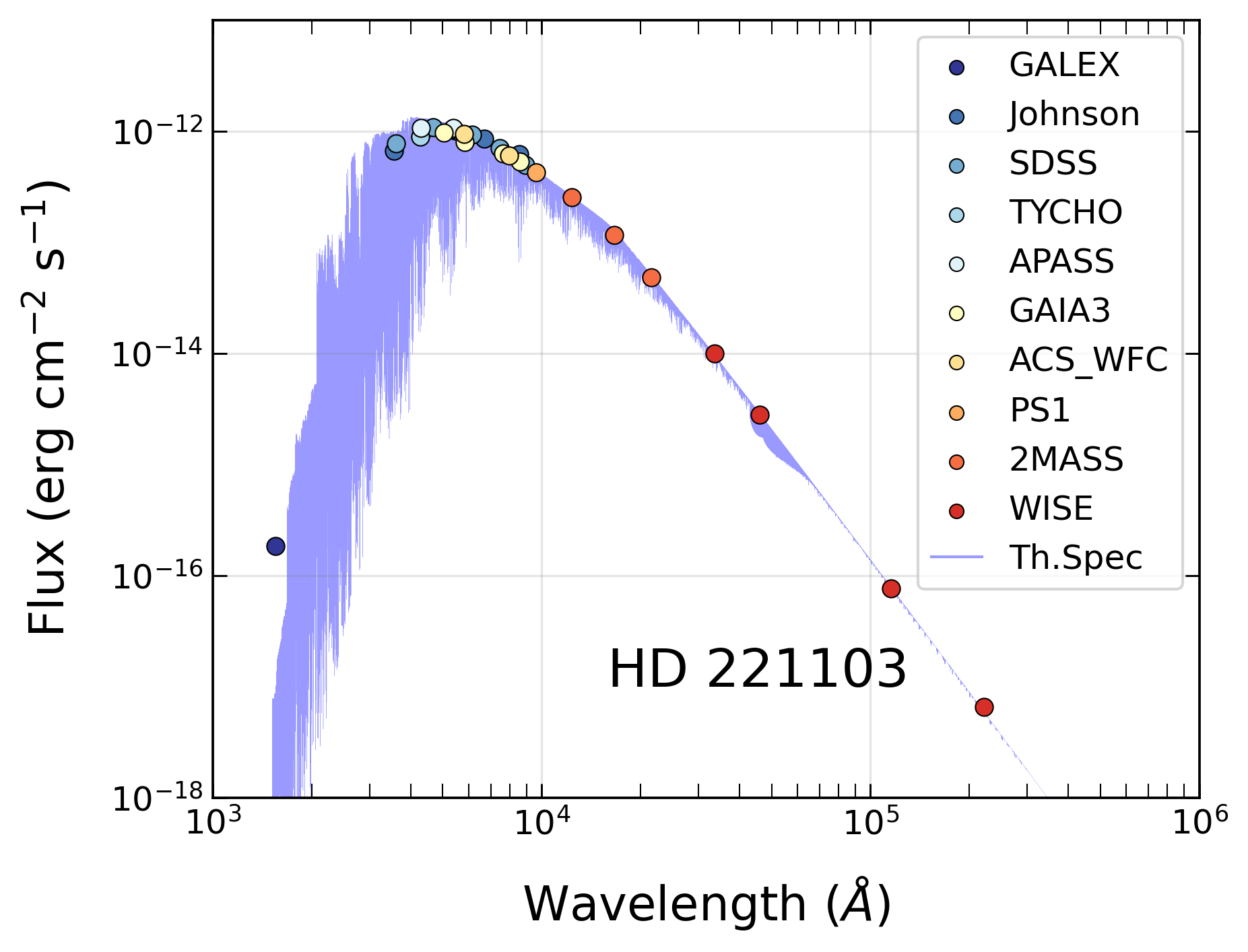} &
\includegraphics[width=0.4\textwidth]{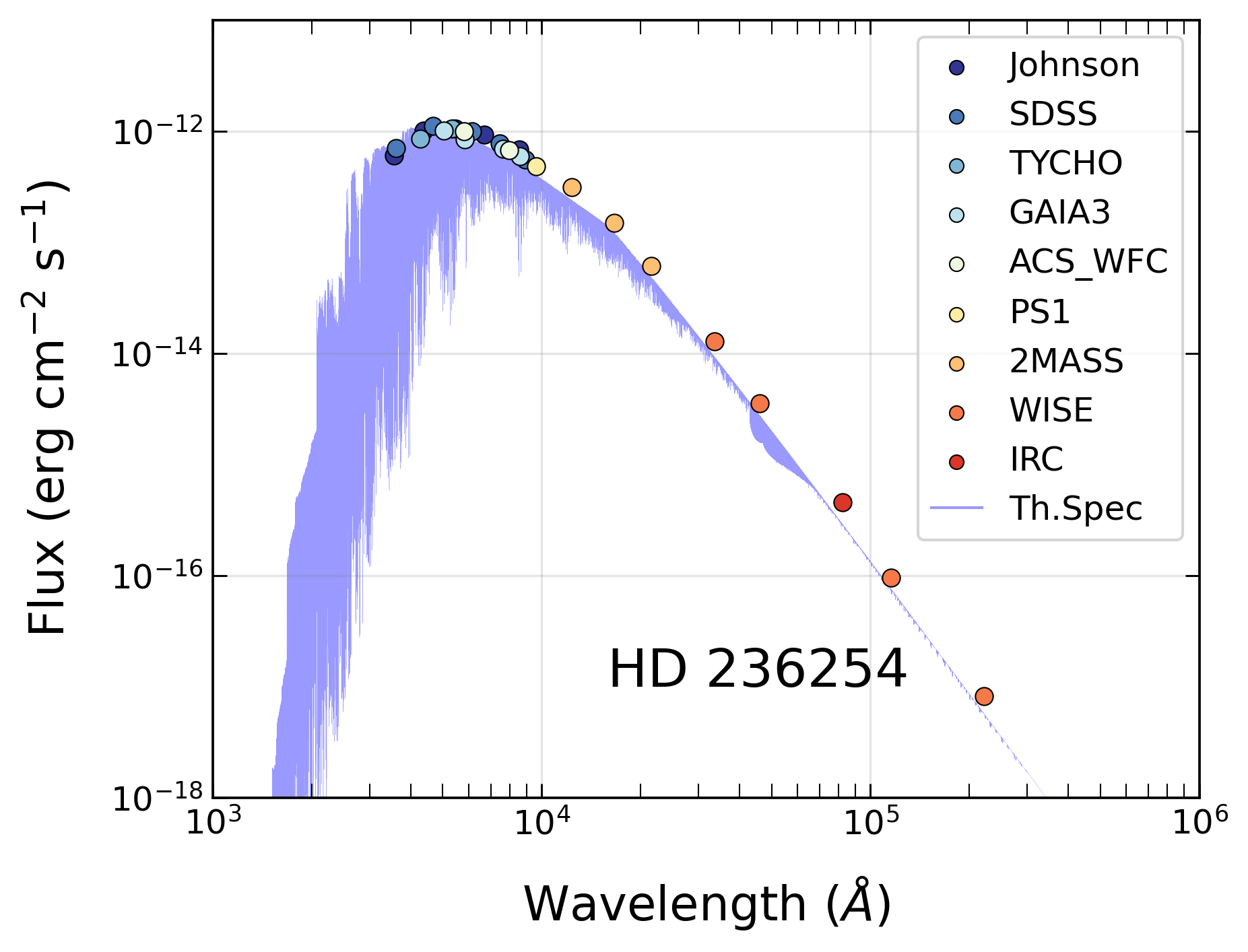}
\end{tabular}
\caption{Spectral energy density fitting for our sample of stars. The colored circles correspond to the
observed photometric data, while the purple solid line is the best-fit synthetic
spectrum.}
\label{fig:SED}
\end{figure*}

\FloatBarrier

\section{Joint Keplerian and Gaussian Process Modulations for HIP 53087's and HIP 93858's Radial Velocity Time-Series}

\begin{figure*}[htbp]
    \centering
    \includegraphics[width = \linewidth]{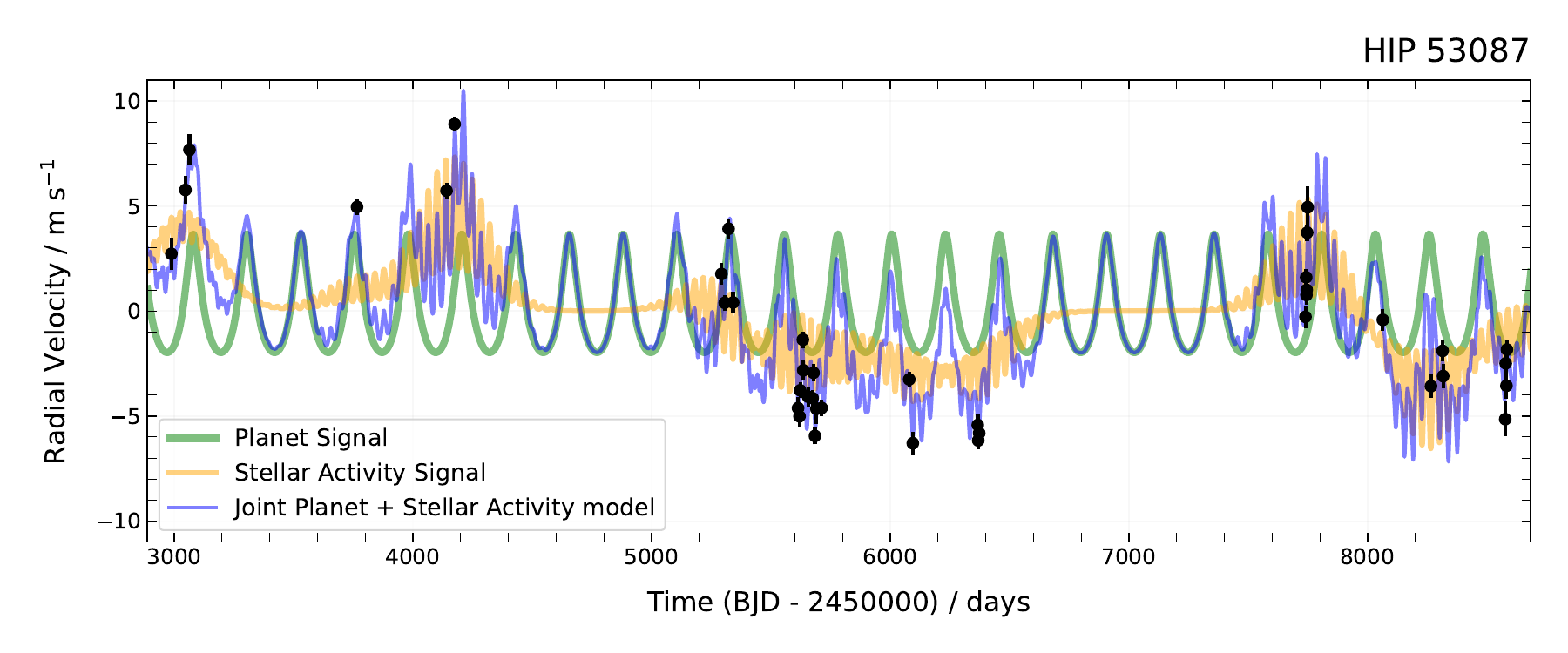}
    \includegraphics[width = \linewidth]{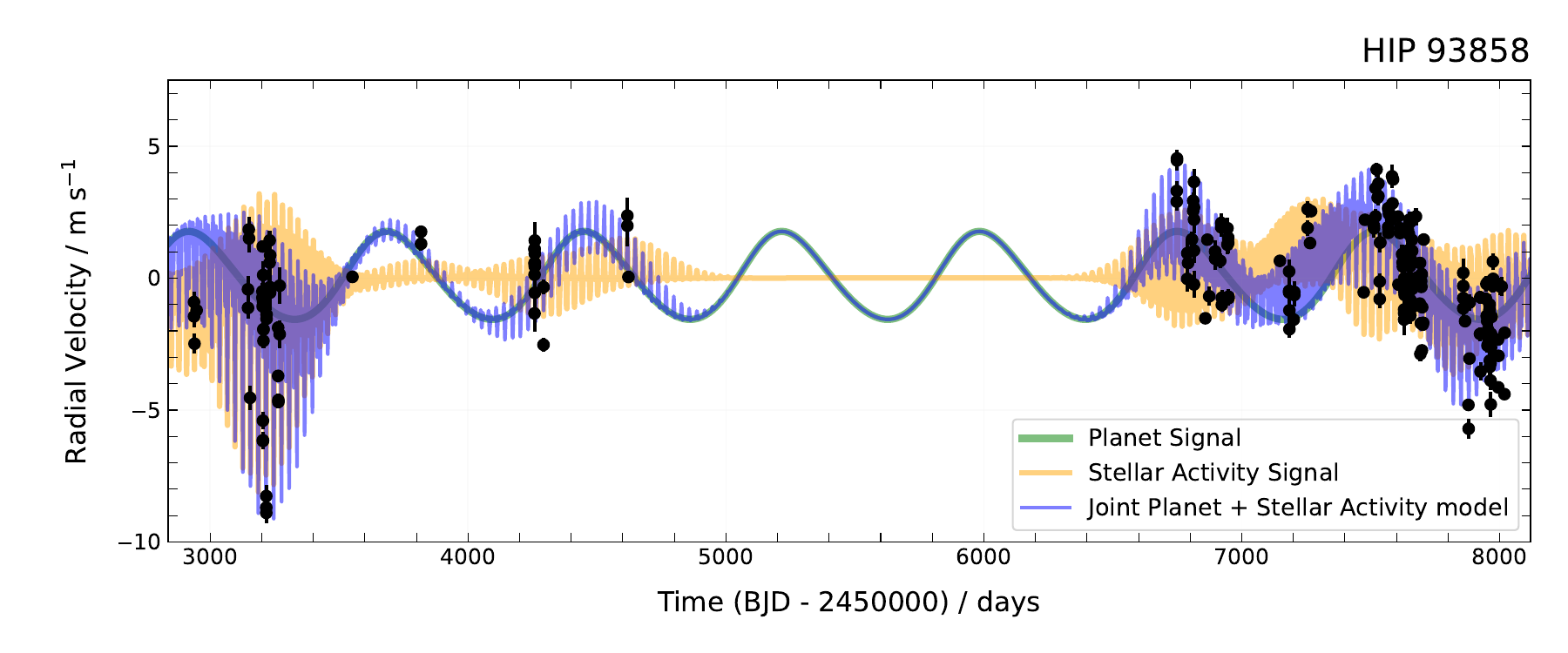}
    \caption{Radial velocity observations of HIP 53087 and HIP 93858, respectively, with the planet model (green curve; over-plotted in some regions by the combined model), the Gaussian Process model (yellow curve), and their combined model (blue curve).}
    \label{fig:phase_rv}
\end{figure*}

\FloatBarrier

\section{Chemical Figures and Abundances Tables}

\begin{figure*}[htbp]
    \centering
    \begin{tabular}{c}
        \includegraphics[width=\columnwidth]{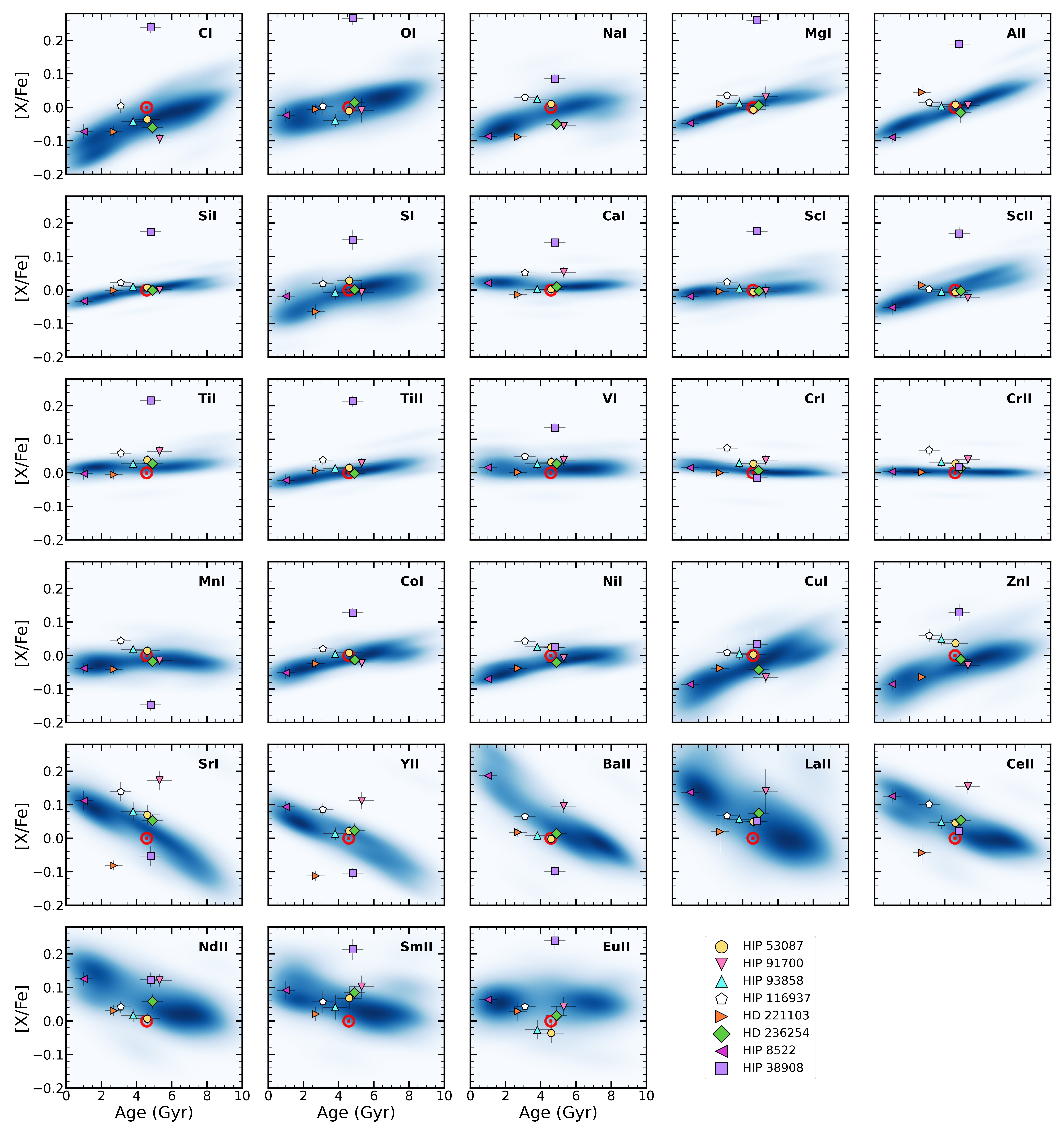}
    \end{tabular}
    \caption{GCE-corrected abundances as a function of isochronal age for the stars in our sample and the reported blue straggler HIP 38908 (square symbol). The blue density regions indicate the distribution of solar twins from \citet{Yana_Galarza:2021MNRAS.504.1873Y}. The Sun is represented in red by its standard symbol.}
    \label{fig:GCE-solartwins}
\end{figure*}

\FloatBarrier

\begin{table*}[htbp]
    \centering
	\caption{$^{a}$Differential chemical abundances for HD 236254 relative to the Sun. $^{b}$GCE-corrected abundances from \citet{Cowley:2022MNRAS.512.3684C}. The last column is the 50\% dust condensation temperature of elements \citep{Lodders:2003ApJ...591.1220L}. $^{c}$ Weighted average of \ion{Fe}{1} and \ion{Fe}{2}. Corresponding tables for the other stars in our sample are available online.}
	\begin{tabular}{lcccr}
		\hline
		\hline
		Element            &   Z   & HD 236254$^{a}$          &   HD 236254$^{b}$        & $T_{\rm{Cond}}$  \\
		                   &       &    $\Delta$[X/H] (dex)  &    $\Delta$[X/H] (dex)  &      (K)         \\
        \hline
        C               & 6     & $+0.020 \pm 0.044$   & $+0.039 \pm 0.044$   & 40    \\
        O               & 8     & $+0.034 \pm 0.019$   & $+0.043 \pm 0.019$   & 180   \\
        Na              & 11    & $+0.009 \pm 0.009$   & $+0.020 \pm 0.009$   & 958   \\
        Mg              & 12    & $+0.115 \pm 0.046$   & $+0.127 \pm 0.046$   & 1336  \\
        Al              & 13    & $+0.054 \pm 0.039$   & $+0.070 \pm 0.039$   & 1653  \\
        Si              & 14    & $+0.066 \pm 0.009$   & $+0.073 \pm 0.009$   & 1310  \\
        S               & 16    & $+0.056 \pm 0.016$   & $+0.067 \pm 0.016$   & 664   \\
        K               & 19    & $+0.001 \pm 0.030$   & $+0.001 \pm 0.030$   & 1006  \\
        Ca              & 20    & $+0.080 \pm 0.012$   & $+0.078 \pm 0.012$   & 1517  \\
        ScI             & 21    & $+0.063 \pm 0.009$   & $+0.070 \pm 0.008$   & 1659  \\
        ScII            & 21    & $+0.066 \pm 0.017$   & $+0.073 \pm 0.008$   & 1659  \\
        TiI             & 22    & $+0.097 \pm 0.013$   & $+0.101 \pm 0.008$   & 1741  \\
        TiII            & 22    & $+0.077 \pm 0.011$   & $+0.081 \pm 0.008$   & 1582  \\
        V               & 23    & $+0.101 \pm 0.015$   & $+0.102 \pm 0.015$   & 1429  \\
        CrI             & 24    & $+0.090 \pm 0.014$   & $+0.088 \pm 0.009$   & 1429  \\
        CrII            & 24    & $+0.070 \pm 0.011$   & $+0.068 \pm 0.009$   & 1296  \\
        Mn              & 25    & $+0.065 \pm 0.012$   & $+0.069 \pm 0.012$   & 1158  \\
        Fe$^{c}$        & 26    & $+0.061 \pm 0.010$   & $+0.061 \pm 0.010$   & 1334  \\
        Co              & 27    & $+0.061 \pm 0.012$   & $+0.070 \pm 0.012$   & 1352  \\
        Ni              & 28    & $+0.041 \pm 0.009$   & $+0.050 \pm 0.009$   & 1353  \\
        Cu              & 29    & $+0.028 \pm 0.020$   & $+0.047 \pm 0.020$   & 1037  \\
        Zn              & 30    & $+0.066 \pm 0.027$   & $+0.078 \pm 0.027$   & 726   \\
        Sr              & 38    & $+0.133 \pm 0.027$   & $+0.104 \pm 0.027$   & 1464  \\
        Y               & 39    & $+0.047 \pm 0.029$   & $+0.020 \pm 0.029$   & 1659  \\
        Ba              & 56    & $+0.095 \pm 0.011$   & $+0.058 \pm 0.011$   & 1455  \\
        La              & 57    & $+0.089 \pm 0.095$   & $+0.063 \pm 0.095$   & 1578  \\
        Ce              & 58    & $+0.130 \pm 0.013$   & $+0.105 \pm 0.013$   & 1478  \\
        Nd              & 60    & $+0.120 \pm 0.028$   & $+0.097 \pm 0.028$   & 1602  \\
        Sm              & 62    & $+0.159 \pm 0.025$   & $+0.149 \pm 0.025$   & 1590  \\
        Eu              & 63    & $+0.086 \pm 0.027$   & $+0.081 \pm 0.027$   & 1356  \\
		\hline
		\hline
	\end{tabular}
	\label{tab:chemical_abundances}
\end{table*}

\FloatBarrier

\section{Extension of the accretion models}

\begin{table*}[htbp]
\centering
\caption{Extension of the episodic accretion models from \citet{Baraffe:2017A&A...597A..19B}, now including Be abundances.}
\begin{tabular}{ccccccccc}
\hline
$M/M_\odot$ & $T_{\rm eff}$ & $L_\ast$ & $\log g$ & $R$ (cm) & $R/R_\odot$ & Li/Li$_0$ & Be/Be$_0$ & time (Gyr) \\
\hline
0.735 & 4683 & -0.566 & 4.506 & $5.517\times 10^{10}$ & 0.793 & 0.001 & 0.8500 & 0.03 \\
0.926 & 5047 & -0.353 & 4.523 & $6.071\times 10^{10}$ & 0.872 & 0.286 & 0.9930 & 0.03 \\
\hline
0.735 & 4362 & -0.801 & 4.617 & $4.852\times 10^{10}$ & 0.697 & 0.001 & 0.8500 & 1.0 \\
0.926 & 5053 & -0.316 & 4.488 & $6.321\times 10^{10}$ & 0.908 & 0.286 & 0.9930 & 1.0 \\
\hline
0.735 & 4400 & -0.766 & 4.597 & $4.967\times 10^{10}$ & 0.714 & 0.000 & 0.8490 & 5.0 \\
0.926 & 5145 & -0.226 & 4.430 & $6.759\times 10^{10}$ & 0.971 & 0.286 & 0.9930 & 5.0 \\
\hline
0.735 & 4422 & -0.747 & 4.587 & $5.022\times 10^{10}$ & 0.722 & 0.000 & 0.8480 & 7.0 \\
0.926 & 5193 & -0.171 & 4.390 & $7.071\times 10^{10}$ & 1.016 & 0.285 & 0.9930 & 7.0 \\
\hline
\end{tabular}
\label{tab:extensionmods}
\end{table*}

\FloatBarrier

\bibliography{sample701}{}
\bibliographystyle{aasjournalv7}



\end{document}